\documentclass[11pt,a4paper]{article}

\usepackage[T1]{fontenc}
\usepackage[utf8]{inputenc}
\usepackage{lmodern}
\usepackage{microtype}
\usepackage{geometry}
\usepackage{amsmath,amssymb,mathtools}
\usepackage{mathrsfs}
\usepackage{booktabs}
\usepackage{array}
\usepackage{tikz}
\usetikzlibrary{positioning}
\usepackage{xcolor}
\usepackage{hyperref}
\hypersetup{
  colorlinks=true,
  linkcolor=blue!55!black,
  citecolor=blue!55!black,
  urlcolor=blue!65!black,
  pdftitle={3d N=4 rank-zero mirror symmetry, TQFT interfaces, and Zagier duality of Nahm sums},
  pdfauthor={Yutaka Yoshida}
}
\numberwithin{equation}{section}

\title{\vspace*{1.8cm}\textbf{3d $\mathcal N=4$ rank-zero mirror symmetry, TQFT interfaces, and Zagier duality of Nahm sums}}
\author{Yutaka Yoshida\\[0.7em]
\small Department of Current Legal Studies, Faculty of Law, Meiji Gakuin University,\\
\small 1-2-37 Shirokanedai, Minato-ku, Tokyo 108-8636, Japan\\[0.45em]
\small Institute for Mathematical Informatics, Meiji Gakuin University,\\
\small 1518 Kamikurata-cho, Totsuka-ku, Yokohama 244-8539, Japan}
\date{}

\begin{document}
\maketitle

\begin{abstract}
Starting from the $(E_8,T_1)$ Nahm sum for the vacuum character of $T_{10}M_{\rm eff}(11,2)$, we construct a three-dimensional (3d) $\mathcal N=2$ $U(1)^8$ Chern--Simons matter theory $\mathcal T$ with level matrix $C_{E_8}$. Its $A$-twisted vacuum half-index reproduces the character exactly, while four distinguished Wilson loop half-indices agree with the remaining characters to the orders checked. Applying particle--vortex duality and gauging yields a second 3d $\mathcal N=2$ Chern--Simons matter theory $\mathcal T^\vee$ with level matrix $C_{E_8}^{-1}$. Exact superconformal-index matching supports the claim that the two theories flow to 3d $\mathcal N=4$ rank-zero mirror SCFTs. For the $(\mathcal D,D_c)$ boundary condition, the mirror map between the $A$-twisted Wilson loop half-indices of $\mathcal T$ and the $B$-twisted Wilson loop half-indices of $\mathcal T^\vee$ coincides componentwise with the Zagier transformation, realizing a Zagier duality between the complete $(E_8,T_1)$ and $(T_1,E_8)$ Nahm systems. Using the Bethe-equation formalism, we show that the corresponding $A/B$-twisted TQFT data agree up to orientation reversal and the $E_8$ framing phase $e^{-2\pi \mathrm{i}/3}$. The duality interface gives the torus bilinear
$Z_{\mathcal I}(\tau)=\sum_i\chi_i(\tau)\chi_i^\vee(\tau)=\chi_{(E_8)_1}(\tau)$,
where $\{\chi_i\}$ and $\{\chi_i^\vee\}$ are the corresponding five-component character bases. Thus Zagier duality is realized as part of a concrete 3d mirror and interface structure.
\end{abstract}

\clearpage
\tableofcontents

\section{Introduction}
\label{sec:introduction}

3d $\mathcal N=4$ mirror symmetry is usually recognized geometrically through the exchange of Coulomb and Higgs branches~\cite{Intriligator:1996ex,Hanany:1996ie,deBoer:1996mp}.  For rank-zero theories this diagnostic largely disappears: both branches are zero-dimensional, even though the infrared fixed point may remain strongly interacting~\cite{Gang:2018huc,Gang:2023rei,Gang:2021hrd}.  In this setting, supersymmetric partition functions, topological line operators, and the $A$- and $B$-twisted theories become especially useful probes of mirror symmetry.

An important source of such theories is the $R$-twisted reduction of four-dimensional $\mathcal N=2$ theories.  The underlying $R$-twisting and BPS-monodromy perspective has an earlier origin in~\cite{Cecotti:2010fi}.  Building on recent developments~\cite{Gaiotto:2024ioj,ArabiArdehali:2024ysy}, many 3d $\mathcal N=2$ Chern--Simons matter theories flowing to $\mathcal N=4$ rank-zero SCFTs have been constructed in this framework~\cite{ArabiArdehali:2024vli,NishinakaYoshida:2025generalizedAD,Go:2025ixu,Kim:2025rog}, while the same circle-reduction framework has recently been extended to theories with nontrivial Coulomb branches~\cite{Hamachika:2026whv}.  The relation to vertex operator algebras is rooted in the 4d SCFT/VOA correspondence~\cite{Beem:2013sza}, which associates a two-dimensional (2d) chiral algebra to a protected sector of a 4d $\mathcal N=2$ SCFT.  The 3d boundary chiral algebras relevant here were introduced and developed in~\cite{Costello:2018fnz,Costello:2018swh}.  Their relation to the 4d SCFT/VOA correspondence in the $R$-twisted reduction framework is part of the broader 4d/3d/2d picture discussed in~\cite{Dedushenko:2023cvd}.  In several related constructions, character or Nahm-sum expressions for boundary half-indices provide a direct guide to ultraviolet 3d Chern--Simons matter descriptions~\cite{GangKimParkStubbs:2024Nahm,KimSong:2024AD}.  In the present work, however, the mirror relation itself is constructed independently by particle--vortex duality and gauging; the comparison with the complete character systems and the Zagier transformation is performed only afterward.

The topological $A/B$-twists provide a complementary infrared description of rank-zero SCFTs.  The correspondence between rank-zero SCFTs and non-unitary TQFTs, together with the extraction of modular and rational-CFT data from supersymmetric observables, was developed in~\cite{Gang:2021hrd,Gang:2023rei}.  For ultraviolet $\mathcal N=2$ gauge-theory descriptions, the Bethe-vacuum and Seifert-manifold formalism gives a practical way to compute handle-gluing and fibering operators, Wilson loop observables, and the resulting TQFT data~\cite{Closset:2016arn,Closset:2017zgf,ClossetKimWillett:2018Seifert}.  Closely related rank-zero mirror phenomena were studied in~\cite{CreutzigGarnerKim:2024mirror}, where the mirror relation between $\mathcal T_{N,k}$ and the parity-reversed theory with $N$ and $k$ exchanged was proposed from the matching of superconformal indices and $A/B$-twisted TQFT data, leading in turn to level-rank dualities of the associated boundary VOAs.  In the construction developed here, the logical order is different: the mirror theory is obtained directly at the ultraviolet field-theory level by applying particle--vortex duality to the elementary chirals and then gauging.  The superconformal index and twisted TQFT data are therefore tests of a mirror pair that has already been constructed by a microscopic 3d duality.

More concretely, we start from the rank-eight $(E_8,T_1)$ Nahm system.  We first construct a 3d $\mathcal N=2$ Chern--Simons matter theory $\mathcal T$ by requiring its $(\mathcal D,D_c)$ half-index to reproduce the vacuum $(E_8,T_1)$ Nahm sum.  This requirement determines a theory with gauge group $U(1)^8$, eight charge-one chiral multiplets, and Chern--Simons matrix
\begin{equation}
  K=C_{E_8}.
  \label{eq:intro-CE8-theory}
\end{equation}
The linear Nahm data are compatible with a unique seven-term monopole superpotential of the type considered here.  After including this deformation, $F$-maximization and the superconformal index provide evidence that $\mathcal T$ flows to an interacting 3d $\mathcal N=4$ rank-zero SCFT, with supersymmetry enhanced from $\mathcal N=2$ in the ultraviolet to $\mathcal N=4$ at the infrared fixed point.  For the same boundary condition, inserting five distinguished Wilson loops produces five half-indices whose Nahm-sum expressions agree with the complete five-character system associated with $T_{10}M_{\rm eff}(11,2)$.

We then derive a second ultraviolet $\mathcal N=2$ Chern--Simons matter theory $\mathcal T^\vee$ directly by applying particle--vortex duality to the eight elementary chirals and gauging the original abelian flavor symmetries with the matrix $C_{E_8}$.  After folding the duality wall, the Schur complement generates the inverse matrix, giving a $U(1)^8$ theory with Chern--Simons matrix
\begin{equation}
  K^\vee=C_{E_8}^{-1}.
  \label{eq:intro-CE8-inverse-theory}
\end{equation}
The same transformation maps the seven dressed monopoles of $\mathcal T$ to the seven bare monopoles of $\mathcal T^\vee$, reverses the surviving topological fugacity, and transmits Wilson loop charges according to
\begin{equation}
  \boldsymbol Q^\vee=C_{E_8}^{-1}\boldsymbol Q.
  \label{eq:intro-Wilson-map}
\end{equation}
The two bulk superconformal indices obey the exact identity
\begin{equation}
  \mathcal I^\vee(q,\eta)=\mathcal I(q,\eta^{-1}),
  \label{eq:intro-SCI-mirror}
\end{equation}
and the two distinguished twisted limits are exchanged.  Together with the agreement of the extremized $R$-symmetry data, these results support the proposal that $\mathcal T$ and $\mathcal T^\vee$ flow to a 3d $\mathcal N=4$ rank-zero mirror pair.

The particle--vortex construction can also be followed directly at the boundary.  Starting from the elementary one-node particle--vortex wall~\cite{Dimofte:2017tpi} and gauging eight copies produces a duality-transformed boundary kernel in which the dual gauge fields are integrated with Neumann boundary conditions, while the original Dirichlet fluxes are packaged into an $E_8$ lattice wavefunction.  Up to the vacuum Casimir factor, this wavefunction is the Cartan-refined vacuum character of the level-one $E_8$ lattice VOA.  Its quasi-periodicity supplies precisely the anomaly needed for the transformed boundary condition, and Wilson insertions are carried through the kernel by the same inverse-Cartan map in \eqref{eq:intro-Wilson-map}.  Thus the $E_8$ lattice sector already appears in the ultraviolet boundary realization of the particle--vortex wall, before the topological interface analysis of Section~\ref{sec:E8-interface}.

Only after deriving the 3d mirror map and its boundary realization do we turn to the arithmetic structure of the two $A$- and $B$-twisted half-index systems.  Nahm sums arise naturally in the study of rational conformal-field-theory characters, and their modularity has been investigated through dilogarithm and Bloch-group identities~\cite{Nahm:2004bloch,Zagier:2007dilogarithm,VlasenkoZwegers:2011nahm}.  The Dynkin-diagram construction $\mathsf A(X,Y)=C(X)\otimes C(Y)^{-1}$ and its generalizations provide a broad class of modular Nahm systems~\cite{SunWang:2026Dynkin}, while their direct realization as half-indices of abelian Chern--Simons matter theories has recently been explored systematically at low rank~\cite{GangKimParkStubbs:2024Nahm}.

A particularly relevant arithmetic operation is Zagier's transformation of Nahm data~\cite{Zagier:2007dilogarithm}.  For a rank-$r$ Nahm triple $(\mathsf A,\mathsf B,\mathsf C)$, it is
\begin{equation}
  (\mathsf A,\mathsf B,\mathsf C)
  \longmapsto
  \left(
    \mathsf A^{-1},
    \mathsf A^{-1}\mathsf B,
    \frac12\mathsf B^{\mathsf T}\mathsf A^{-1}\mathsf B
    -\frac{r}{24}-\mathsf C
  \right).
  \label{eq:intro-Zagier-transform}
\end{equation}
We use \emph{Zagier transformation} for the map of a single Nahm triple in \eqref{eq:intro-Zagier-transform}.  In this paper, \emph{Zagier duality} means the stronger statement that this transformation acts componentwise on the complete five-sector Nahm system: every Nahm triple associated with a character of the $(E_8,T_1)$ system is mapped to the paired triple of the $(T_1,E_8)$ system, including the overall constant term.
For the five Wilson sectors of the present theory, the first two entries are exactly the transformations already produced by particle--vortex duality: $C_{E_8}\mapsto C_{E_8}^{-1}$ in the bulk and $\boldsymbol Q\mapsto C_{E_8}^{-1}\boldsymbol Q$ for the Wilson charges.  The scalar term contains the universal rank-eight contribution $-8/24=-1/3$ and the charge-dependent quadratic shift.  The dual half-indices thereby reproduce the five-component $M_{\rm eff}(2,11)$ system with the $(T_1,E_8)$ quadratic form.  Applying the same transformation to all five sectors gives the Zagier duality between the two complete Nahm systems.  This duality is not an assumption used to construct the mirror pair; it is the arithmetic packaging of a 3d transformation that can be derived independently.

The $A/B$-twists sharpen this correspondence.  The $A$-twisted Bethe equations of $\mathcal T$ and the $B$-twisted Bethe equations of $\mathcal T^\vee$ are related by the exact particle--vortex complement map
\begin{equation}
  \boldsymbol y=\boldsymbol 1-\boldsymbol x.
  \label{eq:intro-Bethe-complement}
\end{equation}
Both theories have five regular Bethe vacua.  Their Wilson eigenvalues, handle-gluing operators, and fibering operators are carried into one another by the complement map, and the resulting based Wilson loop algebra is the even-label $SU(2)_9$ fusion ring, equivalently the $SO(3)_9$ fusion ring.  The modular $S$ data agree after pairing the five sectors, while the fibering operators are inverse up to the universal multiplier
\begin{equation}
  \kappa_{E_8}=e^{-2\pi \mathrm{i}/3}=e^{-2\pi \mathrm{i}\,8/24}.
  \label{eq:intro-E8-framing}
\end{equation}
Consequently, for the oriented degree-$p$ circle bundle $\mathcal M_{g,p}$ over a closed genus-$g$ Riemann surface, the twisted partition functions satisfy
\begin{align}
  Z_{\mathcal T^\vee}^{B}(\mathcal M_{g,p})
  &=e^{-2\pi \mathrm{i}p/3}
    Z_{\mathcal T}^{A}(\mathcal M_{g,-p}),
  \nonumber\\
  Z_{\mathcal T^\vee}^{A}(\mathcal M_{g,p})
  &=e^{-2\pi \mathrm{i}p/3}
    Z_{\mathcal T}^{B}(\mathcal M_{g,-p}).
  \label{eq:intro-Seifert-mirror}
\end{align}
The replacement $p\mapsto -p$ makes the orientation reversal explicit, while the remaining multiplier is the framed response of the invertible bosonic $E_8$ phase with chiral central charge $c_-=8$.  The extracted fusion and modular data admit a natural standard completion by the $m=5,6$ Galois-conjugate $SO(3)_9$ Reshetikhin--Turaev TQFTs, again related by orientation reversal.  This completion is additional mathematical input rather than a unique reconstruction from localization; its full categorical treatment is deferred to a forthcoming companion paper.

The particle--vortex duality has a particularly simple realization on the torus state spaces of the $A$- and $B$-twisted theories.  In the paired Bethe-idempotent basis the interface acts as the identity, while in the Wilson loop basis it is the permutation matrix $P_\sigma$ implementing the five transmitted sectors.  The closed Wilson loops used to prepare solid-torus states and the lines transmitted across the interface are different geometric realizations of the same topological line sectors.  Consequently the line-transmission rule induces a bilinear pairing of the corresponding character wavefunctions.  The resulting torus interface amplitude is
\begin{align}
  Z_{\mathcal I}(\tau)
  &=\boldsymbol\chi_A(\tau)^{\mathsf T}
   P_\sigma\,
   \boldsymbol\chi_B^\vee(\tau)
   \nonumber\\
  &=\chi_0\chi_0^\vee
   +\chi_{5/11}\chi_{6/11}^\vee
   +\chi_{8/11}\chi_{3/11}^\vee
   +\chi_{10/11}\chi_{1/11}^\vee
   +\chi_{12/11}\chi_{10/11}^\vee.
  \label{eq:intro-interface-amplitude}
\end{align}
The five pairings in \eqref{eq:intro-interface-amplitude} are not imposed from the two character systems.  They are determined by the transmission of the distinguished Wilson loop sectors through the particle--vortex duality interface.  The duality-interface construction therefore derives the particle--vortex pairing between the $A$-twisted Wilson loop half-indices of $\mathcal T$ and the $B$-twisted Wilson loop half-indices of $\mathcal T^\vee$.  These half-indices are identified with the five sectors of $T_{10}M_{\mathrm{eff}}(11,2)$ and $M_{\mathrm{eff}}(2,11)$, respectively, and the resulting pairing agrees precisely with the sector pairing of the Zagier duality between the two complete Nahm systems.

The interface amplitude transforms under the modular group according to the twisted $S$ and $T$ data: it is $S$ invariant and carries the $T$ multiplier $e^{-2\pi\mathrm{i}/3}$.  It follows that $\eta(\tau)^8 Z_{\mathcal I}(\tau)$ is a holomorphic modular form of weight four with unit vacuum multiplicity, and hence
\begin{equation}
  Z_{\mathcal I}(\tau)
  =\frac{E_4(\tau)}{\eta(\tau)^8}
  =\chi_{(E_8)_1}(\tau).
  \label{eq:intro-E8-character}
\end{equation}
This is an exact statement about the torus interface amplitude.  It does not by itself identify the microscopic local interface VOA with $V_{E_8,1}$; determining that chiral algebra remains an open problem.

The resulting picture ties together three appearances of $E_8$ that arise at different stages of the analysis.  The Cartan matrix $C_{E_8}$ first enters as the quadratic form of the $(E_8,T_1)$ Nahm sum and hence as the gauging matrix of $\mathcal T$.  The particle--vortex transformation produces an $E_8$ lattice sector in the transformed boundary kernel and the inverse lattice $C_{E_8}^{-1}$ in the mirror bulk theory.  Finally, the $A/B$-twisted interface retains the framing phase of the invertible $E_8$ theory and organizes the five character sectors into the $(E_8)_1$ torus amplitude.  Thus the bulk inverse-Cartan map, the boundary $E_8$ lattice sector, the $A/B$-twisted orientation reversal, and the $(E_8)_1$ torus amplitude arise as different manifestations of the same particle--vortex duality construction.  What remains unexplained is why these independently derived 3d structures assemble into the rigid componentwise Zagier duality of the two complete five-component Nahm systems.

The paper is organized as follows.  Section~\ref{sec:CE8-theory} constructs $\mathcal T$ by interpreting the vacuum $(E_8,T_1)$ Nahm sum as a half-index, determines the monopole superpotential and infrared $R$-symmetry, and studies the superconformal index and Wilson loop half-indices.  Section~\ref{sec:particle-vortex-mirror} constructs $\mathcal T^\vee$ from particle--vortex duality, derives the bulk and line maps, identifies the Zagier duality of the two five-component Nahm systems, and constructs the duality-transformed boundary kernel with its $(E_8)_1$ lattice sector.  Section~\ref{sec:AB-twisted-TQFTs} extracts the $A/B$-twisted Bethe--Seifert data, the $SO(3)_9$ fusion structure, the relative $E_8$ framing phase, and comments on a natural standard ribbon completion of the non-invertible sector.  Section~\ref{sec:E8-interface} studies the duality interface on the torus state space, derives the exact $(E_8)_1$ torus interface amplitude, and checks compatibility with the monopole deformations.  Section~\ref{sec:summary-outlooks} summarizes the results and discusses the microscopic wall, boundary and interface chiral algebras, the full extended TQFT, generalizations to other rank-zero mirror pairs, and the mathematical origin of Zagier duality.

\section{$C_{E_8}$ Chern--Simons matter theory from the $(E_8,T_1)$ Nahm sum}
\label{sec:CE8-theory}

\subsection{From the $(E_8,T_1)$ Nahm sum to the $U(1)^8$ theory}
\label{subsec:from-E8T1-to-3d}

We begin with the Virasoro minimal model $M(11,2)$.  Its central charge is
\begin{equation}
  c[M(11,2)]
  =1-\frac{6(11-2)^2}{11\cdot 2}
  =-\frac{232}{11}.
\end{equation}
Among its five irreducible representations, the smallest conformal weight is
$h_{\min}=-10/11$.  We use the effective normalization
\begin{equation}
  c_{\mathrm{eff}}=c-24h_{\min}=\frac{8}{11},
  \qquad
  h_a^{\mathrm{eff}}=h_a-h_{\min},
  \label{eq:Meff-definition}
\end{equation}
and denote the resulting five-component character system by
$M_{\mathrm{eff}}(11,2)$.  We then consider its $T_{10}$ Hecke transform,
which gives another five-component character system with effective central charge
\begin{equation}
  c_{\mathrm{eff}}\!\left[T_{10}M_{\mathrm{eff}}(11,2)\right]
  =10\,c_{\mathrm{eff}}
  =\frac{80}{11}.
\end{equation}
Our starting point is the vacuum character of this Hecke-transformed system,
$T_{10}M_{\mathrm{eff}}(11,2)$,
\begin{equation}
  \chi_0(q)
  =q^{-10/33}
  \left(
    1+120q+1660q^2+12320q^3+\cdots
  \right).
  \label{eq:T10-vacuum-expansion}
\end{equation}
The $T_{10}M_{\mathrm{eff}}(11,2)$ system and its Hecke realization were studied in
\cite{DuanLeeSun:2022Hecke}.

The same vacuum character admits the $(E_8,T_1)$ Nahm-sum representation
\begin{equation}
  \chi_0(q)
  =q^{-10/33}
  \sum_{\boldsymbol m\in\mathbb Z_{\geq0}^{8}}
  \frac{
    q^{\frac12\boldsymbol m^{\mathsf T}C_{E_8}\boldsymbol m}
  }{
    \prod_{i=1}^{8}(q)_{m_i}
  },
  \qquad
  (q)_n:=\prod_{j=1}^{n}(1-q^j),
  \label{eq:E8T1-vacuum-Nahm}
\end{equation}
with $(q)_0=1$.  Here the notation $(E_8,T_1)$ refers to the ordered pair of Dynkin diagrams that determines the Nahm quadratic form.  For a Dynkin diagram $G$, we denote its Cartan matrix by $C(G)$ and write
\begin{equation}
  C_{E_8}:=C(E_8).
\end{equation}
More generally, the quadratic matrix associated with an ordered pair $(G,G')$ is
\begin{equation}
  \mathsf A(G,G')=C(G)\otimes C(G')^{-1}.
\end{equation}
For the tadpole diagram $T_1=A_2/\mathbb Z_2$, the corresponding one-dimensional Cartan-type matrix is $C(T_1)=(1)$.  Hence
\begin{equation}
  \mathsf A(E_8,T_1)
  =C(E_8)\otimes C(T_1)^{-1}
  =C_{E_8}.
  \label{eq:E8T1-quadratic-form}
\end{equation}
The identification in \eqref{eq:E8T1-vacuum-Nahm} is part of the generalized Dynkin-diagram Nahm-sum construction of
\cite{SunWang:2026Dynkin}.

We now use \eqref{eq:E8T1-vacuum-Nahm} in the opposite direction: rather than starting from a 3d theory and computing its half-index, we regard the Nahm sum as a specialization of a 3d half-index and read off an ultraviolet Lagrangian.  We first fix the boundary-condition notation used throughout the paper.  We denote by $\mathcal D$ the Dirichlet boundary condition for a 3d vector multiplet and by $D_c$ the deformed Dirichlet boundary condition for a charge-one chiral multiplet.  We refer to the combined choice, with $\mathcal D$ imposed on every vector multiplet and $D_c$ on every chiral multiplet, as the $(\mathcal D,D_c)$ boundary condition.

For a $U(1)^r$ gauge theory with Chern--Simons matrix $K$ and one charge-one chiral multiplet for each gauge factor, we keep fugacities $x_i$ for the ultraviolet topological symmetry $U(1)_{\mathrm{top}}^r$.  In the conventions relevant here, the half-index of the $(\mathcal D,D_c)$ boundary condition~\cite{Dimofte:2017tpi} has the schematic form
\begin{equation}
  I\!\!I_{(\mathcal D,D_c)}(q,\boldsymbol x)
  =
  \sum_{\boldsymbol m\in\mathbb Z_{\geq0}^{r}}
  \frac{
    q^{\frac12\boldsymbol m^{\mathsf T}K\boldsymbol m}
  }{
    \prod_{i=1}^{r}(q)_{m_i}
  }
  \prod_{i=1}^{r}x_i^{-m_i}.
  \label{eq:generic-Nahm-half-index}
\end{equation}
Here $m_i$ is the magnetic charge of the $i$-th gauge factor, and $x_i$ is the corresponding topological fugacity in our boundary convention.  A linear term $\boldsymbol B^{\mathsf T}\boldsymbol m$ in a Nahm sum can therefore be implemented by a specialization of the $x_i$.  More generally, the fugacities need not remain independent after a monopole deformation: a monopole superpotential breaks a subspace of $U(1)_{\mathrm{top}}^r$, while an unbroken topological symmetry may mix with the $R$-symmetry.  The character-like half-index is then obtained by a specialization compatible with this symmetry breaking and mixing, as in related constructions of rank-zero theories~\cite{Gaiotto:2024ioj}.

In \eqref{eq:generic-Nahm-half-index}, the rank of the sum is the gauge rank, the quadratic form is the integral Chern--Simons gauging matrix, and each factor $(q)_{m_i}^{-1}$ is supplied by a charge-one chiral multiplet.  This half-index/Nahm-sum dictionary is the one developed for 3d rank-zero theories in \cite{GangKimParkStubbs:2024Nahm}.

Comparison of \eqref{eq:E8T1-vacuum-Nahm} with the unrefined specialization of \eqref{eq:generic-Nahm-half-index} therefore fixes the ultraviolet gauge and matter content:
\begin{equation}
  \mathcal T:\qquad
  G=U(1)^8,
  \qquad
  K=C_{E_8},
  \qquad
  \text{matter}=\{\Phi_i\}_{i=1}^{8},
  \qquad
  Q_j(\Phi_i)=\delta_{ij}.
  \label{eq:CE8-UV-theory}
\end{equation}
Thus the $i$-th chiral multiplet has charge $+1$ under $U(1)_i$ and is neutral under the other seven gauge factors.  In particular, the identity matter-charge matrix is not an additional assumption: it is read directly from the eight independent Pochhammer denominators in \eqref{eq:E8T1-vacuum-Nahm}.  Each charge-one chiral multiplet, together with its canonical parity-anomaly contact term, may equivalently be regarded as a copy of the tetrahedron theory $\mathcal T_\Delta$.

We use the following labeling of the $E_8$ Dynkin diagram:
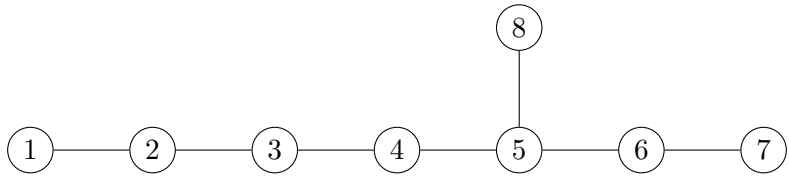
\begin{center}
\begin{tikzpicture}[
  every node/.style={circle,draw,minimum size=6mm,inner sep=0pt},
  node distance=10mm
]
  \node (1) {$1$};
  \node[right=of 1] (2) {$2$};
  \node[right=of 2] (3) {$3$};
  \node[right=of 3] (4) {$4$};
  \node[right=of 4] (5) {$5$};
  \node[right=of 5] (6) {$6$};
  \node[right=of 6] (7) {$7$};
  \node[above=of 5] (8) {$8$};
  \draw (1)--(2)--(3)--(4)--(5)--(6)--(7);
  \draw (5)--(8);
\end{tikzpicture}
\end{center}
We denote by $E(E_8)$ the edge set of this numbered Dynkin graph.
In the corresponding simple-root basis,
\begin{equation}
C_{E_8}=
\begin{pmatrix}
2&-1&0&0&0&0&0&0\\
-1&2&-1&0&0&0&0&0\\
0&-1&2&-1&0&0&0&0\\
0&0&-1&2&-1&0&0&0\\
0&0&0&-1&2&-1&0&-1\\
0&0&0&0&-1&2&-1&0\\
0&0&0&0&0&-1&2&0\\
0&0&0&0&-1&0&0&2
\end{pmatrix}.
\label{eq:E8-Cartan}
\end{equation}
The matrix is symmetric and positive definite, with $\det C_{E_8}=1$, and its quadratic form is even.

There is a minor but important convention concerning parity anomaly.  We use the $\mathcal T_\Delta$ convention in which a charge-one chiral is accompanied by the background contact term $k_{FF}=-1/2$.  After gauging the flavor symmetry of each of the eight $\mathcal T_\Delta$ blocks, these terms become diagonal gauge contact terms.  Accordingly, $K=C_{E_8}$ in \eqref{eq:CE8-UV-theory} denotes the \emph{integral gauging matrix} that appears in the Nahm quadratic form, whereas including the parity-anomaly contribution of the eight chiral multiplets gives the ultraviolet gauge contact-term matrix
\begin{equation}
  K_{\mathrm{UV}}
  =C_{E_8}-\frac12 I_8.
  \label{eq:CE8-UV-level}
\end{equation}
Keeping these two matrices distinct avoids double counting the parity-anomaly contribution in the monopole analysis of the next subsection.

For later use, it is helpful to state precisely the status of the boundary interpretation.  The series in \eqref{eq:E8T1-vacuum-Nahm} is a well-defined ultraviolet half-index for the $(\mathcal D,D_c)$ boundary condition of the eight $\mathcal T_\Delta$ blocks.  At this point we do \emph{not} assume that this boundary condition survives the monopole deformation or that its specialization is already the infrared $A$-twisted boundary condition.  The monopole superpotential will be determined in Section~2.2 and the superconformal $R$-symmetry in Section~2.3.  Only after those steps will the relevant twisted specialization be identified.  This ordering avoids using the infrared $R$-symmetry as an input in reconstructing the ultraviolet theory.

We have therefore reconstructed, directly from the vacuum $(E_8,T_1)$ Nahm sum, a rank-eight abelian Chern--Simons matter theory with one charge-one chiral multiplet at each node.  The next step is to determine a monopole superpotential compatible with the topological-fugacity specialization required by the vacuum Nahm series.  We will see that this requirement leaves a single topological $U(1)$ and that, after $F$-maximization, the corresponding twisted specialization reproduces \eqref{eq:E8T1-vacuum-Nahm}.

\subsection{The monopole superpotential and its uniqueness}
\label{subsec:monopole-superpotential}

We now determine a monopole deformation of the ultraviolet theory reconstructed in Section~2.1.  The relevant monopole operators are strongly constrained by the $E_8$ Cartan matrix.

Let $V_{\boldsymbol m}$ denote a bare monopole operator of GNO charge
\begin{equation}
  \boldsymbol m=(m_1,\ldots,m_8)\in\mathbb Z^8,
  \qquad
  m_i=\frac{1}{2\pi}\int_{S^2}F_i,
  \label{eq:monopole-flux}
\end{equation}
and let $\phi_i$ be the scalar component of the chiral multiplet $\Phi_i$.  A scalar dressed monopole has the form
\begin{equation}
  \mathcal O_{\boldsymbol n,\boldsymbol m}
  =\prod_{i=1}^{8}\phi_i^{n_i}V_{\boldsymbol m},
  \qquad
  \boldsymbol n=(n_1,\ldots,n_8)\in\mathbb Z_{\geq0}^{8}.
  \label{eq:dressed-monopole}
\end{equation}
Our goal is to identify the gauge-invariant scalar half-BPS monopole operators that can appear in a monopole superpotential.  We will classify all dressed monopoles of the form \eqref{eq:dressed-monopole}, determine those with reference $R$-charge $R_*=2$, and show that requiring the superpotential to preserve precisely one topological $U(1)$ uniquely fixes the set of monopole monomials that must appear.  A similar strategy of classifying admissible monopole superpotentials and analyzing their uniqueness was employed in \cite{Hamachika:2026whv}.  In the present theory, this analysis will also provide the monopole deformation compatible with the topological-fugacity specialization discussed in Section~2.1.

There is no nonconstant gauge-invariant polynomial built only from the $\phi_i$, since the matter-charge matrix is the identity.  We use the reference $R$-symmetry of the tetrahedron-theory convention,
\begin{equation}
  R_*(\phi_i)=0.
  \label{eq:Rstar-chiral}
\end{equation}
With the parity-anomaly convention of Section~2.1, gauge invariance and the scalar half-BPS condition for \eqref{eq:dressed-monopole} are
\begin{equation}
  n_i+\sum_{j=1}^{8}(C_{E_8})_{ij}m_j
  -\frac12\bigl(|m_i|+m_i\bigr)=0,
  \qquad
  n_i m_i=0,
  \qquad i=1,\ldots,8.
  \label{eq:monopole-conditions}
\end{equation}
The first equation is the gauge-invariance condition for the dressed monopole, including the electric charge induced by the Chern--Simons couplings, the parity-anomaly contact term, and the electric charge of the chiral dressing.  The second equation is the BPS condition that the chiral scalar at a given node cannot dress a monopole carrying nonzero flux at the same node.  These are the monopole conventions used in the Nahm-sum construction cited in Section~2.1.

Define the positive part of the magnetic charge componentwise by
\begin{equation}
  (\boldsymbol m_+)_i:=\max(m_i,0).
  \label{eq:positive-part}
\end{equation}
Then the first equation in \eqref{eq:monopole-conditions} determines the dressing uniquely:
\begin{equation}
  \boldsymbol n=\boldsymbol m_+-C_{E_8}\boldsymbol m.
  \label{eq:dressing-vector}
\end{equation}
Since $R_*(\phi_i)=0$, the reference $R$-charge is
\begin{equation}
  R_*(\mathcal O_{\boldsymbol n,\boldsymbol m})
  =\frac12\sum_{i=1}^{8}\bigl(m_i+|m_i|\bigr)
  =\sum_{m_i>0}m_i.
  \label{eq:monopole-Rstar}
\end{equation}
Taking the scalar product of \eqref{eq:dressing-vector} with $\boldsymbol m$ and using $n_i m_i=0$ gives
\begin{equation}
  \boldsymbol m^{\mathsf T}C_{E_8}\boldsymbol m
  =\sum_{m_i>0}m_i^2.
  \label{eq:monopole-norm-identity}
\end{equation}

The equations above admit a simple classification in terms of the $E_8$ graph.  First, negative magnetic flux is impossible.  Let
\begin{equation}
  N:=\{i\mid m_i<0\},
  \qquad
  u_i:=-m_i>0\quad(i\in N),
\end{equation}
and let $C_N$ be the principal submatrix of $C_{E_8}$ indexed by $N$.  If $i\in N$, then $n_i=0$, and \eqref{eq:dressing-vector} implies $(C_{E_8}\boldsymbol m)_i=0$.  Since all components outside $N$ are nonnegative, this gives
\begin{equation}
  (C_N\boldsymbol u)_i
  =-\sum_{\substack{j\notin N\\ j\sim i}}m_j
  \leq0,
  \label{eq:negative-support}
\end{equation}
where $i\sim j$ means that $i$ and $j$ are adjacent in the $E_8$ Dynkin graph.  Every principal submatrix of the positive-definite matrix $C_{E_8}$ is positive definite.  Multiplying \eqref{eq:negative-support} by $u_i>0$ and summing over $i\in N$ would therefore imply
\begin{equation}
  0<\boldsymbol u^{\mathsf T}C_N\boldsymbol u\leq0
\end{equation}
if $N$ were nonempty.  Hence
\begin{equation}
  m_i\geq0
  \qquad(i=1,\ldots,8).
  \label{eq:nonnegative-flux}
\end{equation}

Let $S:=\{i\mid m_i>0\}$ be the support of a nonzero magnetic charge.  At every $i\in S$, the BPS condition gives $n_i=0$, and hence
\begin{equation}
  (C_{E_8}\boldsymbol m)_i=m_i.
\end{equation}
Writing $C_{E_8}=2I_8-A(E_8)$, where $A(E_8)$ is the adjacency matrix of the Dynkin graph, this becomes
\begin{equation}
  m_i=\sum_{\substack{j\in S\\j\sim i}}m_j.
  \label{eq:support-equation}
\end{equation}
Consider a connected component $\Gamma$ of the subgraph induced by $S$, and let $\deg_\Gamma(i)$ be the number of neighbors of $i$ inside $\Gamma$.  Summing \eqref{eq:support-equation} over $\Gamma$ gives
\begin{equation}
  \sum_{i\in\Gamma}\bigl(\deg_\Gamma(i)-1\bigr)m_i=0.
  \label{eq:degree-sum}
\end{equation}
An isolated vertex is incompatible with \eqref{eq:support-equation}, so every vertex has degree at least one.  Since all $m_i$ in $\Gamma$ are positive, \eqref{eq:degree-sum} then forces $\deg_\Gamma(i)=1$ at every vertex.  Thus each connected component of the support is a single edge, and \eqref{eq:support-equation} requires equal fluxes at its two endpoints.

It follows that every gauge-invariant dressed-monopole flux is uniquely represented by a weighted induced matching $M$ of the $E_8$ Dynkin graph:
\begin{equation}
  \boldsymbol m
  =\sum_{e=\langle i,j\rangle\in M}
  k_e(\boldsymbol e_i+\boldsymbol e_j),
  \qquad
  k_e\in\mathbb Z_{>0}.
  \label{eq:weighted-induced-matching}
\end{equation}
Here $\boldsymbol e_i$ is the $i$-th standard basis vector of $\mathbb Z^8$, and an induced matching means that the subgraph spanned by the endpoints of the selected edges is precisely their disjoint union.  Conversely, every flux of the form \eqref{eq:weighted-induced-matching} satisfies \eqref{eq:monopole-conditions}, with its dressing fixed by \eqref{eq:dressing-vector}.  Thus \eqref{eq:weighted-induced-matching} is a complete classification of the ultraviolet scalar dressed-monopole chiral operators.

For such an operator, \eqref{eq:monopole-Rstar} and \eqref{eq:monopole-norm-identity} reduce to
\begin{equation}
  R_*(\mathcal O_{\boldsymbol n,\boldsymbol m})
  =2\sum_{e\in M}k_e,
  \qquad
  \boldsymbol m^{\mathsf T}C_{E_8}\boldsymbol m
  =2\sum_{e\in M}k_e^2.
  \label{eq:matching-charges}
\end{equation}
In particular, every nontrivial dressed-monopole chiral has even reference $R$-charge at least two.  The operators with $R_*=2$ arise from a single edge with unit weight,
\begin{equation}
  \boldsymbol m=\boldsymbol e_i+\boldsymbol e_j,
  \qquad
  \langle i,j\rangle\in E(E_8).
  \label{eq:edge-flux}
\end{equation}
For the node ordering in Section~2.1, the seven operators are
\begin{align}
  \mathcal O_{12}&=\phi_3V_{(1,1,0,0,0,0,0,0)},
  &
  \mathcal O_{23}&=\phi_1\phi_4V_{(0,1,1,0,0,0,0,0)},\nonumber\\
  \mathcal O_{34}&=\phi_2\phi_5V_{(0,0,1,1,0,0,0,0)},
  &
  \mathcal O_{45}&=\phi_3\phi_6\phi_8V_{(0,0,0,1,1,0,0,0)},\nonumber\\
  \mathcal O_{56}&=\phi_4\phi_7\phi_8V_{(0,0,0,0,1,1,0,0)},
  &
  \mathcal O_{67}&=\phi_5V_{(0,0,0,0,0,1,1,0)},\nonumber\\
  \mathcal O_{58}&=\phi_4\phi_6V_{(0,0,0,0,1,0,0,1)}.
  \label{eq:seven-monopoles}
\end{align}
The subscripts indicate the corresponding edge of the numbered $E_8$ diagram.  Their seven magnetic charges $\boldsymbol e_i+\boldsymbol e_j$ are linearly independent: successively removing leaves of the $E_8$ tree forces the coefficient of each edge to vanish in any linear relation.

We therefore introduce the monopole superpotential
\begin{equation}
  W
  =\sum_{\langle i,j\rangle\in E(E_8)}
  \lambda_{ij}\mathcal O_{ij},
  \qquad
  \lambda_{ij}\neq0.
  \label{eq:monopole-superpotential}
\end{equation}
The sense in which its support is unique is now immediate.  A superpotential built only from gauge-invariant $R_*=2$ monopoles leaves a topological symmetry of dimension $8-r$, where $r$ is the rank of the magnetic charges appearing in its support.  Requiring exactly one topological $U(1)$ therefore requires $r=7$.  Since \eqref{eq:seven-monopoles} is the complete set of nontrivial $R_*=2$ dressed monopoles and their seven magnetic charges already have rank seven, all seven terms must be present.  Thus the set of superpotential monomials in \eqref{eq:monopole-superpotential} is uniquely fixed by these requirements, although the nonzero coefficients $\lambda_{ij}$ are not.

Finally, we determine the symmetry left unbroken by \eqref{eq:monopole-superpotential}.  Let $T_i$ denote the topological symmetry of the $i$-th gauge factor, with current $j_i=(2\pi)^{-1}\star F_i$, and collect these generators into $\boldsymbol T=(T_1,\ldots,T_8)$.  We then write
\begin{equation}
  \mathsf T=\boldsymbol a\cdot\boldsymbol T.
\end{equation}
A monopole of charge $\boldsymbol e_i+\boldsymbol e_j$ is neutral under $\mathsf T$ precisely when
\begin{equation}
  a_i+a_j=0
  \qquad
  \text{for every edge }\langle i,j\rangle\in E(E_8).
  \label{eq:bipartite-condition}
\end{equation}
Because the $E_8$ graph is connected and bipartite, the solution space is one-dimensional.  With our node ordering, a primitive generator is
\begin{equation}
  \boldsymbol a=(1,-1,1,-1,1,-1,1,-1).
  \label{eq:residual-topological-vector}
\end{equation}
Thus the superpotential breaks seven combinations of the original topological symmetries and preserves a single $U(1)_{\mathsf T}$.

A general trial mixing compatible with the superpotential is consequently
\begin{equation}
  R_\lambda
  =R_*+\lambda\,\mathsf T
  =R_*+\lambda\,\boldsymbol a\cdot\boldsymbol T,
  \qquad \lambda\in\mathbb R.
  \label{eq:one-dimensional-mixing}
\end{equation}
This is the unique one-dimensional linear topological grading left after the monopole deformation.  In the half-index variables of Section~2.1, the mixing in \eqref{eq:one-dimensional-mixing} produces the corresponding linear factor in the $q$-grading.  The value of $\lambda$ is not fixed by the ultraviolet monopole analysis; it will be determined by $F$-maximization in Section~2.3.

The weighted-induced-matching classification also shows that every ultraviolet dressed-monopole chiral is neutral under the surviving $U(1)_{\mathsf T}$.  Indeed, $a_i+a_j=0$ on every edge, and hence $\boldsymbol a\cdot\boldsymbol m=0$ for every flux of the form \eqref{eq:weighted-induced-matching}.  Therefore
\begin{equation}
  R_\lambda(\mathcal O_{\boldsymbol n,\boldsymbol m})
  =R_*(\mathcal O_{\boldsymbol n,\boldsymbol m})
  =2\sum_{e\in M}k_e.
  \label{eq:monopole-R-on-mixing-line}
\end{equation}
In particular, there is no gauge-invariant scalar dressed-monopole chiral operator with
\begin{equation}
  0<R_\lambda<2
  \label{eq:no-low-R-monopole}
\end{equation}
anywhere along the allowed mixing line.  This statement concerns the ultraviolet dressed-monopole chiral ring; it does not by itself exclude an accidental sector that could emerge along the renormalization-group flow.

As will be confirmed after $F$-maximization in the next subsection, the corresponding $A$-twisted specialization reproduces the vacuum $(E_8,T_1)$ Nahm sum.

\subsection{$F$-maximization and the infrared $R$-symmetry}
\label{subsec:Fmax}

The monopole superpotential has reduced the ultraviolet topological symmetry to the single combination $U(1)_{\mathsf T}$ generated by \eqref{eq:residual-topological-vector}.  We now determine the candidate infrared superconformal $R$-symmetry along the allowed mixing line \eqref{eq:one-dimensional-mixing}.  This step fixes the mixing within the abelian symmetry space visible in the ultraviolet theory.  It does not by itself exclude an accidental infrared symmetry or prove supersymmetry enhancement.

In the absence of accidental abelian symmetries, the superconformal $R$-symmetry locally maximizes the three-sphere free energy~\cite{Jafferis:2010un},
\begin{equation}
  F=-\log\left|Z_{S^3}\right|.
  \label{eq:F-def}
\end{equation}
Restricting to the one-parameter family allowed by the monopole superpotential,
\begin{equation}
  R_\lambda=R_*+\lambda\,\mathsf T,
  \qquad
  \boldsymbol\mu(\lambda)=\lambda\boldsymbol a,
  \qquad \lambda\in\mathbb R,
  \label{eq:Fmax-family}
\end{equation}
we define
\begin{equation}
  F(\lambda):=-\log\left|Z_{S^3}(\boldsymbol\mu=\lambda\boldsymbol a)\right|.
  \label{eq:F-lambda}
\end{equation}
For the numerical evaluation we use the localized Seifert/Bethe-vacuum representation of the round-sphere partition function~\cite{Closset:2017zgf},
\begin{equation}
  Z_{S^3}(\lambda)
  =\sum_{\widehat u\in\mathcal S_{\mathrm{BE}}(\lambda)}
  \mathcal H(\widehat u;\lambda)^{-1}
  \mathcal F(\widehat u;\lambda),
  \label{eq:S3-Bethe-formula}
\end{equation}
where $\mathcal S_{\mathrm{BE}}(\lambda)$ is the finite set of regular Bethe vacua and $\mathcal H$ and $\mathcal F$ are the handle-gluing and fibering operators.  Their explicit form will be given when the twisted Bethe description is developed below.

Numerical extremization gives
\begin{equation}
  \lambda_{\mathrm{num}}=0.9999999586\ldots,
  \qquad
  \boldsymbol\mu(\lambda_{\mathrm{num}})
  =\lambda_{\mathrm{num}}\boldsymbol a.
  \label{eq:Fmax-result}
\end{equation}
This strongly motivates the exact candidate
\begin{equation}
  \lambda_0=1,
  \qquad
  \boldsymbol\mu_0:=\boldsymbol\mu(1)=\boldsymbol a
  =(1,-1,1,-1,1,-1,1,-1)
  \label{eq:exact-mixing-candidate}
\end{equation}
for the infrared superconformal mixing.  Evaluating the numerical profile at this point gives
\begin{equation}
  F(1)=1.772597493875324\ldots,
  \qquad
  F''(1)=-2.1337176\ldots<0.
  \label{eq:Fmax-curvature}
\end{equation}
Together with \eqref{eq:Fmax-result}, the negative curvature shows that the numerical profile has a strict local maximum indistinguishable from $\lambda=1$ at the quoted precision.  This is not an analytic proof that $F'(1)=0$, and the scan does not exclude an accidental symmetry outside the ultraviolet mixing space.

The corresponding sphere partition function is
\begin{equation}
  \left|Z_{S^3}(\boldsymbol\mu_0)\right|
  =0.169891124049132\ldots
  \stackrel{\mathrm{num.}}{=}
  \frac{2}{\sqrt{11}}\sin\frac{\pi}{11}.
  \label{eq:S3-value}
\end{equation}
The two sides agree to the quoted numerical precision.  The trigonometric expression in \eqref{eq:S3-value} should be regarded as a numerical recognition rather than an analytic evaluation of the localization integral.

The candidate mixing vector is integral,
\begin{equation}
  \boldsymbol\mu_0=\boldsymbol a\in\mathbb Z^8
  \subset \left(\frac12\mathbb Z\right)^8.
  \label{eq:half-integral-mixing}
\end{equation}
The integrality of $\boldsymbol\mu_0$ is compatible with the $R$-charge quantization used in the half-index conventions.  This is a useful consistency check for the infrared interpretation, but it is not sufficient to establish supersymmetry enhancement.  The superconformal-index test will be given in the next subsection.

We can now return to the topological-fugacity specialization introduced in Section~2.1.  The seven monopole terms constrain the eight ultraviolet topological fugacities to the one-dimensional subtorus generated by $\boldsymbol a$,
\begin{equation}
  x_i=z^{a_i},
  \qquad i=1,\ldots,8.
  \label{eq:topological-subtorus}
\end{equation}
For the superconformal grading determined by \eqref{eq:exact-mixing-candidate}, the mixing factor is equivalently implemented by $z=-q^{1/2}$ in the boundary convention of \eqref{eq:generic-Nahm-half-index}.  Assuming the infrared $\mathcal N=4$ enhancement, the topological $A$-twisted half-index is obtained from the superconformal grading by the $\nu=-1$ specialization, namely by shifting the $R$-symmetry by $-\mathsf T$~\cite{GangKimParkStubbs:2024Nahm}.  In the present theory this shifts the mixing vector from $\boldsymbol\mu_0$ to $\boldsymbol\mu_0-\boldsymbol a$.  Since
\begin{equation}
  \boldsymbol\mu_0-\boldsymbol a=0,
  \label{eq:A-twist-cancellation}
\end{equation}
the $A$-twist cancels the linear topological grading completely.  At the unrefined point the fugacities therefore reduce to
\begin{equation}
  x_1=\cdots=x_8=1,
  \label{eq:A-twist-fugacity-one}
\end{equation}
and \eqref{eq:generic-Nahm-half-index} becomes
\begin{equation}
  I\!\!I_{(\mathcal D,D_c)}^{A}(q)
  =\sum_{\boldsymbol m\in\mathbb Z_{\geq0}^{8}}
  \frac{q^{\frac12\boldsymbol m^{\mathsf T}C_{E_8}\boldsymbol m}}
  {\prod_{i=1}^{8}(q)_{m_i}}.
  \label{eq:A-twisted-vacuum-half-index}
\end{equation}
Thus, up to the overall factor $q^{-10/33}$, the original unrefined $(E_8,T_1)$ Nahm sum is precisely the $A$-twisted $(\mathcal D,D_c)$ half-index selected by the superconformal mixing found through $F$-maximization.  This closes the circle between the Nahm-sum input of Section~2.1, the monopole deformation of Section~2.2, and the candidate infrared $R$-symmetry determined here.

\subsection{The superconformal index and evidence for a rank-zero fixed point}
\label{subsec:SCI-rank-zero}

We now use the bulk superconformal index to test the infrared interpretation.  First, the superconformal index contains the charged conserved-supercurrent contribution expected when the ultraviolet $\mathcal N=2$ supersymmetry enhances to $\mathcal N=4$.  Second, the two distinguished specializations associated with the $A$- and $B$-twists probe the Coulomb- and Higgs-branch operator rings.  Both tests use the candidate mixing vector $\boldsymbol\mu_0=\boldsymbol a$ determined in Section~2.3.

We use the $(-1)^R$ convention and define the 3d superconformal index as~\cite{Kim:2009wb,Imamura:2011su}
\begin{equation}
  \mathcal I(q,\eta)
  :=\operatorname{Tr}_{\mathcal H_{S^2}}
  (-1)^{R_{\boldsymbol\mu_0}}
  q^{j_3+R_{\boldsymbol\mu_0}/2}\eta^{\mathsf T},
  \label{eq:SCI-definition}
\end{equation}
where $j_3$ is the Cartan generator of spatial rotations and $\eta$ is the fugacity for the surviving topological symmetry $U(1)_{\mathsf T}$.  A GNO sector $\boldsymbol m\in\mathbb Z^8$ carries topological charges $T_i(\boldsymbol m)=m_i$.

Let $s_i$ be the gauge fugacity of the $i$-th $U(1)$ factor.  For the charge assignment $Q_j(\Phi_i)=\delta_{ij}$, localization gives
\begin{equation}
  \mathcal I(q,\eta)
  =\sum_{\boldsymbol m\in\mathbb Z^8}
  \eta^{\boldsymbol a\cdot\boldsymbol m}
  \left(-q^{1/2}\right)^{\boldsymbol\mu_0\cdot\boldsymbol m}
  \oint\prod_{i=1}^{8}\frac{ds_i}{2\pi \mathrm{i} s_i}
  \prod_{i=1}^{8}s_i^{(C_{E_8}\boldsymbol m)_i}
  \mathcal I_\Delta(m_i,s_i).
  \label{eq:SCI-holonomy}
\end{equation}
Here $\mathcal I_\Delta(m,s)$ is the fugacity-basis index of a charge-one chiral multiplet.  Its Laurent expansion defines the charge-basis tetrahedron index,
\begin{equation}
  \mathcal I_\Delta(m,s)
  =\sum_{e\in\mathbb Z}\mathcal I_\Delta(m,e)s^e,
  \label{eq:tetrahedron-Laurent}
\end{equation}
with
\begin{equation}
  \mathcal I_\Delta(m,e)
  =\sum_{n=\max(0,-e)}^\infty
  \frac{(-1)^n q^{\frac12 n(n+1)-(n+e/2)m}}
  {(q)_n(q)_{n+e}}.
  \label{eq:tetrahedron-index}
\end{equation}
Substituting \eqref{eq:tetrahedron-Laurent} into \eqref{eq:SCI-holonomy}, each gauge integral becomes
\begin{equation}
  \oint\frac{ds_i}{2\pi \mathrm{i} s_i}
  s_i^{(C_{E_8}\boldsymbol m)_i+e_i}
  =\delta_{e_i,-(C_{E_8}\boldsymbol m)_i}.
  \label{eq:SCI-Gauss-law-projection}
\end{equation}
Thus the holonomy integrals impose
\begin{equation}
  e_i=-\bigl(C_{E_8}\boldsymbol m\bigr)_i.
  \label{eq:SCI-Gauss-law}
\end{equation}
Since the matter-charge matrix is the identity, all electric charges are fixed uniquely.  At the candidate superconformal mixing $\boldsymbol\mu_0=\boldsymbol a$, the superconformal index is therefore
\begin{equation}
  \mathcal I(q,\eta)
  =\sum_{\boldsymbol m\in\mathbb Z^8}
  \eta^{\boldsymbol a\cdot\boldsymbol m}
  \left(-q^{1/2}\right)^{\boldsymbol\mu_0\cdot\boldsymbol m}
  \prod_{i=1}^{8}
  \mathcal I_\Delta\!\left(
    m_i,-\bigl(C_{E_8}\boldsymbol m\bigr)_i
  \right).
  \label{eq:full-SCI}
\end{equation}
A degree-bounded enumeration of the magnetic sectors gives
\begin{equation}
\begin{aligned}
  \mathcal I(q,\eta)
  ={}&1-q-(\eta+\eta^{-1})q^{3/2}-2q^2
  -\eta^{-1}q^{5/2}\\
  &+(\eta^2-1)q^3+O(q^{7/2}).
\end{aligned}
  \label{eq:SCI-expansion}
\end{equation}
All magnetic sectors that can contribute through $q^3$ are included in this expansion.

Relative to an $\mathcal N=2$ superconformal algebra, an $\mathcal N=4$ algebra contains two additional conserved-supercurrent multiplets.  In the convention of \eqref{eq:SCI-definition} they contribute the charged pair
\begin{equation}
  -(\eta+\eta^{-1})q^{3/2},
  \label{eq:extra-supercurrent-signal}
\end{equation}
which is precisely the term appearing in \eqref{eq:SCI-expansion}.  The pair of contributions \eqref{eq:extra-supercurrent-signal} provides nontrivial evidence for supersymmetry enhancement in the infrared~\cite{GangKimParkStubbs:2024Nahm}.  Their quantum numbers are precisely those expected from the additional supercurrents required to complete the ultraviolet supersymmetry to $\mathcal N=4$.  Together with the determination of the infrared $R$-symmetry in the previous subsection, the superconformal index therefore strongly supports the emergence of an interacting $\mathcal N=4$ fixed point.

Assuming the $\mathcal N=4$ enhancement, the infrared $R$-symmetry group is $SU(2)_H\times SU(2)_C$.  The residual $U(1)_{\mathsf T}$ combines with the superconformal $R$-symmetry into its Cartan.  We normalize the Cartan generators $J_3^H$ and $J_3^C$ so that
\begin{equation}
  R_{\boldsymbol\mu_0}=J_3^C+J_3^H,
  \qquad
  \mathsf T=J_3^C-J_3^H.
  \label{eq:N4-R-decomposition}
\end{equation}
For
\begin{equation}
  R_\nu:=R_{\boldsymbol\mu_0}+\nu\mathsf T,
\end{equation}
we then have
\begin{equation}
  R_{-1}=2J_3^H,
  \qquad
  R_{+1}=2J_3^C.
  \label{eq:AB-R-symmetries}
\end{equation}
The corresponding specializations of the superconformal index are
\begin{equation}
  \mathcal I_\nu(q)
  :=\mathcal I\!\left(q,\eta=\left(-q^{1/2}\right)^\nu\right),
  \qquad
  \mathcal I_A(q):=\mathcal I_{-1}(q),
  \qquad
  \mathcal I_B(q):=\mathcal I_{+1}(q).
  \label{eq:AB-index-specializations}
\end{equation}
Since $\boldsymbol\mu_0=\boldsymbol a$, the two specializations replace the mixing vector by
\begin{equation}
  \boldsymbol\mu_0-\boldsymbol a=0,
  \qquad
  \boldsymbol\mu_0+\boldsymbol a=2\boldsymbol a.
  \label{eq:AB-effective-mixings}
\end{equation}
Under the enhancement assumption, these are the Coulomb- and Higgs-branch limits of the 3d $\mathcal N=4$ superconformal index, respectively~\cite{Closset:2016arn,Razamat:2014pta}.

The componentwise-cutoff computations stabilize to
\begin{align}
  \mathcal I_A(q)&=1+O(q^{11/2}),
  \label{eq:A-index-rank-zero}\\
  \mathcal I_B(q)&=1+O(q^{9/2}).
  \label{eq:B-index-rank-zero}
\end{align}
These specializations support the interpretation that both branches are points and hence that the infrared fixed point is rank zero.  Combining the $F$-maximization of Section~2.3, the charged-supercurrent signal \eqref{eq:extra-supercurrent-signal}, and the two rank-zero specializations above, we propose that the monopole-deformed $C_{E_8}$ Chern--Simons matter theory flows to an interacting rank-zero $\mathcal N=4$ SCFT.

\subsection{Wilson loop half-indices and the five-character sector basis}
\label{subsec:Wilson-half-indices}

The complete five-character system of $T_{10}M_{\mathrm{eff}}(11,2)$ is
\[
  \left\{
  \chi_0,\,
  \chi_{5/11},\,
  \chi_{8/11},\,
  \chi_{10/11},\,
  \chi_{12/11}
  \right\}.
\]
The vacuum character $\chi_0$ was reproduced in Section~2.3 from the $(\mathcal D,D_c)$ half-index with the $A$-twist $R$-charge assignment.  We now identify four additional supersymmetric Wilson loop sectors whose normalized half-indices agree with the remaining four characters through the displayed order.  Together with the exact vacuum identity, this identifies a distinguished five-sector line basis that will be used in the particle--vortex duality construction of Section~3.

We consider a supersymmetric Wilson loop inserted at the tip of the cigar and wrapping $S^1$.  We denote its charge vector by $\boldsymbol Q\in\mathbb Z^8$ and choose the sign convention such that, in the monopole sector $\boldsymbol m$, its insertion contributes $q^{-\boldsymbol Q\cdot\boldsymbol m}$.  From this point on, whenever a Wilson loop is inserted into the $(\mathcal D,D_c)$ boundary condition, we suppress the boundary label to avoid cumbersome notation and write the Wilson loop half-index as $I\!\!I^{A}[W_{\boldsymbol Q}]$.  At the $A$-twist $R$-charge specialization $\boldsymbol\mu_0-\boldsymbol a=0$, the Wilson loop half-index is therefore
\begin{equation}
  I\!\!I^{A}[W_{\boldsymbol Q}](q)
  =\sum_{\boldsymbol m\in\mathbb Z_{\geq0}^{8}}
  \frac{
  q^{\frac12\boldsymbol m^{\mathsf T}C_{E_8}\boldsymbol m
  -\boldsymbol Q\cdot\boldsymbol m}
  }{
  \prod_{i=1}^{8}(q)_{m_i}
  }.
  \label{eq:Wilson-half-index}
\end{equation}
Thus the Wilson loop charge vector $\boldsymbol Q$ is precisely the vector appearing in the negative linear term $-\boldsymbol Q\cdot\boldsymbol m$ in the exponent of $q$ in the numerator of the Nahm sum.  We label the five sectors by their effective conformal weights
\begin{equation}
  \mathcal H
  =\left\{0,\frac5{11},\frac8{11},\frac{10}{11},\frac{12}{11}\right\}.
  \label{eq:five-weight-set}
\end{equation}
For each $h\in\mathcal H$, define
\begin{equation}
  \delta_h
  :=\min_{\boldsymbol m\in\mathbb Z_{\geq0}^{8}}
  \left(
  \frac12\boldsymbol m^{\mathsf T}C_{E_8}\boldsymbol m
  -\boldsymbol Q_h\cdot\boldsymbol m
  \right).
  \label{eq:Wilson-minimum}
\end{equation}
Since $c_{\mathrm{eff}}=80/11$, the character with effective weight $h$ starts at $q^{h-10/33}$.  At this stage it is convenient to record the $q$-shift required for the character comparison as
\begin{equation}
  \gamma_h
  :=h-\frac{10}{33}-\delta_h.
  \label{eq:Wilson-normalization}
\end{equation}
The quantities $\gamma_h$ in \eqref{eq:Wilson-normalization} should not be interpreted as five independent normalization parameters attached by hand to the Wilson loops.  Here they are only a bookkeeping device for comparing the raw half-indices with the candidate character gradings.  In Section~\ref{subsec:dual-Wilson-Zagier} we will show that the same five shifts are related uniquely across the particle--vortex duality by the scalar contact-term transformation; in particular, once the vacuum Casimir normalizations, the paired sector grading, and the common framing convention are fixed, the non-vacuum shifts cannot be adjusted independently.  The fractional parts of the sector gradings will also be recovered independently from the topological spins of the $A$-twisted theory in Section~\ref{sec:AB-twisted-TQFTs}.
A convenient set of Wilson-charge representatives is
\begin{center}
\small
\begin{tabular}{cclcc}
\toprule
$h$ & $\boldsymbol Q_h$ & & $\delta_h$ & $\gamma_h$\\
\midrule
$0$ & $(0,0,0,0,0,0,0,0)$ && $0$ & $-10/33$\\
$5/11$ & $(2,-2,2,-2,2,-1,1,-2)$ && $-3$ & $104/33$\\
$8/11$ & $(1,-1,1,-1,2,-2,2,-2)$ && $-2$ & $80/33$\\
$10/11$ & $(0,0,1,-1,1,-1,1,-1)$ && $0$ & $20/33$\\
$12/11$ & $(1,-1,1,-1,1,-1,1,-1)$ && $0$ & $26/33$\\
\bottomrule
\end{tabular}
\end{center}
For the vacuum sector, $\boldsymbol Q_0=0$ and \eqref{eq:Wilson-half-index} reduces to $I\!\!I_{(\mathcal D,D_c)}^{A}$ in \eqref{eq:A-twisted-vacuum-half-index}.  For the four non-vacuum sectors, direct expansion gives
\begin{align}
I\!\!I^{A}[W_{\boldsymbol Q_{5/11}}](q)
={}&q^{-3}\bigl(8+245q+2360q^2+15100q^3+75600q^4\nonumber\\
&\hspace{25mm}{}+321552q^5+1210000q^6+O(q^7)\bigr),\nonumber\\
I\!\!I^{A}[W_{\boldsymbol Q_{8/11}}](q)
={}&q^{-2}\bigl(35+680q+5628q^2+33440q^3+159020q^4\nonumber\\
&\hspace{25mm}{}+652240q^5+2386140q^6+O(q^7)\bigr),\nonumber\\
I\!\!I^{A}[W_{\boldsymbol Q_{10/11}}](q)
={}&84+1240q+9570q^2+54160q^3+249880q^4\nonumber\\
&\hspace{25mm}{}+1001296q^5+3599710q^6+O(q^7),\nonumber\\
I\!\!I^{A}[W_{\boldsymbol Q_{12/11}}](q)
={}&50+560q+4180q^2+22352q^3+100695q^4\nonumber\\
&\hspace{25mm}{}+393960q^5+1393820q^6+O(q^7).
\label{eq:Wilson-series}
\end{align}
With the provisional shifts $q^{\gamma_h}$ in \eqref{eq:Wilson-normalization}, these series reproduce the four non-vacuum characters of $T_{10}M_{\mathrm{eff}}(11,2)$.  Together with the vacuum sector, the comparison may be summarized as
\begin{align}
  \chi_0(q)
  &=q^{-10/33}I\!\!I_{(\mathcal D,D_c)}^{A}(q),
  &
  \chi_{5/11}(q)
  &=q^{104/33}I\!\!I^{A}[W_{\boldsymbol Q_{5/11}}](q),\nonumber\\
  \chi_{8/11}(q)
  &=q^{80/33}I\!\!I^{A}[W_{\boldsymbol Q_{8/11}}](q),
  &
  \chi_{10/11}(q)
  &=q^{20/33}I\!\!I^{A}[W_{\boldsymbol Q_{10/11}}](q),\nonumber\\
  \chi_{12/11}(q)
  &=q^{26/33}I\!\!I^{A}[W_{\boldsymbol Q_{12/11}}](q).
  \label{eq:five-character-Wilson-correspondence}
\end{align}
The $C_{E_8}$ theory therefore realizes the $(E_8,T_1)$ Nahm sum for the vacuum character exactly and supplies four distinguished Wilson loop sectors whose normalized half-indices reproduce the remaining characters through the displayed order.

\paragraph{Remark on the five Wilson loop sectors and the $A$-twisted torus state space.}
At the level of the half-index calculation alone, \eqref{eq:five-character-Wilson-correspondence} identifies one vacuum sector and four distinguished Wilson loop sectors with the five components of the $T_{10}M_{\mathrm{eff}}(11,2)$ character system, but it does not yet explain why precisely five independent topological sectors should appear in the infrared.  This point will be clarified in Section~\ref{sec:AB-twisted-TQFTs}.  The $A$-twist of the $C_{E_8}$ theory has exactly five regular Bethe vacua, and Section~\ref{sec:AB-twisted-TQFTs} shows that they give a five-dimensional torus state space $\mathscr H_{T^2}^{A}(\mathcal T)$.  The five Wilson loop sectors chosen above furnish a Wilson loop basis\footnote{Here the Wilson loop basis means the basis furnished by the five distinguished topological line sectors represented by the Wilson loops $W_{\boldsymbol Q_h}$ with $h\in\mathcal H$.  Their multiplication is identified in Section~\ref{sec:AB-twisted-TQFTs} with the formal Verlinde algebra of the five-character system.} of this state space, while the five Bethe vacua label the Bethe-idempotent basis introduced in Section~\ref{sec:AB-twisted-TQFTs}.

This also gives a useful interpretation of the half-indices found here.  In the 3d $\mathcal N=4$ analogue of the Chern--Simons/WZW solid-torus construction discussed in Section~\ref{subsec:susy-CS-WZW-solid-torus} and summarized in Table~\ref{tab:CS-WZW-susy-comparison}, inserting $W_{\boldsymbol Q_h}$ along the core of the solid torus prepares a state $|W_{\boldsymbol Q_h}\rangle_A\in\mathscr H_{T^2}^{A}(\mathcal T)$, and the Wilson loop half-index is its holomorphic solid-torus amplitude.  After the Section~4 analysis, the five normalized half-indices in \eqref{eq:five-character-Wilson-correspondence} can therefore be viewed as the holomorphic wavefunctions associated with the five torus states.  This interpretation concerns the semisimple torus-state and modular data and does not assume a microscopic identification of the boundary VOA.

In the next section we apply particle--vortex duality to the full theory and derive the dual $C_{E_8}^{-1}$ Chern--Simons matter Lagrangian.

\section{Particle--vortex construction of the 3d $\mathcal N=4$ mirror pair}
\label{sec:particle-vortex-mirror}

In Section~2, we constructed the ultraviolet theory $\mathcal T$ directly from the $(E_8,T_1)$ Nahm sum and found evidence that its monopole superpotential deformation flows to an interacting rank-zero $\mathcal N=4$ fixed point.  We now construct a second ultraviolet Lagrangian by applying particle--vortex duality to $\mathcal T$.  We first determine the dual gauge and matter content and then derive the image of the seven monopole-superpotential terms.  After deriving the dual bulk theory and the Wilson line map from particle--vortex duality on $\mathbb R^3$, we compute the Wilson loop half-indices of the two theories and compare their quadratic, linear, and scalar data.  The arithmetic interpretation of this comparison will be introduced only after the 3d transformation has been derived.  The infrared comparison will be developed in the following subsections.

\subsection{The dual $C_{E_8}^{-1}$ Chern--Simons matter Lagrangian from particle--vortex duality}
\label{subsec:dual-Lagrangian-PV}

We begin by recalling the ultraviolet Lagrangian data of $\mathcal T$.  It contains eight $U(1)$ vector multiplets with gauge fields $A_i$ and eight chiral multiplets $\Phi_i$, where $\Phi_i$ has charge $+1$ under the $i$-th gauge factor and is neutral under the remaining seven factors.  Separating the matter kinetic terms from the Chern--Simons and contact terms, the Lagrangian may be written schematically as
\begin{equation}
  \mathcal L_{\mathcal T}
  =
  \sum_{i=1}^{8}\mathcal L_{\mathrm{matter}}[\Phi_i,A_i]
  +\frac{1}{4\pi}
  \sum_{i,j=1}^{8}
  (K_{\mathrm{UV}})_{ij} A_i\,dA_j
  +\mathcal L_W,
  \label{eq:T-Lagrangian}
\end{equation}
with
\begin{equation}
  K_{\mathrm{UV}}
  =C_{E_8}-\frac12 I_8.
  \label{eq:T-UV-matrix-recap}
\end{equation}
The diagonal $-\frac12 I_8$ is the canonical parity-anomaly contact term of the eight charge-one chirals, while $C_{E_8}$ is the integral gauge Chern--Simons matrix of $\mathcal T$.  The term $\mathcal L_W$ is the supersymmetric interaction generated by the monopole superpotential \eqref{eq:monopole-superpotential}.

For the purpose of applying particle--vortex duality, it is useful to undo the gauging conceptually and regard the eight chiral multiplets, together with their canonical contact terms, as eight tetrahedron theories $\mathcal T_\Delta$.  Before gauging the eight flavor symmetries, we temporarily regard the fields $A_i$ as background gauge fields for these flavor symmetries.  In this description, $C_{E_8}$ specifies the background flavor Chern--Simons couplings among the $A_i$.  Promoting the $A_i$ to dynamical gauge fields turns these background couplings into the gauge Chern--Simons matrix $C_{E_8}$ appearing in \eqref{eq:T-UV-matrix-recap}.

We now apply particle--vortex duality to the eight tetrahedron-theory factors before performing this gauging.  The elementary 3d $\mathcal N=2$ particle--vortex duality \cite{Dimofte:2011py} replaces the flavor current of a charge-one tetrahedron chiral by the topological current of a dual $U(1)$ gauge theory; the corresponding duality wall and its boundary realization were described in \cite{Dimofte:2017tpi}.  For each node $i$, let $B_i$ denote the gauge field introduced in the particle--vortex description and let $\Phi_i^\vee$ be its charge-one chiral multiplet.

This transformation also admits a folded-interface interpretation.  One may place the original and dual descriptions on the two sides of a particle--vortex duality wall and then fold the dual side across the wall.  In the folded description the dual theory is orientation reversed.  For the derivation below, this viewpoint will be used only to keep track of the relative signs of Chern--Simons and contact terms.  We will return to its protected interface interpretation after studying the twisted TQFT data.

After particle--vortex duality, the background field $A_i$ couples to the topological current of $B_i$ through a BF interaction.  We then promote the $A_i$ to dynamical gauge fields, thereby implementing the same $C_{E_8}$ gauging that defines $\mathcal T$.  Applying this operation independently at all eight nodes gives, before integrating out the $A_i$, the following topological part of the folded Lagrangian:
\begin{equation}
  \mathcal L_{\mathrm{top}}^{\mathrm{fold}}
  =\frac{1}{4\pi}
  \left[
  \sum_{i,j=1}^{8}(C_{E_8})_{ij}A_i\,dA_j
  +2\epsilon\sum_{i=1}^{8}A_i\,dB_i
  +\frac12\sum_{i=1}^{8}B_i\,dB_i
  \right],
  \qquad \epsilon=\pm1.
  \label{eq:PV-folded-Lagrangian}
\end{equation}
Here $\epsilon$ records the orientation convention for the BF coupling.  Introducing the vector notation
\begin{equation}
  \boldsymbol A=(A_1,\ldots,A_8)^{\mathsf T},
  \qquad
  \boldsymbol B=(B_1,\ldots,B_8)^{\mathsf T},
  \label{eq:AB-vector-notation}
\end{equation}
the same quadratic couplings can be written as
\begin{equation}
  \mathcal L_{\mathrm{top}}^{\mathrm{fold}}
  =
  \frac{1}{4\pi}
  \begin{pmatrix}
    \boldsymbol A^{\mathsf T} & \boldsymbol B^{\mathsf T}
  \end{pmatrix}
  \begin{pmatrix}
    C_{E_8} & \epsilon I_8\\
    \epsilon I_8 & \frac12 I_8
  \end{pmatrix}
  d
  \begin{pmatrix}
    \boldsymbol A\\
    \boldsymbol B
  \end{pmatrix}.
  \label{eq:PV-folded-matrix-Lagrangian}
\end{equation}

The appearance of the inverse Cartan matrix can now be seen directly from the equations of motion.  Varying \eqref{eq:PV-folded-Lagrangian} with respect to the dynamical gauge fields $\boldsymbol A$ gives
\begin{equation}
  C_{E_8}\,d\boldsymbol A
  +\epsilon\,d\boldsymbol B
  =0.
  \label{eq:A-equation-of-motion}
\end{equation}
Since $C_{E_8}$ is invertible, this implies
\begin{equation}
  d\boldsymbol A
  =-\epsilon\,C_{E_8}^{-1}d\boldsymbol B.
  \label{eq:A-solved-by-B}
\end{equation}
Substituting this relation back into the quadratic action yields
\begin{equation}
  \mathcal L_{\mathrm{top}}^{\mathrm{fold}}
  =\frac{1}{4\pi}
  \boldsymbol B^{\mathsf T}
  \left(
    \frac12 I_8-C_{E_8}^{-1}
  \right)
  d\boldsymbol B.
  \label{eq:PV-folded-effective-Lagrangian}
\end{equation}
Equivalently, the matrix $\frac12 I_8-C_{E_8}^{-1}$ is the Schur complement of the $C_{E_8}$ block in \eqref{eq:PV-folded-matrix-Lagrangian}.

The relative sign in \eqref{eq:PV-folded-effective-Lagrangian} is a consequence of the folded description.  The dual theory is orientation reversed after folding, so unfolding it reverses the sign of its Chern--Simons and contact terms.  The ultraviolet contact-term matrix of the physical dual theory is therefore
\begin{equation}
  K_{\mathrm{UV}}^\vee
  =C_{E_8}^{-1}-\frac12 I_8.
  \label{eq:dual-UV-contact}
\end{equation}
The diagonal $-\frac12 I_8$ is again the canonical parity-anomaly contact term of eight charge-one chirals.  Hence the integral gauge Chern--Simons matrix of the physical dual theory is
\begin{equation}
  K^\vee=C_{E_8}^{-1}.
  \label{eq:dual-integral-K}
\end{equation}
Since $\det C_{E_8}=1$, this inverse is itself integral:
\begin{equation}
C_{E_8}^{-1}
=
\begin{pmatrix}
2&3&4&5&6&4&2&3\\
3&6&8&10&12&8&4&6\\
4&8&12&15&18&12&6&9\\
5&10&15&20&24&16&8&12\\
6&12&18&24&30&20&10&15\\
4&8&12&16&20&14&7&10\\
2&4&6&8&10&7&4&5\\
3&6&9&12&15&10&5&8
\end{pmatrix}.
\label{eq:E8-inverse}
\end{equation}
Thus particle--vortex duality produces an ordinary 3d $\mathcal N=2$ abelian Chern--Simons matter Lagrangian
\begin{equation}
  \mathcal L_{\mathcal T^\vee}
  =
  \sum_{i=1}^{8}\mathcal L_{\mathrm{matter}}[\Phi_i^\vee,B_i]
  +\frac{1}{4\pi}
  \sum_{i,j=1}^{8}
  \left(C_{E_8}^{-1}-\frac12 I_8\right)_{ij}
  B_i\,dB_j
  +\mathcal L_{W^\vee}.
  \label{eq:dual-Lagrangian}
\end{equation}
In particular,
\begin{equation}
  \mathcal T^\vee:\qquad
  G^\vee=U(1)^8,
  \qquad
  K^\vee=C_{E_8}^{-1},
  \qquad
  \text{matter}=\{\Phi_i^\vee\}_{i=1}^{8},
  \qquad
  Q_j^\vee(\Phi_i^\vee)=\delta_{ij}.
  \label{eq:dual-UV-theory}
\end{equation}
Therefore the inverse $E_8$ Cartan matrix arises directly from particle--vortex duality.

We next determine the particle--vortex image of the monopole superpotential.  The seven operators in \eqref{eq:seven-monopoles} are associated with the seven edges $\langle i,j\rangle$ of the numbered $E_8$ Dynkin graph.  Their magnetic charges are
\begin{equation}
  \boldsymbol m_{ij}
  =\boldsymbol e_i+\boldsymbol e_j,
  \qquad
  \langle i,j\rangle\in E(E_8),
  \label{eq:edge-magnetic-charges}
\end{equation}
and, because these fluxes are nonnegative, their dressing vectors follow from \eqref{eq:dressing-vector} as
\begin{equation}
  \boldsymbol n^{(ij)}
  =\boldsymbol m_{ij}-C_{E_8}\boldsymbol m_{ij}.
  \label{eq:edge-original-dressing}
\end{equation}
The BPS condition $n_k^{(ij)}(m_{ij})_k=0$ implies that the support of the dressing is disjoint from the two nodes carrying positive magnetic flux.

The magnetic-flux map follows directly from the same BF equation that produced the inverse Cartan matrix.  Integrating \eqref{eq:A-equation-of-motion} over a two-sphere surrounding a monopole insertion and defining
\begin{equation}
  \boldsymbol m
  :=\frac{1}{2\pi}\int_{S^2}d\boldsymbol A,
  \qquad
  \boldsymbol\ell
  :=\frac{1}{2\pi}\int_{S^2}d\boldsymbol B,
  \label{eq:original-dual-flux-definitions}
\end{equation}
we obtain
\begin{equation}
  C_{E_8}\boldsymbol m+\epsilon\boldsymbol\ell=0.
  \label{eq:BF-flux-equation}
\end{equation}
We choose the orientation convention $\epsilon=-1$, for which the map takes the simple form
\begin{equation}
  \boldsymbol\ell=C_{E_8}\boldsymbol m.
  \label{eq:PV-magnetic-map}
\end{equation}
The opposite convention reverses all dual magnetic charges and is physically equivalent after the corresponding orientation reversal.

For an edge monopole, define
\begin{equation}
  \boldsymbol\ell_{ij}
  :=C_{E_8}\boldsymbol m_{ij}
  =C_{E_8}(\boldsymbol e_i+\boldsymbol e_j).
  \label{eq:dual-edge-flux}
\end{equation}
Combining this with \eqref{eq:edge-original-dressing} gives the useful identity
\begin{equation}
  \boldsymbol\ell_{ij}
  =\boldsymbol m_{ij}-\boldsymbol n^{(ij)}.
  \label{eq:dual-flux-as-flux-minus-dressing}
\end{equation}
Both $\boldsymbol m_{ij}$ and $\boldsymbol n^{(ij)}$ are nonnegative, and their supports are disjoint.  Therefore the positive and negative parts of the dual magnetic charge are
\begin{equation}
  (\boldsymbol\ell_{ij})_+
  =\boldsymbol m_{ij},
  \qquad
  (\boldsymbol\ell_{ij})_-
  =\boldsymbol n^{(ij)},
  \label{eq:dual-flux-positive-negative}
\end{equation}
where $(\ell_-)_k:=\max(-\ell_k,0)$.  Thus the chiral dressing of the original monopole is converted by particle--vortex duality into the negative magnetic components of the dual monopole flux.

As a concrete example, consider the edge $\langle4,5\rangle$.  In the original theory,
\begin{equation}
  \boldsymbol m_{45}
  =(0,0,0,1,1,0,0,0),
  \qquad
  \mathcal O_{45}
  =\phi_3\phi_6\phi_8
  V_{(0,0,0,1,1,0,0,0)},
  \label{eq:O45-example}
\end{equation}
so that
\begin{equation}
  \boldsymbol n^{(45)}
  =(0,0,1,0,0,1,0,1).
  \label{eq:O45-dressing-example}
\end{equation}
Equation \eqref{eq:dual-flux-as-flux-minus-dressing} then gives
\begin{equation}
  \begin{aligned}
  \boldsymbol\ell_{45}
  &=\boldsymbol m_{45}-\boldsymbol n^{(45)}\\
  &=(0,0,-1,1,1,-1,0,-1)\\
  &=C_{E_8}(\boldsymbol e_4+\boldsymbol e_5).
  \end{aligned}
  \label{eq:O45-dual-flux-example}
\end{equation}
The factors $\phi_3\phi_6\phi_8$ dressing the original monopole are therefore reflected in the negative flux components at nodes $3$, $6$, and $8$ of the dual monopole.

For completeness, the seven dual magnetic charges are
\begin{align}
\boldsymbol\ell_{12}&=(1,1,-1,0,0,0,0,0),\nonumber\\
\boldsymbol\ell_{23}&=(-1,1,1,-1,0,0,0,0),\nonumber\\
\boldsymbol\ell_{34}&=(0,-1,1,1,-1,0,0,0),\nonumber\\
\boldsymbol\ell_{45}&=(0,0,-1,1,1,-1,0,-1),\nonumber\\
\boldsymbol\ell_{56}&=(0,0,0,-1,1,1,-1,-1),\nonumber\\
\boldsymbol\ell_{67}&=(0,0,0,0,-1,1,1,0),\nonumber\\
\boldsymbol\ell_{58}&=(0,0,0,-1,1,-1,0,1).
\label{eq:seven-dual-fluxes}
\end{align}

It remains to verify that these operators are bare monopoles in the dual theory.  A scalar monopole of magnetic charge $\boldsymbol\ell$ in the $C_{E_8}^{-1}$ theory can in general be dressed as
\begin{equation}
  \mathcal O^\vee_{\boldsymbol n^\vee,\boldsymbol\ell}
  =\prod_{k=1}^{8}(\phi_k^\vee)^{n_k^\vee}
  V^\vee_{\boldsymbol\ell}.
  \label{eq:dual-general-dressed-monopole}
\end{equation}
With the same parity-anomaly convention as in Section~2.2, gauge invariance determines the dressing by
\begin{equation}
  \boldsymbol n^\vee
  =\boldsymbol\ell_+-C_{E_8}^{-1}\boldsymbol\ell.
  \label{eq:dual-monopole-dressing}
\end{equation}
For the seven charges \eqref{eq:dual-edge-flux}, equations \eqref{eq:PV-magnetic-map} and \eqref{eq:dual-flux-positive-negative} give
\begin{equation}
  \begin{aligned}
  \boldsymbol n^\vee_{ij}
  &=(\boldsymbol\ell_{ij})_+
  -C_{E_8}^{-1}\boldsymbol\ell_{ij}\\
  &=\boldsymbol m_{ij}
  -C_{E_8}^{-1}C_{E_8}\boldsymbol m_{ij}\\
  &=0.
  \end{aligned}
  \label{eq:dual-monopoles-bare}
\end{equation}
Thus no chiral dressing is required in the dual description.

The reference $R$-charge provides an additional check.  With $R_*^\vee(\phi_i^\vee)=0$, the same tetrahedron-theory convention used in Section~2.2 gives
\begin{equation}
  R_*^\vee(V^\vee_{\boldsymbol\ell})
  =\sum_{\ell_i>0}\ell_i.
  \label{eq:dual-reference-R}
\end{equation}
Since $(\boldsymbol\ell_{ij})_+=\boldsymbol m_{ij}=\boldsymbol e_i+\boldsymbol e_j$, every dual monopole obtained above has
\begin{equation}
  R_*^\vee(V^\vee_{\boldsymbol\ell_{ij}})=2.
  \label{eq:dual-edge-R-two}
\end{equation}
They therefore have precisely the reference $R$-charge required for superpotential terms.

We conclude that the seven dressed edge monopoles of $\mathcal T$ are mapped to seven bare monopoles of $\mathcal T^\vee$,
\begin{equation}
  \mathcal O_{ij}
  \longmapsto
  V^\vee_{\boldsymbol\ell_{ij}},
  \qquad
  \boldsymbol\ell_{ij}
  =C_{E_8}(\boldsymbol e_i+\boldsymbol e_j).
  \label{eq:monopole-operator-map}
\end{equation}
Accordingly, the dual monopole superpotential is
\begin{equation}
  W^\vee
  =\sum_{\langle i,j\rangle\in E(E_8)}
  \lambda_{ij}^\vee\,
  V^\vee_{C_{E_8}(\boldsymbol e_i+\boldsymbol e_j)},
  \qquad
  \lambda_{ij}^\vee\neq0.
  \label{eq:dual-monopole-superpotential}
\end{equation}
The coefficients depend on the normalization of the monopole operators, while the seven-term support and their magnetic charges are fixed by the particle--vortex operator map.  Hence the analysis above determines the complete ultraviolet Lagrangian data of the dual monopole-superpotential-deformed theory: the gauge group $U(1)^8$, the integral Chern--Simons matrix $C_{E_8}^{-1}$, eight charge-one chiral multiplets, and the seven bare-monopole superpotential terms.

The particle--vortex construction above also admits a protected duality-interface interpretation.  In the folded description the eight elementary particle--vortex transformations are combined with the $C_{E_8}$ gauging into a wall relating $\mathcal T$ and $\mathcal T^\vee$.  After developing the $A$- and $B$-twisted TQFT data, we will return to this interface and show how the paired Wilson line sectors realize the relative invertible $E_8$ phase and the associated level-one $E_8$ character.

\subsection{$F$-maximization and the infrared $R$-symmetry}
\label{subsec:dual-Fmax}

Having determined the ultraviolet Lagrangian and the monopole superpotential of $\mathcal T^\vee$, we next determine the $R$-symmetry selected along its renormalization-group flow.  We first identify the topological symmetry that remains unbroken by $W^\vee$.

Let $T_i^\vee$ denote the topological symmetry associated with the $i$-th gauge factor of $\mathcal T^\vee$, and write $\boldsymbol T^\vee=(T_1^\vee,\ldots,T_8^\vee)$.  A linear combination
\begin{equation}
  \mathsf T^\vee
  =\boldsymbol a^\vee\cdot\boldsymbol T^\vee
  \label{eq:dual-residual-topological-generator}
\end{equation}
is preserved by the superpotential if each of the seven bare monopoles in \eqref{eq:dual-monopole-superpotential} is neutral.  Since their magnetic charges are
\begin{equation}
  \boldsymbol\ell_{ij}
  =C_{E_8}(\boldsymbol e_i+\boldsymbol e_j),
  \qquad
  \langle i,j\rangle\in E(E_8),
  \label{eq:dual-edge-charge-recap}
\end{equation}
this requires
\begin{equation}
  \boldsymbol a^\vee\cdot\boldsymbol\ell_{ij}=0
  \qquad
  (\langle i,j\rangle\in E(E_8)).
  \label{eq:dual-residual-orthogonality}
\end{equation}
Using the symmetry of $C_{E_8}$, the condition may be rewritten as
\begin{equation}
  (C_{E_8}\boldsymbol a^\vee)_i
  +(C_{E_8}\boldsymbol a^\vee)_j=0
  \qquad
  (\langle i,j\rangle\in E(E_8)).
  \label{eq:dual-residual-edge-condition}
\end{equation}
Because the $E_8$ Dynkin graph is connected and bipartite, $C_{E_8}\boldsymbol a^\vee$ must be proportional to the alternating vector of the original theory,
\begin{equation}
  \boldsymbol a=(1,-1,1,-1,1,-1,1,-1).
\end{equation}
Choosing the primitive integer representative and fixing the overall sign by
\begin{equation}
  C_{E_8}\boldsymbol a^\vee=-\boldsymbol a,
  \label{eq:dual-residual-Cartan-relation}
\end{equation}
we obtain
\begin{equation}
  \boldsymbol a^\vee=(1,3,4,6,7,5,2,4).
  \label{eq:dual-residual-vector}
\end{equation}
Thus the seven monopole-superpotential terms break $U(1)^8_{\mathrm{top}}$ to the single topological symmetry $U(1)_{\mathsf T^\vee}$.  Equation \eqref{eq:dual-residual-Cartan-relation} already exhibits the Cartan/inverse-Cartan relation between the surviving topological symmetries of the two particle--vortex descriptions.

We now determine the candidate infrared superconformal $R$-symmetry by $F$-maximization.  Since the monopole superpotential leaves only $U(1)_{\mathsf T^\vee}$, the allowed one-parameter family of trial $R$-symmetries is
\begin{equation}
  R_\lambda^\vee
  =R_*^\vee+\lambda\,\mathsf T^\vee,
  \qquad
  \boldsymbol\mu^\vee(\lambda)
  =\lambda\boldsymbol a^\vee,
  \qquad
  \lambda\in\mathbb R.
  \label{eq:dual-Fmax-family}
\end{equation}
In the absence of accidental abelian symmetries, the superconformal $R$-symmetry locally maximizes the three-sphere free energy~\cite{Jafferis:2010un},
\begin{equation}
  F^\vee(\lambda)
  :=-\log\left|Z_{S^3}^\vee(\lambda)\right|.
  \label{eq:dual-F-def}
\end{equation}
For the numerical evaluation we use the same localized Seifert/Bethe-vacuum representation as in Section~2.3~\cite{Closset:2017zgf},
\begin{equation}
  Z_{S^3}^\vee(\lambda)
  =\sum_{\widehat u^\vee\in\mathcal S_{\mathrm{BE}}^\vee(\lambda)}
  \mathcal H^\vee(\widehat u^\vee;\lambda)^{-1}
  \mathcal F^\vee(\widehat u^\vee;\lambda),
  \label{eq:dual-S3-Bethe-formula}
\end{equation}
where $\mathcal S_{\mathrm{BE}}^\vee(\lambda)$ is the finite set of regular Bethe vacua of the $C_{E_8}^{-1}$ theory, and $\mathcal H^\vee$ and $\mathcal F^\vee$ are its handle-gluing and fibering operators.

Numerical extremization gives $\lambda_{\mathrm{num}}^\vee\simeq-1$ to the precision of the computation.  This motivates the exact candidate
\begin{equation}
  \lambda_0^\vee=-1,
  \qquad
  \boldsymbol\mu_0^\vee
  :=\boldsymbol\mu^\vee(-1)
  =-\boldsymbol a^\vee
  =(-1,-3,-4,-6,-7,-5,-2,-4)
  \label{eq:dual-exact-mixing-candidate}
\end{equation}
for the infrared superconformal mixing.  Evaluating the numerical profile at this point gives
\begin{equation}
  F^\vee(-1)=1.772597493874873\ldots,
  \qquad
  \left.\frac{d^2F^\vee}{d\lambda^2}\right|_{\lambda=-1}
  =-2.1337174\ldots<0.
  \label{eq:dual-Fmax-curvature}
\end{equation}
Together with the numerical location of the extremum near $\lambda=-1$, the negative curvature shows that the numerical profile has a strict local maximum indistinguishable from $\lambda=-1$ at the quoted precision.  The corresponding sphere partition function is
\begin{equation}
  \left|Z_{S^3}^\vee(\boldsymbol\mu_0^\vee)\right|
  =0.169891124049209\ldots
  \stackrel{\mathrm{num.}}{=}
  \frac{2}{\sqrt{11}}\sin\frac{\pi}{11}.
  \label{eq:dual-S3-value}
\end{equation}
The trigonometric expression is, as on the original side, a numerical recognition rather than an analytic evaluation of the localization integral.

The dual result agrees with the original-theory $F$-maximization of Section~2.3 within the numerical precision of the computation.  In particular,
\begin{equation}
  \left|Z_{S^3}^\vee(\boldsymbol\mu_0^\vee)\right|
  =\left|Z_{S^3}(\boldsymbol\mu_0)\right|
  \label{eq:dual-original-S3-match}
\end{equation}
numerically.  This equality provides a further nontrivial check of the particle--vortex relation between the two ultraviolet descriptions.

The two candidate infrared mixing vectors are related by the same Cartan transformation that appeared in the ultraviolet duality.  Using
\begin{equation}
  \boldsymbol\mu_0=\boldsymbol a,
  \qquad
  \boldsymbol\mu_0^\vee=-\boldsymbol a^\vee,
\end{equation}
and \eqref{eq:dual-residual-Cartan-relation}, we obtain
\begin{equation}
  C_{E_8}\boldsymbol\mu_0^\vee
  =\boldsymbol\mu_0,
  \qquad
  \boldsymbol\mu_0^\vee
  =C_{E_8}^{-1}\boldsymbol\mu_0.
  \label{eq:dual-original-mixing-map}
\end{equation}
Thus the Cartan/inverse-Cartan transformation relates not only the gauge Chern--Simons matrices but also the infrared $R$-symmetry mixing vectors.

Finally,
\begin{equation}
  \boldsymbol\mu_0^\vee\in\mathbb Z^8
  \subset\left(\frac12\mathbb Z\right)^8,
  \label{eq:dual-half-integral-mixing}
\end{equation}
so the dual mixing is compatible with the $R$-charge quantization used in the half-index conventions.  The superconformal index provides the next independent test of the infrared $\mathcal N=4$ mirror relation.

\subsection{Supersymmetric indices: tests of mirror symmetry}
\label{subsec:dual-SCI-mirror-map}

We now compare the superconformal index of the dual theory with that of the original theory.  The comparison can be carried out directly at the level of the localization sums and gives an exact relation to all orders in $q$.  We use the charge-basis tetrahedron index $\mathcal I_\Delta(m,e)$ defined in \eqref{eq:tetrahedron-index} and the same $(-1)^R$ convention as in Section~2.4.

Let
\begin{equation}
  \boldsymbol m\in\mathbb Z^8,
  \qquad
  \boldsymbol\ell\in\mathbb Z^8
  \label{eq:original-dual-flux-lattices}
\end{equation}
denote the GNO magnetic fluxes of $\mathcal T$ and $\mathcal T^\vee$, respectively.  At the candidate superconformal mixing $\boldsymbol\mu_0^\vee=-\boldsymbol a^\vee$ determined in Section~3.2, localization gives
\begin{equation}
  \mathcal I^\vee(q,\eta)
  =\sum_{\boldsymbol\ell\in\mathbb Z^8}
  \eta^{\boldsymbol a^\vee\cdot\boldsymbol\ell}
  \left(-q^{1/2}\right)^{-\boldsymbol a^\vee\cdot\boldsymbol\ell}
  \prod_{i=1}^{8}
  \mathcal I_\Delta\!\left(
    \ell_i,
    -\bigl(C_{E_8}^{-1}\boldsymbol\ell\bigr)_i
  \right).
  \label{eq:dual-full-SCI}
\end{equation}
The exponent of $\eta$ is the charge under the surviving topological symmetry $\mathsf T^\vee=\boldsymbol a^\vee\cdot\boldsymbol T^\vee$, while the power of $-q^{1/2}$ implements the mixing $\boldsymbol\mu_0^\vee=-\boldsymbol a^\vee$.

The particle--vortex construction of Section~3.1 provides the natural change of magnetic variables
\begin{equation}
  \boldsymbol\ell=C_{E_8}\boldsymbol m.
  \label{eq:index-flux-map}
\end{equation}
Because $C_{E_8}$ is an integral unimodular matrix, \eqref{eq:index-flux-map} is a bijection of $\mathbb Z^8$.  Using the symmetry of $C_{E_8}$ together with \eqref{eq:dual-residual-Cartan-relation}, we obtain
\begin{equation}
  C_{E_8}^{-1}\boldsymbol\ell=\boldsymbol m,
  \qquad
  \boldsymbol a^\vee\cdot\boldsymbol\ell
  =-\boldsymbol a\cdot\boldsymbol m.
  \label{eq:index-charge-map}
\end{equation}
Thus the surviving topological charges are identified with opposite sign.  This is the origin of the inversion of the topological fugacity in the mirror map.

The second ingredient is the exact polarization-exchange identity of the tetrahedron index~\cite{Dimofte:2011py},
\begin{equation}
  \mathcal I_\Delta(m,e)
  =\mathcal I_\Delta(-e,-m).
  \label{eq:tetrahedron-polarization-exchange}
\end{equation}
In the present variables it implies, component by component,
\begin{equation}
  \mathcal I_\Delta\!\left(
    (C_{E_8}\boldsymbol m)_i,-m_i
  \right)
  =
  \mathcal I_\Delta\!\left(
    m_i,-(C_{E_8}\boldsymbol m)_i
  \right).
  \label{eq:tetrahedron-factor-map}
\end{equation}
Substituting \eqref{eq:index-flux-map}--\eqref{eq:tetrahedron-factor-map} into \eqref{eq:dual-full-SCI} gives
\begin{align}
  \mathcal I^\vee(q,\eta)
  &={}
  \sum_{\boldsymbol m\in\mathbb Z^8}
  \eta^{-\boldsymbol a\cdot\boldsymbol m}
  \left(-q^{1/2}\right)^{\boldsymbol a\cdot\boldsymbol m}
  \prod_{i=1}^{8}
  \mathcal I_\Delta\!\left(
    m_i,-(C_{E_8}\boldsymbol m)_i
  \right)
  \nonumber\\
  &=\mathcal I(q,\eta^{-1}),
  \label{eq:exact-index-mirror-relation}
\end{align}
where in the last equality we used $\boldsymbol\mu_0=\boldsymbol a$ in the original-theory expression \eqref{eq:full-SCI}.  Equation \eqref{eq:exact-index-mirror-relation} is therefore an exact identity of formal $q^{1/2}$-series with Laurent-polynomial coefficients in $\eta$, rather than a finite-order comparison.

Using the original-theory expansion \eqref{eq:SCI-expansion}, the dual superconformal index begins as
\begin{equation}
\begin{aligned}
  \mathcal I^\vee(q,\eta)
  ={}&1-q-(\eta+\eta^{-1})q^{3/2}-2q^2
  -\eta q^{5/2}\\
  &+(\eta^{-2}-1)q^3+O(q^{7/2}).
\end{aligned}
  \label{eq:dual-SCI-expansion}
\end{equation}
In particular, the same charged pair
\begin{equation}
  -(\eta+\eta^{-1})q^{3/2}
  \label{eq:dual-extra-supercurrent-signal}
\end{equation}
appears in the dual theory.  As in Section~2.4, this is precisely the contribution expected from the two additional conserved-supercurrent multiplets needed for enhancement from $\mathcal N=2$ to $\mathcal N=4$~\cite{GangKimParkStubbs:2024Nahm}.  The exact index relation therefore shows that the supersymmetry-enhancement signal is shared by the two particle--vortex descriptions.

The same identity also determines the mirror map between the $A$- and $B$-twisted specializations.  Define, in direct analogy with \eqref{eq:AB-index-specializations},
\begin{equation}
  \mathcal I_\nu^\vee(q)
  :=\mathcal I^\vee\!\left(
    q,\eta=\left(-q^{1/2}\right)^\nu
  \right),
  \qquad
  \mathcal I_A^\vee:=\mathcal I_{-1}^\vee,
  \qquad
  \mathcal I_B^\vee:=\mathcal I_{+1}^\vee.
  \label{eq:dual-AB-index-specializations}
\end{equation}
Equation \eqref{eq:exact-index-mirror-relation} immediately gives
\begin{equation}
  \mathcal I_\nu^\vee(q)=\mathcal I_{-\nu}(q),
  \label{eq:dual-index-nu-map}
\end{equation}
and hence
\begin{equation}
  \mathcal I_A^\vee(q)=\mathcal I_B(q),
  \qquad
  \mathcal I_B^\vee(q)=\mathcal I_A(q).
  \label{eq:AB-index-exchange}
\end{equation}
Thus the two distinguished topological specializations are exchanged exactly, as required by 3d $\mathcal N=4$ mirror symmetry.  In particular, the rank-zero behavior found for both specializations of the original theory in \eqref{eq:A-index-rank-zero}--\eqref{eq:B-index-rank-zero} is inherited by the dual theory with the $A$ and $B$ labels exchanged.

The mirror map visible in the superconformal index is therefore
\begin{equation}
  \boldsymbol\ell=C_{E_8}\boldsymbol m,
  \qquad
  \mathsf T^\vee=-\mathsf T,
  \qquad
  \eta^\vee=\eta^{-1},
  \qquad
  A\longleftrightarrow B.
  \label{eq:SCI-mirror-map-summary}
\end{equation}
Together with the particle--vortex derivation of the dual Lagrangian in Section~3.1 and the matching sphere partition functions in Section~3.2, the exact all-order identity \eqref{eq:exact-index-mirror-relation} provides strong evidence that the two monopole-superpotential-deformed theories flow to a mirror pair of interacting rank-zero $\mathcal N=4$ SCFTs.

\subsection{Wilson loop half-indices and the emergence of Zagier duality}
\label{subsec:dual-Wilson-Zagier}

In this subsection, Wilson loop half-indices provide Nahm-sum expressions for the non-vacuum characters, completing the five-character systems of $T_{10}M_{\mathrm{eff}}(11,2)$ and $M_{\mathrm{eff}}(2,11)$.  We then compare the two complete systems under 3d particle--vortex duality.  We will call their resulting componentwise relation \emph{Zagier duality}; at the level of each individual sector, it is implemented by the Zagier transformation of the corresponding Nahm triple introduced below.

The order of the argument is important.  The dual bulk theory was obtained in Section~\ref{subsec:dual-Lagrangian-PV} by applying particle--vortex duality to the 3d $U(1)^8$ Chern--Simons matter theory on $\mathbb R^3$.  In particular, the integral Chern--Simons matrix is transformed as
\begin{equation}
  C_{E_8}\longmapsto C_{E_8}^{-1}.
  \label{eq:bulk-CS-inverse-map-recap}
\end{equation}
This transformation is obtained directly at the level of the 3d bulk theory, before computing any boundary half-index.

The same particle--vortex transformation determines the map of Wilson line charges.  A Wilson line of electric charge $\boldsymbol Q$ enters the topological action as a linear source for the gauge field.  In the Gaussian step of the particle--vortex transformation, the same inverse matrix that acts on the quadratic Chern--Simons coupling acts on this linear source.  With the Wilson loop convention used in Section~\ref{subsec:Wilson-half-indices}, this gives
\begin{equation}
  \boldsymbol Q^\vee=C_{E_8}^{-1}\boldsymbol Q.
  \label{eq:Wilson-charge-pv-map}
\end{equation}
Because $C_{E_8}$ is unimodular, $C_{E_8}^{-1}$ is integral, so an integral Wilson charge is again mapped to an integral Wilson charge.  The same transmission law will be recovered independently from the particle--vortex complement map and the Bethe-vacuum relations in Section~\ref{sec:AB-twisted-TQFTs}, and from the duality wall in Section~\ref{sec:E8-interface}.

Having determined the dual bulk theory and the Wilson-charge map, we now compute the half-indices on the two sides.  Recall first the $A$-twisted Wilson loop half-index of the original theory $\mathcal T$ from Section~\ref{subsec:Wilson-half-indices}:
\begin{equation}
  I\!\!I^{A}[W_{\boldsymbol Q}](q)
  =\sum_{\boldsymbol m\in\mathbb Z_{\geq0}^{8}}
  \frac{
  q^{\frac12\boldsymbol m^{\mathsf T}C_{E_8}\boldsymbol m
  -\boldsymbol Q^{\mathsf T}\boldsymbol m}
  }{
  \prod_{i=1}^{8}(q)_{m_i}
  }.
  \label{eq:original-Wilson-half-recalled}
\end{equation}

We next compute the corresponding half-index directly in the particle--vortex dual theory $\mathcal T^\vee$.  It is useful first to make explicit which $R$-charge specialization defines the $B$-twisted half-index.  Assuming the infrared $\mathcal N=4$ enhancement, we normalize the Cartan generators of the dual $SU(2)_H^\vee\times SU(2)_C^\vee$ symmetry by
\begin{equation}
  R_{\boldsymbol\mu_0^\vee}^{\vee}
  =J_3^{C,\vee}+J_3^{H,\vee},
  \qquad
  \mathsf T^\vee
  =J_3^{C,\vee}-J_3^{H,\vee},
  \label{eq:dual-N4-R-decomposition-half}
\end{equation}
in direct analogy with \eqref{eq:N4-R-decomposition}.  For
\begin{equation}
  R_\nu^\vee
  :=R_{\boldsymbol\mu_0^\vee}^{\vee}+\nu\mathsf T^\vee,
  \qquad
  \boldsymbol\mu_\nu^\vee
  :=\boldsymbol\mu_0^\vee+\nu\boldsymbol a^\vee,
  \label{eq:dual-half-Rnu}
\end{equation}
we then have
\begin{equation}
  R_{-1}^\vee=2J_3^{H,\vee},
  \qquad
  R_{+1}^\vee=2J_3^{C,\vee}.
  \label{eq:dual-half-AB-R}
\end{equation}
Thus the $B$-twisted choice is $\nu=+1$.  Since the superconformal mixing determined in Section~3.2 is $\boldsymbol\mu_0^\vee=-\boldsymbol a^\vee$, its effective mixing vector is
\begin{equation}
  \boldsymbol\mu_B^\vee
  =\boldsymbol\mu_0^\vee+\boldsymbol a^\vee
  =0.
  \label{eq:dual-B-half-mixing-cancellation}
\end{equation}
Equivalently, the reference $R$-charge with zero topological mixing, $R_{\boldsymbol 0}^\vee$, is precisely the $B$-twisted $R$-charge of the dual theory.  Thus the unrefined $C_{E_8}^{-1}$ Nahm sum obtained below is precisely the $B$-twisted $(\mathcal D,D_c)$ half-index of $\mathcal T^\vee$, selected from the superconformal grading by the $\nu=+1$ specialization.

The half-index version of the specialization in \eqref{eq:dual-AB-index-specializations} can be written directly at the level of the boundary topological fugacities.  For a Wilson loop of charge $\boldsymbol Q^\vee$ wrapping the core circle of $S^1\times D^2$, define
\begin{equation}
  I\!\!I^{\vee,\nu}[W^\vee_{\boldsymbol Q^\vee}](q)
  :=I\!\!I^{\vee}[W^\vee_{\boldsymbol Q^\vee}]
  \!\left(q,\boldsymbol x^\vee(\nu)\right),
  \qquad
  x_i^\vee(\nu)
  :=\left(-q^{1/2}\right)^{(\boldsymbol\mu_0^\vee+\nu\boldsymbol a^\vee)_i},
  \label{eq:dual-half-index-specialization}
\end{equation}
with
\begin{equation}
  I\!\!I^{\vee A}[W^\vee_{\boldsymbol Q^\vee}]
  :=I\!\!I^{\vee,\nu=-1}[W^\vee_{\boldsymbol Q^\vee}],
  \qquad
  I\!\!I^{\vee B}[W^\vee_{\boldsymbol Q^\vee}]
  :=I\!\!I^{\vee,\nu=+1}[W^\vee_{\boldsymbol Q^\vee}].
  \label{eq:dual-half-AB-specializations}
\end{equation}
At the $B$-twisted point, \eqref{eq:dual-B-half-mixing-cancellation} gives $x_i^\vee(+1)=1$ for all $i$.  We impose on $\mathcal T^\vee$ the same $(\mathcal D,D_c)$ choice---Dirichlet $\mathcal D$ for all eight dual vector multiplets and deformed Dirichlet $D_c$ for all eight dual chirals---independently of the boundary condition chosen on $\mathcal T$.  As on the original side, we suppress the boundary label whenever a Wilson loop is inserted.  With the same Wilson loop sign convention, the resulting $B$-twisted Wilson loop half-index is
\begin{equation}
  I\!\!I^{\vee B}[W^\vee_{\boldsymbol Q^\vee}](q)
  =
  \sum_{\boldsymbol\ell\in\mathbb Z_{\geq0}^{8}}
  \frac{
  q^{\frac12\boldsymbol\ell^{\mathsf T}C_{E_8}^{-1}\boldsymbol\ell
  -(\boldsymbol Q^\vee)^{\mathsf T}\boldsymbol\ell}
  }{
  \prod_{i=1}^{8}(q)_{\ell_i}
  }.
  \label{eq:dual-Wilson-half-index}
\end{equation}
Since $C(T_1)=(1)$, the quadratic form in \eqref{eq:dual-Wilson-half-index} is
$C(T_1)\otimes C(E_8)^{-1}=C_{E_8}^{-1}$, precisely the quadratic form of the $(T_1,E_8)$ Nahm sum.
Equation~\eqref{eq:dual-Wilson-half-index} is obtained by localizing the dual 3d theory itself.  In particular, it is not obtained by applying a transformation to the localized half-index \eqref{eq:original-Wilson-half-recalled} of the original theory.  For zero Wilson charge we write the vacuum half-index explicitly as
\begin{equation}
  I\!\!I_{(\mathcal D,D_c)}^{\vee B}(q)
  :=I\!\!I^{\vee B}[W^\vee_{\boldsymbol Q_0^\vee}](q),
  \qquad \boldsymbol Q_0^\vee=0.
  \label{eq:dual-vacuum-DDc-half-index}
\end{equation}

Let
\begin{equation}
  \mathcal H
  =\left\{0,\frac5{11},\frac8{11},\frac{10}{11},\frac{12}{11}\right\}
\end{equation}
denote the five Wilson sectors of the original theory introduced in Section~\ref{subsec:Wilson-half-indices}.  Applying the particle--vortex Wilson line map \eqref{eq:Wilson-charge-pv-map} to their charge representatives gives
\begin{equation}
  \widetilde{\boldsymbol Q}_{h}^{\,\vee}
  :=C_{E_8}^{-1}\boldsymbol Q_h,
  \qquad h\in\mathcal H.
  \label{eq:dual-charges-from-pv}
\end{equation}
In the ordering
\begin{equation}
  h=0,\frac5{11},\frac8{11},\frac{10}{11},\frac{12}{11},
\end{equation}
the five transformed charges are
\begin{align}
  \widetilde{\boldsymbol Q}_{0}^{\,\vee}
  &=(0,0,0,0,0,0,0,0),\nonumber\\
  \widetilde{\boldsymbol Q}_{5/11}^{\,\vee}
  &=(0,-2,-2,-4,-4,-3,-1,-3),\nonumber\\
  \widetilde{\boldsymbol Q}_{8/11}^{\,\vee}
  &=(0,-1,-1,-2,-2,-2,0,-2),\nonumber\\
  \widetilde{\boldsymbol Q}_{10/11}^{\,\vee}
  &=(0,0,0,-1,-1,-1,0,-1),\nonumber\\
  \widetilde{\boldsymbol Q}_{12/11}^{\,\vee}
  &=(-1,-3,-4,-6,-7,-5,-2,-4).
  \label{eq:dual-Wilson-charge-list}
\end{align}

Direct expansion of the corresponding dual half-indices gives
\begin{align}
I\!\!I^{\vee B}[W^\vee_{\widetilde{\boldsymbol Q}_{0}^{\,\vee}}](q)
={}&1+q+2q^2+3q^3+5q^4+6q^5+9q^6+O(q^7),\nonumber\\
I\!\!I^{\vee B}[W^\vee_{\widetilde{\boldsymbol Q}_{5/11}^{\,\vee}}](q)
={}&1+q+q^2+2q^3+3q^4+4q^5+6q^6+O(q^7),\nonumber\\
I\!\!I^{\vee B}[W^\vee_{\widetilde{\boldsymbol Q}_{8/11}^{\,\vee}}](q)
={}&1+q+2q^2+2q^3+4q^4+5q^5+8q^6+O(q^7),\nonumber\\
I\!\!I^{\vee B}[W^\vee_{\widetilde{\boldsymbol Q}_{10/11}^{\,\vee}}](q)
={}&1+q+2q^2+3q^3+4q^4+6q^5+9q^6+O(q^7),\nonumber\\
I\!\!I^{\vee B}[W^\vee_{\widetilde{\boldsymbol Q}_{12/11}^{\,\vee}}](q)
={}&1+q^2+q^3+2q^4+2q^5+4q^6+O(q^7).
  \label{eq:dual-Wilson-series}
\end{align}
The vacuum series, after multiplication by $q^{-1/33}$, is the known all-order vacuum character of $M_{\mathrm{eff}}(2,11)$.  Comparing the remaining four series with the other four characters of the same five-component system identifies the particle--vortex-transformed Wilson charges, through relative order $q^6$, with the effective weights
\begin{equation}
  0,\qquad \frac6{11},\qquad \frac3{11},\qquad \frac1{11},\qquad \frac{10}{11},
\end{equation}
respectively.  Thus the Wilson line map induced by particle--vortex duality determines the sector pairing
\begin{equation}
  0\mapsto0,
  \qquad
  \frac5{11}\mapsto\frac6{11},
  \qquad
  \frac8{11}\mapsto\frac3{11},
  \qquad
  \frac{10}{11}\mapsto\frac1{11},
  \qquad
  \frac{12}{11}\mapsto\frac{10}{11}.
  \label{eq:particle-vortex-sector-map}
\end{equation}
We denote the paired dual weight by $h^\vee$ and from now on write
\begin{equation}
  \boldsymbol Q_{h^\vee}^\vee
  :=\widetilde{\boldsymbol Q}_{h}^{\,\vee}
  =C_{E_8}^{-1}\boldsymbol Q_h.
  \label{eq:paired-dual-Wilson-charge}
\end{equation}
Equivalently, the set of dual effective weights is
\begin{equation}
  \mathcal H^\vee
  =\left\{0,\frac1{11},\frac3{11},\frac6{11},\frac{10}{11}\right\}.
  \label{eq:dual-five-weight-set}
\end{equation}

It remains to determine the overall powers of $q$.  All entries of $C_{E_8}^{-1}$ are nonnegative, while the four nonzero vectors $\boldsymbol Q_{h^\vee}^\vee$ above are componentwise nonpositive.  Hence
\begin{equation}
  \frac12\boldsymbol\ell^{\mathsf T}C_{E_8}^{-1}\boldsymbol\ell
  -(\boldsymbol Q_{h^\vee}^\vee)^{\mathsf T}\boldsymbol\ell
  \geq0,
  \qquad
  \boldsymbol\ell\in\mathbb Z_{\geq0}^{8},
\end{equation}
with equality at $\boldsymbol\ell=\boldsymbol0$.  The lattice minimum is therefore zero in every dual Wilson sector.  The vacuum character fixes the common dual Casimir normalization to $-1/33$.  Once a transformed Wilson sector has been identified with effective grade $h^\vee$, its overall power is consequently
\begin{equation}
  \gamma_{h^\vee}^\vee
  =h^\vee-\frac1{33}.
  \label{eq:dual-Wilson-normalization}
\end{equation}
Thus
\begin{equation}
  \left(\gamma_{h^\vee}^\vee\right)
  =\frac1{33}(-1,17,8,2,29)
\end{equation}
in the paired ordering
\begin{equation}
  h^\vee=0,\frac6{11},\frac3{11},\frac1{11},\frac{10}{11}.
\end{equation}
We denote by
\begin{equation}
  \left\{
  \chi_0^\vee,\,
  \chi_{6/11}^\vee,\,
  \chi_{3/11}^\vee,\,
  \chi_{1/11}^\vee,\,
  \chi_{10/11}^\vee
  \right\}
  \label{eq:dual-complete-character-system}
\end{equation}
the complete five-character system of $M_{\mathrm{eff}}(2,11)$, written in the ordering paired with the five Wilson sectors of $T_{10}M_{\mathrm{eff}}(11,2)$ under particle--vortex duality.  The superscript $\vee$ is used to distinguish this character system from the $T_{10}M_{\mathrm{eff}}(11,2)$ characters $\chi_h$ introduced in Section~\ref{subsec:Wilson-half-indices}.

Restoring these overall powers gives
\begin{align}
  \chi_0^\vee(q)
  &=q^{-1/33}I\!\!I_{(\mathcal D,D_c)}^{\vee B}(q),\nonumber\\
  \chi_{6/11}^\vee(q)
  &=q^{17/33}I\!\!I^{\vee B}[W^\vee_{\boldsymbol Q_{6/11}^\vee}](q),\nonumber\\
  \chi_{3/11}^\vee(q)
  &=q^{8/33}I\!\!I^{\vee B}[W^\vee_{\boldsymbol Q_{3/11}^\vee}](q),\nonumber\\
  \chi_{1/11}^\vee(q)
  &=q^{2/33}I\!\!I^{\vee B}[W^\vee_{\boldsymbol Q_{1/11}^\vee}](q),\nonumber\\
  \chi_{10/11}^\vee(q)
  &=q^{29/33}I\!\!I^{\vee B}[W^\vee_{\boldsymbol Q_{10/11}^\vee}](q).
  \label{eq:dual-Wilson-character-match}
\end{align}
We now introduce the arithmetic transformation with which these physically derived data should be compared.  A rank-$r$ Nahm sum associated with a triple $(\mathsf A,\mathsf B,\mathsf C)$ is the $q$-hypergeometric series
\begin{equation}
  F_{\mathsf A,\mathsf B,\mathsf C}(q)
  :=
  \sum_{\boldsymbol n\in\mathbb Z_{\geq0}^{r}}
  \frac{
  q^{\frac12\boldsymbol n^{\mathsf T}\mathsf A\boldsymbol n
  +\mathsf B^{\mathsf T}\boldsymbol n+\mathsf C}
  }{
  \prod_{i=1}^{r}(q)_{n_i}
  },
  \label{eq:general-Nahm-sum}
\end{equation}
where $\mathsf A\in\operatorname{Mat}_{r\times r}(\mathbb Q)$ is symmetric and positive definite, $\mathsf B\in\mathbb Q^r$, and $\mathsf C\in\mathbb Q$.  The matrix $\mathsf A$, vector $\mathsf B$, and scalar $\mathsf C$ are, respectively, the quadratic, linear, and constant data of the exponent.  Nahm's modularity problem asks for which rational triples $(\mathsf A,\mathsf B,\mathsf C)$ the series \eqref{eq:general-Nahm-sum} is modular.  The problem was motivated in part by fermionic sum representations of rational 2d conformal-field-theory characters and their relation to dilogarithm and Bloch-group data~\cite{Nahm:2004bloch,VlasenkoZwegers:2011nahm}.

In his systematic study of this problem, Zagier observed a striking duality of the Nahm data~\cite{Zagier:2007dilogarithm}.  The proposed transformation is
\begin{equation}
  (\mathsf A,\mathsf B,\mathsf C)
  \longmapsto
  (\mathsf A^\vee,\mathsf B^\vee,\mathsf C^\vee)
  =
  \left(
  \mathsf A^{-1},
  \mathsf A^{-1}\mathsf B,
  \frac12\mathsf B^{\mathsf T}\mathsf A^{-1}\mathsf B
  -\frac{r}{24}-\mathsf C
  \right).
  \label{eq:general-Zagier-transform}
\end{equation}
Several features of Nahm's modularity problem motivated this proposal.  The matrix condition entering Nahm's conjectural modularity criterion is invariant under $\mathsf A\mapsto\mathsf A^{-1}$; the asymptotic expansions of a Nahm sum and the series built from the transformed data are closely related; on the conformal-field-theory side the involution $\mathsf A\leftrightarrow\mathsf A^{-1}$ is related to familiar coset and level-rank duality phenomena; and Zagier's low-rank tables contained modular examples paired by the transformation \eqref{eq:general-Zagier-transform}~\cite{Zagier:2007dilogarithm,Wang:2024zagier}.  These observations led to the conjecture that modularity should be preserved under \eqref{eq:general-Zagier-transform}.  The latter statement is not valid for arbitrary Nahm triples: explicit rank-four counterexamples are now known~\cite{Wang:2024zagier}.  We therefore do not assume a general modularity-preservation theorem here.  We use ``Zagier transformation'' only for the explicit transformation of one Nahm triple in \eqref{eq:general-Zagier-transform}.  When two complete character systems admit sector-by-sector Nahm triples and every sector on one side is mapped to its paired sector on the other by this transformation, including the constant term, we will call the resulting relation between the two complete systems \emph{Zagier duality}.  The claim below is that the $(E_8,T_1)$ and $(T_1,E_8)$ five-component systems are related by Zagier duality, and that the required sector pairing is generated independently by the 3d particle--vortex mirror map.

For the original Wilson loop half-index \eqref{eq:original-Wilson-half-recalled}, including its overall $q$-normalization, the corresponding rank-eight Nahm triple is
\begin{equation}
  (\mathsf A,\mathsf B,\mathsf C)
  =\left(C_{E_8},-\boldsymbol Q_h,\gamma_h\right).
  \label{eq:Nahm-data-from-Wilson}
\end{equation}
The dual half-index \eqref{eq:dual-Wilson-half-index} has
\begin{equation}
  (\mathsf A^\vee,\mathsf B^\vee)
  =\left(C_{E_8}^{-1},-\boldsymbol Q_{h^\vee}^\vee\right).
\end{equation}
Using the particle--vortex Wilson line map \eqref{eq:paired-dual-Wilson-charge}, we find
\begin{equation}
  \mathsf A^\vee=\mathsf A^{-1},
  \qquad
  \mathsf B^\vee=\mathsf A^{-1}\mathsf B.
  \label{eq:pv-Zagier-matrix-map}
\end{equation}
Thus the inverse quadratic form and the transformed linear term appearing in the Wilson loop half-indices coincide precisely with the first two components of the Zagier transformation.  Importantly, these relations were not imposed on a localized expression: they were derived first from particle--vortex duality of the 3d bulk theory and its Wilson lines, and only now recognized as the arithmetic transformation \eqref{eq:general-Zagier-transform}.

For later use in the discussion of duality-transformed boundary conditions, if a flux sector is followed through the particle--vortex map $\boldsymbol\ell=C_{E_8}\boldsymbol m$, equations \eqref{eq:Wilson-charge-pv-map} and \eqref{eq:bulk-CS-inverse-map-recap} imply the algebraic identity
\begin{equation}
  \frac12\boldsymbol m^{\mathsf T}C_{E_8}\boldsymbol m
  -\boldsymbol Q^{\mathsf T}\boldsymbol m
  =
  \frac12\boldsymbol\ell^{\mathsf T}C_{E_8}^{-1}\boldsymbol\ell
  -(\boldsymbol Q^\vee)^{\mathsf T}\boldsymbol\ell.
  \label{eq:pv-Zagier-exponent}
\end{equation}
This identity is a consequence of the already derived bulk and line maps; it is not used to define the independently localized dual half-index \eqref{eq:dual-Wilson-half-index}.

It remains to compare the constant term.  For rank eight, the third component of \eqref{eq:general-Zagier-transform} predicts
\begin{equation}
  \mathsf C^\vee
  =\frac12\mathsf B^{\mathsf T}\mathsf A^{-1}\mathsf B
  -\frac13-\mathsf C.
  \label{eq:rank-eight-Zagier-transform}
\end{equation}
Substituting \eqref{eq:Nahm-data-from-Wilson} gives
\begin{equation}
  \gamma_{h^\vee}^\vee
  =\frac12\boldsymbol Q_h^{\mathsf T}C_{E_8}^{-1}\boldsymbol Q_h
  -\frac13-\gamma_h,
  \label{eq:componentwise-charge-constant-map}
\end{equation}
with $h^\vee$ paired with $h$ as above.  In the ordering
\begin{equation}
  h=0,\frac5{11},\frac8{11},\frac{10}{11},\frac{12}{11},
\end{equation}
the quadratic contributions are
\begin{equation}
  \left(
  \frac12\boldsymbol Q_h^{\mathsf T}C_{E_8}^{-1}\boldsymbol Q_h
  \right)
  =(0,4,3,1,2),
  \label{eq:Wilson-quadratic-values}
\end{equation}
and \eqref{eq:componentwise-charge-constant-map} reproduces the five values of $\gamma_{h^\vee}^\vee$ obtained above.  Therefore the particle--vortex correspondence of the five Wilson loop half-indices agrees componentwise with the full rank-eight Zagier transformation, including its scalar term.  This establishes the Zagier duality between the complete $(E_8,T_1)$ and $(T_1,E_8)$ Nahm systems in the sense defined above.

\paragraph{Normalization of the paired Wilson sectors.}
We can now return to the shifts $\gamma_h$ introduced provisionally in Section~\ref{subsec:Wilson-half-indices}.  Solving \eqref{eq:componentwise-charge-constant-map} for the original-side shift gives
\begin{equation}
  \gamma_h
  =\frac12\boldsymbol Q_h^{\mathsf T}C_{E_8}^{-1}\boldsymbol Q_h
  -\frac13-\gamma_{h^\vee}^\vee.
  \label{eq:original-normalization-from-duality}
\end{equation}
Using \eqref{eq:dual-Wilson-normalization} together with \eqref{eq:Wilson-quadratic-values}, one obtains
\begin{equation}
  (\gamma_h)
  =\frac1{33}(-10,104,80,20,26),
  \label{eq:original-normalization-vector-from-duality}
\end{equation}
which is exactly the vector recorded independently in Section~\ref{subsec:Wilson-half-indices}.  Thus the five original $q$-shifts are not five freely fitted constants.  Once the vacuum Casimir normalizations on the two sides, the paired Wilson-sector grading, and the scalar contact/framing convention are fixed, particle--vortex duality relates all five shifts by \eqref{eq:original-normalization-from-duality}.

The same statement can be expressed directly in terms of the character grades.  Combining
\begin{equation}
  \gamma_h=h-\frac{10}{33}-\delta_h,
  \qquad
  \gamma_{h^\vee}^\vee=h^\vee-\frac1{33},
\end{equation}
with \eqref{eq:componentwise-charge-constant-map} gives
\begin{equation}
  h+h^\vee
  =\delta_h
  +\frac12\boldsymbol Q_h^{\mathsf T}C_{E_8}^{-1}\boldsymbol Q_h.
  \label{eq:paired-grade-from-charge-data}
\end{equation}
For the five sectors, the right-hand side is
\begin{equation}
  (0,1,1,1,2),
  \label{eq:paired-grade-integers-from-charge-data}
\end{equation}
so the integer sums of the paired grades follow directly from the raw half-index minima and the 3d quadratic charge data.  Equivalently, inserting \eqref{eq:original-normalization-vector-from-duality} back into
\begin{equation}
  h=\gamma_h+\delta_h+\frac{10}{33}
  \label{eq:original-grade-from-normalization}
\end{equation}
recovers
\begin{equation}
  h=0,\frac5{11},\frac8{11},\frac{10}{11},\frac{12}{11}.
  \label{eq:original-grades-recovered}
\end{equation}

It is useful to separate what is fixed by 3d topological data from what remains a boundary-grading choice.  As shown in Section~\ref{sec:AB-twisted-TQFTs}, the normalized modular $T$ eigenvalues determine the topological spins and hence fix $h$ and $h^\vee$ modulo integers.  Equation~\eqref{eq:paired-grade-from-charge-data} fixes the integer sum of each paired lift, but topological data alone do not distinguish, for example, $1/11$ from $12/11$ on a single side.  The remaining integer lift is part of the boundary character grading.  Once that grading is chosen on one side, \eqref{eq:original-normalization-from-duality} fixes the corresponding shift on the other side.  In this precise sense, the Wilson-sector normalizations are constrained by the vacuum normalization and the duality/interface data rather than adjusted independently sector by sector.

Therefore the complete five Nahm triples obey the exact componentwise identity
\begin{equation}
  (\mathsf A^\vee,\mathsf B^\vee,\mathsf C^\vee)
  =
  \left(
  C_{E_8}^{-1},
  -\boldsymbol Q_{h^\vee}^\vee,
  \gamma_{h^\vee}^\vee
  \right)
  =
  \left(
  \mathsf A^{-1},
  \mathsf A^{-1}\mathsf B,
  \frac12\mathsf B^{\mathsf T}\mathsf A^{-1}\mathsf B
  -\frac13-\mathsf C
  \right),
  \qquad
  h\in\mathcal H,
  \label{eq:componentwise-Zagier-triples}
\end{equation}
where $(\mathsf A,\mathsf B,\mathsf C)=(C_{E_8},-\boldsymbol Q_h,\gamma_h)$.  Equation~\eqref{eq:componentwise-Zagier-triples} is therefore best viewed as the arithmetic packaging of the 3d transformation.  Its first entry is generated by particle--vortex duality, and its second entry is the corresponding Wilson line transmission law $\boldsymbol Q\mapsto C_{E_8}^{-1}\boldsymbol Q$.  The remaining scalar transformation contains the universal rank-eight shift $-8/24=-1/3$ and fixes the character normalization.  We will return in Section~\ref{subsec:folded-lines-E8-phase} to the physical meaning of this scalar shift, where the twisted fibering data identify it with the relative invertible $E_8$ framing sector.

\subsection{Duality-transformed boundaries and the $(E_8)_1$ lattice sector}
\label{subsec:mirror-boundary-half-index}

The comparison in Section~\ref{subsec:dual-Wilson-Zagier} used the $(\mathcal D,D_c)$ boundary condition independently in the two particle--vortex dual bulk theories.  Their half-indices are not equal in general; rather, their Nahm triples are related componentwise by the Zagier transformation.  We now address a different question: what protected boundary data are obtained when the $(\mathcal D,D_c)$ boundary condition is related to the particle--vortex description itself?  The answer is most transparent if one first reviews the elementary one-node boundary identity and only then gauges eight copies with the matrix $C_{E_8}$.

There is an orientation issue that must be kept explicit.  Dimofte--Gaiotto--Paquette distinguish a right boundary, for a theory occupying $x^\perp\leq0$, from a left boundary, for a theory occupying $x^\perp\geq0$, while preserving the same 2d $\mathcal N=(0,2)$ algebra~\cite{Dimofte:2017tpi}.  The sign of Chern--Simons anomaly inflow changes between the two sides, while the sign of a fixed $(0,2)$ matter contribution does not.  We therefore keep track of right and left boundary data separately below.  This orientation dependence is a property of the wall geometry; it does not mean that an additional parity transformation has been inserted into the bulk mirror duality.

We also recall the standard localization distinction between gauge boundary conditions.  A Neumann boundary condition $\mathcal N$ for a 3d vector multiplet leaves the boundary gauge symmetry dynamical and produces a holonomy integral.  A Dirichlet boundary condition $\mathcal D$ leaves a boundary flavor symmetry and produces a sum over boundary monopole sectors~\cite{YoshidaSugiyama:2014S1D2,Dimofte:2017tpi}.  Thus a monopole sum and a holonomy integral should not be assigned simultaneously to the same 3d vector multiplet.  In the construction below the two operations arise from different gauge fields before the old Dirichlet gauge variables are eliminated.

\paragraph{Elementary vortex--Wilson sectors.}
Let $T$ denote a free chiral multiplet and let
\begin{equation}
  T'=U(1)_{+1/2}+\text{one charge-one chiral}
  \label{eq:DGP-elementary-PV-pair}
\end{equation}
be the particle--vortex presentation used in~\cite{Dimofte:2017tpi}.  Their boundary duality contains
\begin{equation}
  D_c\quad\longleftrightarrow\quad(\mathcal N,N),
  \label{eq:DGP-Dc-NN-map}
\end{equation}
where the first $\mathcal N$ is Neumann for the $U(1)$ vector multiplet and the second $N$ is Neumann for the charge-one chiral.  More generally, the duality acts sector by sector on line defects.  If the $D_c$ boundary is accompanied by a flavor-vortex sector of nonnegative charge $m$, the dual gauge description contains the corresponding Wilson defect.  In the $S^1\times D^2$ half-index this defect is represented by a Wilson loop wrapping the core circle $S^1\times\{0\}$, and it appears as a power of the gauge holonomy.

The elementary half-index identity can be written in the form
\begin{equation}
  (q)_\infty
  \oint\frac{ds}{2\pi \mathrm{i} s}\,
  \frac{s^{-m}}{(s;q)_\infty}
  =
  \frac{(q)_\infty}{(q)_m}
  =
  (q^{m+1};q)_\infty,
  \qquad m\in\mathbb Z_{\geq0}.
  \label{eq:DGP-vortex-Wilson-sector}
\end{equation}
This is precisely the DGP vortex--Wilson boundary identity, after choosing the sign convention in which the dual Wilson loop insertion is $s^{-m}$.  The first equality follows directly from
\begin{equation}
  \frac1{(s;q)_\infty}
  =\sum_{r\geq0}\frac{s^r}{(q)_r},
  \label{eq:DGP-one-node-q-binomial}
\end{equation}
so the contour simply projects to $r=m$.  Dividing by the vacuum factor $(q)_\infty$ gives the particularly useful identity
\begin{equation}
  \frac1{(q)_m}
  =
  \oint\frac{ds}{2\pi \mathrm{i} s}\,
  \frac{s^{-m}}{(s;q)_\infty},
  \qquad m\geq0.
  \label{eq:DGP-one-node-coefficient}
\end{equation}
Thus the Pochhammer denominator appearing in a Nahm sum has an elementary particle--vortex interpretation: it is the Neumann-gauge projection in a definite Wilson loop sector of the dual one-node theory.

The explicit DGP duality wall contains more microscopic information.  It is directional, with $T$ and $T'$ occupying opposite half-spaces, and it contains a 2d $\mathcal N=(0,2)$ Fermi multiplet $\Gamma$ with
\begin{equation}
  J_\Gamma=\Phi\Phi'.
  \label{eq:DGP-interface-J}
\end{equation}
For a vortex configuration $\Phi\sim c z^m$, the local coupling becomes $J_\Gamma\sim c z^m\Phi'$.  This shows that a literal collision of the perturbative wall with $D_c$ contains nontrivial localized boundary dynamics.  We will not attempt to reconstruct that collision mode by mode.  Instead, the exact sector identity~\eqref{eq:DGP-vortex-Wilson-sector} will be the protected input used below.  This distinction is important because the perturbative particle--vortex wall is uni-directional, and the opposite orientation can require nonperturbative boundary-monopole couplings~\cite{Dimofte:2017tpi}.

\paragraph{Eightfold gauging and the $E_8$ lattice wavefunction.}
We now take eight copies of the elementary boundary identity and gauge the eight flavor symmetries on the original side with the matrix $C_{E_8}$.  Before eliminating the original gauge variables, the Dirichlet boundary condition on these eight gauge multiplets produces the magnetic lattice sum, while the eight particle--vortex-dual gauge multiplets carry Neumann boundary conditions and therefore produce eight holonomy integrals.  These sums and integrals belong to different 3d gauge fields.

For a Wilson loop of charge $\boldsymbol Q$ wrapping the core circle in the original theory, the $A$-twisted half-index obtained in Section~\ref{subsec:Wilson-half-indices} is
\begin{equation}
  I\!\!I^{A}[W_{\boldsymbol Q}](q)
  =
  \sum_{\boldsymbol m\in\mathbb Z_{\geq0}^8}
  \frac{
  q^{\frac12\boldsymbol m^{\mathsf T}C_{E_8}\boldsymbol m
  -\boldsymbol Q^{\mathsf T}\boldsymbol m}
  }{
  \prod_{i=1}^8(q)_{m_i}
  }.
  \label{eq:section35-original-Wilson-half-index}
\end{equation}
Using the one-node identity~\eqref{eq:DGP-one-node-coefficient} for each factor gives
\begin{equation}
\begin{aligned}
  I\!\!I^{A}[W_{\boldsymbol Q}](q)
  ={}&
  \sum_{\boldsymbol m\in\mathbb Z_{\geq0}^8}
  q^{\frac12\boldsymbol m^{\mathsf T}C_{E_8}\boldsymbol m
  -\boldsymbol Q^{\mathsf T}\boldsymbol m}
  \oint
  \prod_{i=1}^8\frac{ds_i}{2\pi \mathrm{i} s_i}\,
  \boldsymbol s^{-\boldsymbol m}
  \prod_{i=1}^8\frac1{(s_i;q)_\infty}.
  \label{eq:eightfold-DGP-before-resummation}
\end{aligned}
\end{equation}
Here $\boldsymbol s^{-\boldsymbol m}=\prod_i s_i^{-m_i}$.  Because the Neumann-chiral expansion contains only nonnegative powers of every $s_i$, the contour gives zero if any component $m_i$ is negative.  We may therefore extend the magnetic sum in~\eqref{eq:eightfold-DGP-before-resummation} from $\mathbb Z_{\geq0}^8$ to the full lattice $\mathbb Z^8$ without changing the answer.  Exchanging the sum and the integrals then gives
\begin{equation}
\begin{aligned}
  I\!\!I^{A}[W_{\boldsymbol Q}](q)
  ={}&
  (q)_\infty^8
  \oint
  \prod_{i=1}^8\frac{ds_i}{2\pi \mathrm{i} s_i}\,
  \mathcal Z_{E_8,\boldsymbol Q^\vee}(\boldsymbol s;q)
  \prod_{i=1}^8\frac1{(s_i;q)_\infty},
  \label{eq:eightfold-DGP-resummed-kernel}
\end{aligned}
\end{equation}
where
\begin{equation}
  \mathcal Z_{E_8}(\boldsymbol s;q)
  :=
  \frac1{(q)_\infty^8}
  \sum_{\boldsymbol n\in\mathbb Z^8}
  q^{\frac12\boldsymbol n^{\mathsf T}C_{E_8}\boldsymbol n}
  \boldsymbol s^{-\boldsymbol n}
  \label{eq:E8-edge-refined-character}
\end{equation}
The lattice wavefunction in \eqref{eq:E8-edge-refined-character} is obtained directly by resumming the Dirichlet magnetic-flux sectors of the gauged particle--vortex boundary construction.  Multiplying by the vacuum Casimir factor $q^{-1/3}$ gives the Cartan-refined vacuum character of the level-one $E_8$ lattice VOA, equivalently $V_{E_8,1}$~\cite{FrenkelKac:1980}.  In particular, at trivial Cartan fugacity,
\begin{equation}
  q^{-1/3}\mathcal Z_{E_8}(\boldsymbol 1;q)
  =\frac{E_4(q)}{\eta(q)^8}
  =\chi_{(E_8)_1}(q).
  \label{eq:E8-edge-unrefined-character}
\end{equation}
Thus the lattice factor appearing in the duality-transformed boundary kernel is the refined version of the same $(E_8)_1$ character that will reappear in the torus interface amplitude of Section~\ref{subsec:interface-E8-bilinear}.
The transmitted charge, its quadratic normalization, and the shifted lattice factor are
\begin{equation}
  \boldsymbol Q^\vee=C_{E_8}^{-1}\boldsymbol Q,
  \qquad
  d_{\boldsymbol Q}:=
  \frac12\boldsymbol Q^{\mathsf T}C_{E_8}^{-1}\boldsymbol Q,
  \qquad
  \mathcal Z_{E_8,\boldsymbol Q^\vee}
  :=q^{-d_{\boldsymbol Q}}
  \boldsymbol s^{-\boldsymbol Q^\vee}
  \mathcal Z_{E_8}.
  \label{eq:E8-edge-Wilson-sector}
\end{equation}
The last relation follows by completing the square:
\begin{equation}
\begin{aligned}
 &\sum_{\boldsymbol m\in\mathbb Z^8}
 q^{\frac12\boldsymbol m^{\mathsf T}C_{E_8}\boldsymbol m
 -\boldsymbol Q^{\mathsf T}\boldsymbol m}
 \boldsymbol s^{-\boldsymbol m}
 \\
 &\hspace{2cm}=
 q^{-d_{\boldsymbol Q}}
 \boldsymbol s^{-\boldsymbol Q^\vee}
 \sum_{\boldsymbol n\in\mathbb Z^8}
 q^{\frac12\boldsymbol n^{\mathsf T}C_{E_8}\boldsymbol n}
 \boldsymbol s^{-\boldsymbol n}.
  \label{eq:E8-Wilson-complete-square}
\end{aligned}
\end{equation}
Unimodularity of $C_{E_8}$ is used here only to ensure that $\boldsymbol Q^\vee$ is integral and that the lattice shift is legitimate.

Equation~\eqref{eq:eightfold-DGP-resummed-kernel} gives a direct protected origin for the $E_8$ theta factor.  It is the resummed Dirichlet-flux wavefunction of the original $C_{E_8}$ gauge fields after the elementary DGP vortex sectors have been converted into Wilson loop sectors of the Neumann dual gauge fields.  In this first-order description the lattice sum and the holonomy integral therefore have a clear microscopic separation.  Once the original Dirichlet gauge variables are eliminated, their boundary wavefunction remains as the chiral lattice factor $\mathcal Z_{E_8}$.

The factor $\boldsymbol s^{-\boldsymbol Q^\vee}$ in~\eqref{eq:E8-edge-Wilson-sector} should be interpreted accordingly as the insertion of the transmitted Wilson loop of charge $\boldsymbol Q^\vee$ in the dual Neumann gauge theory.  The loop wraps the core circle $S^1\times\{0\}$; it is not a separate boundary endpoint operator.  The additional factor $q^{-d_{\boldsymbol Q}}$ is the quadratic normalization produced by the lattice shift, rather than an independent line-charge factor.

\paragraph{Boundary anomaly and the chiral gauged-WZW interpretation.}
The same $E_8$ lattice factor is selected independently by the boundary anomaly.  The topological quadratic form of the eightfold gauged wall is
\begin{equation}
  \begin{pmatrix}
    C_{E_8} & \epsilon I_8\\
    \epsilon I_8 & \frac12 I_8
  \end{pmatrix},
  \qquad \epsilon=\pm1,
  \label{eq:boundary-wall-folded-matrix}
\end{equation}
and eliminating the original $C_{E_8}$ block gives the folded Schur complement
\begin{equation}
  \frac12 I_8-C_{E_8}^{-1}
  =-K_{\mathrm{UV}}^\vee,
  \qquad
  K_{\mathrm{UV}}^\vee=C_{E_8}^{-1}-\frac12 I_8.
  \label{eq:boundary-wall-Schur-complement}
\end{equation}
The physical dual theory has contact matrix $K_{\mathrm{UV}}^\vee$, but it occupies the $x^\perp\geq0$ side of the oriented wall.  Hence the Chern--Simons inflow at its left $(\mathcal N,N)$ boundary is $-K_{\mathrm{UV}}^\vee$.  The eight Neumann charge-one chirals contribute an additional $-\frac12 I_8$, while the abelian gauginos carry no pure $U(1)^8$ gauge charge.  The net gauge-anomaly matrix is therefore
\begin{equation}
  \mathcal A_{\mathrm{gauge}}^{\vee,L}
  =-
  \left(C_{E_8}^{-1}-\frac12 I_8\right)
  -\frac12 I_8
  =-C_{E_8}^{-1}.
  \label{eq:left-dual-boundary-anomaly}
\end{equation}

A boundary chiral gauged-WZW sector can be used to restore gauge invariance for a Chern--Simons theory with a Neumann-type gauge boundary, as described in~\cite{YoshidaSugiyama:2014S1D2}.  In the present case a chiral sector with anomaly $+C_{E_8}^{-1}$ cancels~\eqref{eq:left-dual-boundary-anomaly}.  The lattice wavefunction~\eqref{eq:E8-edge-refined-character} has exactly this anomaly.  Indeed, for $\boldsymbol\lambda\in\mathbb Z^8$,
\begin{equation}
  \mathcal Z_{E_8}(q^{\boldsymbol\lambda}\boldsymbol s;q)
  =
  q^{-\frac12\boldsymbol\lambda^{\mathsf T}C_{E_8}^{-1}\boldsymbol\lambda}
  \boldsymbol s^{-C_{E_8}^{-1}\boldsymbol\lambda}
  \mathcal Z_{E_8}(\boldsymbol s;q).
  \label{eq:E8-edge-quasiperiodicity}
\end{equation}
Thus its Jacobi/anomaly quadratic form is $C_{E_8}^{-1}$.  Since the underlying lattice is even, positive definite, and unimodular of rank eight, the corresponding chiral lattice theory is the level-one $E_8$ lattice theory.  We denote this effective boundary sector by $\mathcal W_{E_8}$.

At the level of boundary anomalies and protected holomorphic blocks, the gauged particle--vortex construction therefore selects the oriented pair
\begin{equation}
  (\mathcal D,D_c)_{\mathcal T}^{\mathrm{right}}
  \quad\longleftrightarrow\quad
  \mathcal B_A^\vee
  :=
  \left[(\mathcal N,N)+\mathcal W_{E_8}\right]_{\mathcal T^\vee}^{\mathrm{left}}.
  \label{eq:oriented-boundary-pair-A}
\end{equation}
The eight elementary DGP Fermis $\Gamma_i$ should not be identified with $\mathcal W_{E_8}$.  The former are microscopic degrees of freedom of the elementary particle--vortex walls and mediate their operator maps.  The latter is the rank-eight chiral lattice sector that remains after the $C_{E_8}$ gauging and is simultaneously visible in the resummed Dirichlet-flux wavefunction and in the boundary anomaly.

For later reference, we use half-index notation for the right-hand side of~\eqref{eq:eightfold-DGP-resummed-kernel}, with the boundary subscript distinguishing this folded construction from the independently localized $(\mathcal D,D_c)$ half-index:
\begin{equation}
\begin{aligned}
  I\!\!I_{\mathcal B_A^\vee}^{\vee B}
  [W^\vee_{\boldsymbol Q^\vee}](q)
  :={}&
  (q)_\infty^8
  \oint
  \prod_{i=1}^8\frac{ds_i}{2\pi \mathrm{i} s_i}\,
  \mathcal Z_{E_8,\boldsymbol Q^\vee}(\boldsymbol s;q)
  \prod_{i=1}^8\frac1{(s_i;q)_\infty}.
  \label{eq:E8-enriched-Neumann-half-index}
\end{aligned}
\end{equation}
By construction,
\begin{equation}
  I\!\!I_{\mathcal B_A^\vee}^{\vee B}
  [W^\vee_{\boldsymbol Q^\vee}](q)
  =I\!\!I^{A}[W_{\boldsymbol Q}](q),
  \qquad
  \boldsymbol Q^\vee=C_{E_8}^{-1}\boldsymbol Q,
  \label{eq:mirror-half-index-A-exact}
\end{equation}
for every integral Wilson loop charge for which the corresponding transmitted charge is defined.  In particular,
\begin{equation}
  I\!\!I_{\mathcal B_A^\vee}^{\vee B}(q)
  =I\!\!I_{(\mathcal D,D_c)}^{A}(q).
  \label{eq:mirror-half-index-A-vacuum}
\end{equation}
This is an exact all-order identity.  Its derivation now has two complementary interpretations: the theta kernel is obtained directly by resumming the old Dirichlet flux sectors using the one-node DGP vortex--Wilson identity, and its quasi-periodicity independently supplies precisely the anomaly needed by the left Neumann gauge boundary of the dual theory.

\paragraph{Reverse orientation and microscopic limitations.}
There is an analogous algebraic kernel with $C_{E_8}$ and $C_{E_8}^{-1}$ exchanged.  We denote the corresponding orientation-reversed folded boundary data by $\mathcal B_B$.  Since $C_{E_8}^{-1}$ is itself an even integral unimodular positive-definite matrix, define
\begin{equation}
  \mathcal Z_{E_8}^{\vee}(\boldsymbol t;q)
  :=
  \frac{1}{(q)_\infty^8}
  \sum_{\boldsymbol\ell\in\mathbb Z^8}
  q^{\frac12\boldsymbol\ell^{\mathsf T}C_{E_8}^{-1}\boldsymbol\ell}
  \boldsymbol t^{-\boldsymbol\ell}.
  \label{eq:dual-E8-edge-refined-character}
\end{equation}
Its quasi-periodicity has anomaly matrix $C_{E_8}$, and the same coefficient extraction gives
\begin{equation}
  I\!\!I_{\mathcal B_B}^{A}
  [W_{\boldsymbol Q}](q)
  =I\!\!I^{\vee B}[W^\vee_{\boldsymbol Q^\vee}](q),
  \qquad
  \boldsymbol Q=C_{E_8}\boldsymbol Q^\vee.
  \label{eq:mirror-half-index-B-exact}
\end{equation}
The microscopic status of the two directions is not symmetric.  Ref.~\cite{Dimofte:2017tpi} emphasizes that the perturbative particle--vortex interfaces are uni-directional and that constructing the opposite orientation can lead to nonperturbative interface superpotentials involving boundary monopole operators.  We therefore regard~\eqref{eq:mirror-half-index-B-exact} as the exact orientation-reversed protected kernel suggested by duality, without claiming an equally explicit perturbative wall Lagrangian for that direction.

It is also useful to state precisely what has not been established in the forward direction.  The exact identity~\eqref{eq:mirror-half-index-A-exact} does not by itself constitute a complete microscopic derivation of the boundary condition obtained by colliding the full eightfold DGP wall with the $(\mathcal D,D_c)$ boundary.  The explicit wall contains the eight Fermis $\Gamma_i$ with $J_{\Gamma_i}=\Phi_i\Phi_i^\vee$.  In a vortex sector $\Phi_i\sim c_i z^{m_i}$ these become position-dependent boundary couplings, so a literal collision involves localized boundary dynamics in addition to the protected sector map.  We do not need to solve that boundary RG problem here.  The statements used in the rest of the paper are only: the elementary sector identity~\eqref{eq:DGP-vortex-Wilson-sector}, the eightfold resummation~\eqref{eq:eightfold-DGP-resummed-kernel}, the oriented anomaly cancellation~\eqref{eq:left-dual-boundary-anomaly}, and the exact kernel equality~\eqref{eq:mirror-half-index-A-exact}.

Finally, the lattice sector also explains the universal scalar pieces that appeared in Section~\ref{subsec:dual-Wilson-Zagier}.  Restoring the vacuum Casimir factor multiplies~\eqref{eq:E8-edge-refined-character} by
\begin{equation}
  q^{-8/24}=q^{-1/3},
  \label{eq:E8-edge-Casimir-shift}
\end{equation}
while completing the square in a transmitted Wilson sector gives the quadratic normalization
\begin{equation}
  d_{\boldsymbol Q}
  =\frac12\boldsymbol Q^{\mathsf T}C_{E_8}^{-1}\boldsymbol Q.
  \label{eq:edge-Wilson-quadratic-shift}
\end{equation}
These are precisely the universal contributions $d_{\boldsymbol Q}-1/3$ in the scalar component of the rank-eight Zagier transformation.  Their appearance here is consistent with the relative $E_8$ framing sector extracted independently from the twisted-TQFT fibering data.

The remaining microscopic questions are deliberately left outside the scope of this subsection.  A full derivation of the boundary condition produced by colliding the gauged particle--vortex wall with $(\mathcal D,D_c)$ would require tracking the interacting $\Gamma_i$ sectors and possible nonperturbative boundary-monopole couplings.  In addition, the physical theories of Sections~2 and~3 contain seven monopole-superpotential deformations, whose compatibility with an explicit ultraviolet boundary Lagrangian would require further boundary interactions.  We do not assume either microscopic completion.  Accordingly, we use $I\!\!I_{\mathcal B_A^\vee}^{\vee B}[W^\vee_{\boldsymbol Q^\vee}]$ as notation for the folded holomorphic boundary half-index selected by the gauged particle--vortex construction.  The boundary label $\mathcal B_A^\vee$ is essential: this object should not be confused with the independently localized $(\mathcal D,D_c)$ half-index of the dual theory.  The exact protected identity~\eqref{eq:mirror-half-index-A-exact} and the anomaly analysis are the results needed below.

\subsection{Comments on boundary VOAs}
\label{subsec:boundary-VOA-comments}

We conclude this section with a comment on the vertex-algebraic interpretation of the character systems appearing in the half-indices.  Both five-component systems originate from the Virasoro minimal model $M(2,11)\simeq M(11,2)$.  The system $M_{\rm eff}(2,11)$ is obtained by the effective shift of the conformal weights and central charge, while the five-component system $T_{10}M_{\rm eff}(11,2)$ is obtained by the corresponding $T_{10}$ Hecke transformation.  The appearance of these modular character systems in boundary half-indices does not, by itself, imply the existence of vertex operator algebras whose complete sets of irreducible characters are exactly the corresponding five functions.

The two sides need not have symmetric realizations as ordinary rational VOAs.  In the present paper we do not assume that the $T_{10}M_{\rm eff}(11,2)$ system realized by the Wilson loop half-indices of the $C_{E_8}$ Chern--Simons matter theory is the complete irreducible-character vector of an ordinary strongly rational VOA.  Possible alternatives include a vertex operator superalgebra realization, an embedding into a larger VOA for which these five functions span a distinguished modular subrepresentation, or an interpretation in terms of branching or relative characters.

By contrast, the $M_{\rm eff}(2,11)$ system realized on the $C_{E_8}^{-1}$ side admits a direct vertex-superalgebraic realization~\cite{LeeSun:2026MagicTriangle}.  Its five effective characters are realized by supercharacters of the simple affine vertex operator superalgebra
\begin{equation}
  L_4\bigl(\mathfrak{osp}(1|2)\bigr),
  \label{eq:osp12-boundary-VOA}
\end{equation}
whose Sugawara central charge is
\begin{equation}
  c=\frac{8}{11}.
  \label{eq:osp12-central-charge}
\end{equation}
Thus the $C_{E_8}^{-1}$ half-index character system has a direct affine-superalgebraic interpretation, whereas the corresponding boundary-VOA interpretation of the $T_{10}M_{\rm eff}(11,2)$ system is left open here.

These vertex-algebraic questions, including the precise chiral realization of the $T_{10}$ system and the boundary conditions that select the relevant modules, are not needed for the mirror and TQFT arguments below.  We therefore use $T_{10}M_{\rm eff}(11,2)$ and $M_{\rm eff}(2,11)$ primarily as names for the corresponding five-component character systems, without assuming an ordinary strongly rational boundary-VOA realization on both sides.  A detailed study of these boundary VOAs, including the precise VOA-theoretic nature of the $M_{\rm eff}(2,11)$ character system and the associated module-selecting boundary conditions, will be presented in a forthcoming companion paper.

\section{$A/B$-twisted TQFTs of the mirror pair}
\label{sec:AB-twisted-TQFTs}

We now compare the semisimple TQFT data obtained from the topological $A$- and $B$-twists of the two theories constructed in Sections~2 and~3.  The extraction of nonunitary TQFT data from rank-zero 3d $\mathcal N=4$ SCFTs has been developed in~\cite{Gang:2021hrd,Gang:2023rei,GangKimParkStubbs:2024Nahm}, while its relation to boundary VOAs and mirror symmetry was further developed in~\cite{CreutzigGarnerKim:2024mirror}.  The particle--vortex duality exchanges the two twists in the manner required by infrared mirror symmetry.  It induces the transformation
\begin{equation}
  \boldsymbol y=\boldsymbol 1-\boldsymbol x
  \label{eq:particle-vortex-complement-map-intro}
\end{equation}
between the Bethe roots of the two theories.  We refer to \eqref{eq:particle-vortex-complement-map-intro} as the \emph{particle--vortex complement map}.  It gives a one-to-one correspondence between the regular Bethe vacua and hence between the finite sets of topological sectors.  The dictionary used in this section is summarized in Table~\ref{tab:TQFT-dictionary}~\cite{NekrasovShatashvili:2015BetheGauge,Benini:2015noa,Okuda:2012nx,Okuda:2013fea,Closset:2016arn,Closset:2017zgf,ClossetKimWillett:2018Seifert}.  We use the term \emph{formal Verlinde algebra} for the commutative algebra obtained from a modular $S$-matrix through the Verlinde formula, without assuming the existence of an underlying rational VOA.

\begin{table}[htbp]
\centering
\small
\setlength{\tabcolsep}{6pt}
\renewcommand{\arraystretch}{1.08}
\caption{Dictionary between localization data of the $A/B$-twisted 3d theories and the corresponding semisimple TQFT data.}
\label{tab:TQFT-dictionary}
\begin{tabular}{p{0.37\linewidth}p{0.53\linewidth}}
\toprule
3d $A/B$-twisted datum & semisimple TQFT / modular datum \\
\midrule
regular Bethe vacua & simple topological sectors \\
Wilson loop operators & ratios $S_{\alpha\beta}/S_{0\beta}$ \\
handle-gluing operator & vacuum-row data $(S_{0\beta})^{-2}$ \\
Wilson loop algebra in the character basis & formal Verlinde algebra determined by the modular $S$-matrix \\
rephased based Wilson loop algebra & $SO(3)_9$ fusion ring \\
fibering operator & diagonal $T$ data, up to a common framing multiplier \\
particle--vortex complement map & $\boldsymbol y=\boldsymbol 1-\boldsymbol x$; exchange of the $A$- and $B$-twisted TQFT data \\
\bottomrule
\end{tabular}
\end{table}

The table should be understood as a dictionary for the finite-dimensional semisimple sector extracted from regular Bethe vacua.  In particular, it does not by itself identify the full braided line category or a fully extended TQFT.  The five-component character systems identified in Sections~2 and~3 also carry independent 2d modular data.  Our strategy is therefore to reconstruct the modular data from the 3d twisted theories and compare the result with the modular representation of the corresponding character systems.  We begin with the Wilson loop and handle-gluing data, which determine the modular $S$-matrix.  The fibering operator and the modular $T$ data will be discussed separately.

\subsection{Wilson loops and modular $S$-matrix}
\label{subsec:Wilson-S-matrix}

Let $\mathcal W(\boldsymbol u)$ and $\Omega(\boldsymbol u)$ denote the effective twisted superpotential and effective dilaton of the theory compactified on $S^1$~\cite{NekrasovShatashvili:2015BetheGauge,Closset:2016arn,Closset:2017zgf}.  The gauge-flux operators are
\begin{equation}
  \Pi_a(\boldsymbol u)
  :=\exp\!\left(
  2\pi\mathrm{i}\,
  \frac{\partial\mathcal W}{\partial u_a}
  \right),
  \qquad a=1,\ldots,r,
  \label{eq:general-flux-operator}
\end{equation}
where $r$ is the rank of the gauge group and $a=1,\ldots,r$ labels the Cartan gauge variables.  The regular Bethe vacua are the gauge-inequivalent regular solutions of $\Pi_a=1$.  The handle-gluing operator is defined by~\cite{NekrasovShatashvili:2015BetheGauge,Okuda:2012nx,Okuda:2013fea,Closset:2017zgf}
\begin{equation}
  \mathcal H(\boldsymbol u)
  :=e^{2\pi\mathrm{i}\Omega(\boldsymbol u)}
  \det_{a,b}\!\left[
  \frac{\partial^2\mathcal W(\boldsymbol u)}
  {\partial u_a\partial u_b}
  \right].
  \label{eq:handle-gluing-definition}
\end{equation}
Thus $\mathcal H$ is the Hessian determinant of the effective twisted superpotential dressed by the effective-dilaton contribution.

Let $\widehat u_\beta$ denote a regular Bethe vacuum and let $W_\alpha$ denote a supersymmetric Wilson loop operator wrapping the $S^1$ fiber.  After the chosen $A$- or $B$-twist it becomes a topological line operator.  Its value on a Bethe vacuum is related to the modular $S$-matrix by~\cite{Gang:2021hrd,Gang:2023rei,CreutzigGarnerKim:2024mirror,GangKimParkStubbs:2024Nahm}
\begin{equation}
  W_\alpha(\widehat u_\beta)
  =\frac{S_{\alpha\beta}}{S_{0\beta}}.
  \label{eq:Wilson-S-ratio}
\end{equation}
The vacuum row is fixed by the handle-gluing operator through
\begin{equation}
  \mathcal H(\widehat u_\beta)
  =\bigl(S_{0\beta}\bigr)^{-2}.
  \label{eq:handle-S0-relation}
\end{equation}
Thus the Bethe equations, together with the Wilson loop and handle-gluing operators, determine the columns of $S$.  Equation~\eqref{eq:handle-S0-relation} fixes $|S_{0\beta}|$; the relative signs can then be fixed by requiring the reconstructed matrix to be symmetric, with the remaining overall sign chosen so that $S_{00}>0$.

We first carry out this reconstruction for the $A$-twisted $C_{E_8}$ theory.  We then derive the corresponding $B$-twisted data directly from the Bethe algebra of the $C_{E_8}^{-1}$ theory and compare the two descriptions through the particle--vortex complement map.

\subsubsection{The $A$-twisted $C_{E_8}$ theory}
\label{subsubsec:A-twisted-CE8-S}

Set the background topological fugacity to one and write
\begin{equation}
  x_i=e^{2\pi \mathrm{i} u_i},
  \qquad i=1,\ldots,8.
\end{equation}
A convenient logarithmic lift of the effective twisted superpotential at the $A$-twisted point is
\begin{equation}
  \mathcal W_A(\boldsymbol u)
  =\frac12\boldsymbol u^{\mathsf T}C_{E_8}\boldsymbol u
  +\frac{1}{(2\pi\mathrm{i})^2}
  \sum_{i=1}^{8}\operatorname{Li}_2(x_i),
  \label{eq:CE8-A-twisted-superpotential}
\end{equation}
while the corresponding effective-dilaton factor is
\begin{equation}
  e^{2\pi\mathrm{i}\Omega_A(\boldsymbol u)}
  =\prod_{i=1}^{8}(1-x_i).
  \label{eq:CE8-A-effective-dilaton}
\end{equation}
Terms in $\mathcal W_A$ that are independent of $\boldsymbol u$ do not affect the Bethe equations or the handle-gluing operator and are omitted.  Using \eqref{eq:general-flux-operator}, the exponentiated Bethe equations are
\begin{equation}
  1-x_i
  =\prod_{j=1}^{8}x_j^{(C_{E_8})_{ij}},
  \qquad
  i=1,\ldots,8,
  \qquad
  x_i\neq0,1.
  \label{eq:CE8-Bethe-equations}
\end{equation}
The modular data can be reconstructed without solving these equations numerically.  After clearing denominators, eliminating $x_2,\ldots,x_8$, and removing the singular components introduced by the polynomialization, the regular Bethe-vacuum locus can be parametrized by the single coordinate
\begin{equation}
  t:=x_1
\end{equation}
subject to the quintic relation
\begin{equation}
  P(t)
  =t^5-2t^4-5t^3+13t^2-7t+1=0.
  \label{eq:CE8-Bethe-quintic}
\end{equation}
The polynomial $P$ is irreducible over $\mathbb Q$ and has discriminant $11^4$, so in particular it is square-free.  All eight holonomies reduce to polynomials of degree at most four in $t$, and none of $x_i$ or $1-x_i$ vanishes at a root of $P$.

We define the \emph{Bethe algebra} to be the algebra of polynomial functions on the regular Bethe vacua, with the Bethe equations imposed.  In the primitive coordinate $t=x_1$, it is therefore
\begin{equation}
  \mathcal B_{C_{E_8}}
  :=\mathbb Q[t]/(P(t)),
  \qquad
  \mathcal B_{C_{E_8}}\otimes_{\mathbb Q}\mathbb C
  \simeq
  \bigoplus_{\beta=0}^{4}\mathbb C.
  \label{eq:CE8-Bethe-algebra}
\end{equation}
The second relation follows because $P$ has five distinct roots, which are precisely the five regular Bethe vacua.  Since $P$ is irreducible over $\mathbb Q$, the algebra $\mathcal B_{C_{E_8}}$ is a field over $\mathbb Q$ and therefore has no nontrivial idempotents over $\mathbb Q$ itself.  The idempotent decomposition appears after extending scalars to $\mathbb C$.  If the five roots are denoted by $t^{(\beta)}$, the Chinese remainder theorem gives
\begin{equation}
  \mathcal B_{C_{E_8}}\otimes_{\mathbb Q}\mathbb C
  \simeq
  \bigoplus_{\beta=0}^{4}\mathbb C e_\beta,
  \qquad
  e_\beta(t)
  =\prod_{\gamma\neq\beta}
  \frac{t-t^{(\gamma)}}{t^{(\beta)}-t^{(\gamma)}}.
  \label{eq:CE8-Bethe-idempotents}
\end{equation}
These primitive idempotents satisfy
\begin{equation}
  e_\beta\!\left(t^{(\gamma)}\right)=\delta_{\beta\gamma},
  \qquad
  e_\beta e_\gamma=\delta_{\beta\gamma}e_\beta,
  \qquad
  \sum_{\beta=0}^{4}e_\beta=1.
  \label{eq:CE8-Bethe-idempotent-relations}
\end{equation}
Thus each regular Bethe vacuum corresponds to one primitive idempotent of the complexified Bethe algebra.  We refer to the basis $\{e_\beta\}_{\beta=0}^{4}$ as the \emph{Bethe-idempotent basis}.

These roots admit a simple exact parametrization.  For $s=1,\ldots,5$, define
\begin{equation}
  t_s
  :=1-\frac{\sin(6\pi s/11)}{\sin(10\pi s/11)}.
  \label{eq:CE8-Bethe-roots-unordered}
\end{equation}
The five numbers $t_s$ are exactly the five roots of $P(t)$.  A derivation from the eleventh cyclotomic polynomial is given in Appendix~\ref{app:cyclotomic-Bethe-roots}.

To describe the number field generated by these roots, let
\begin{equation}
  \zeta_{11}:=e^{2\pi\mathrm{i}/11}
  \label{eq:zeta11-definition}
\end{equation}
be a primitive eleventh root of unity.  We denote by
\begin{equation}
  \mathbb Q(\zeta_{11})^+
  :=\mathbb Q(\zeta_{11}+\zeta_{11}^{-1})
  =\mathbb Q\!\left(2\cos\frac{2\pi}{11}\right)
  \label{eq:real-cyclotomic-field-definition}
\end{equation}
the maximal real subfield of the eleventh cyclotomic field.  Since $11$ is prime, $\mathbb Q(\zeta_{11})$ has degree ten over $\mathbb Q$.  Complex conjugation sends $\zeta_{11}$ to $\zeta_{11}^{-1}$, and its fixed field is $\mathbb Q(\zeta_{11})^+$.  Hence
\begin{equation}
  [\mathbb Q(\zeta_{11})^+:\mathbb Q]=5.
  \label{eq:real-cyclotomic-field-degree}
\end{equation}
As shown in Appendix~\ref{app:cyclotomic-Bethe-roots}, each $t_s$ belongs to $\mathbb Q(\zeta_{11})^+$.  Since $P(t)$ is irreducible of degree five, the Bethe algebra $\mathcal B_{C_{E_8}}$ is itself a degree-five field over $\mathbb Q$.  It follows that
\begin{equation}
  \mathcal B_{C_{E_8}}
  =\mathbb Q[t]/(P(t))
  \simeq
  \mathbb Q(\zeta_{11})^+.
  \label{eq:CE8-Bethe-cyclotomic-field}
\end{equation}
Thus the five Bethe vacua are the five real Galois conjugates of a single element of the real cyclotomic field.

We now order these five vacua according to the Wilson loop character sectors of Section~2.5,
\begin{equation}
  h_\beta
  =0,\ \frac5{11},\ \frac8{11},\ \frac{10}{11},\ \frac{12}{11},
\end{equation}
and introduce
\begin{equation}
  (s_\beta)=(5,2,3,4,1).
  \label{eq:CE8-s-labels}
\end{equation}
With this ordering,
\begin{equation}
  t^{(\beta)}
  =1-\frac{\sin(6\pi s_\beta/11)}{\sin(10\pi s_\beta/11)}.
  \label{eq:CE8-Bethe-roots-exact}
\end{equation}
The sector labels in \eqref{eq:CE8-s-labels} are fixed by the Wilson loop character ordering of Section~2.5; no numerical matching of the five roots is needed below.

For the Wilson loop of charge $\boldsymbol Q_h$ used in Section~2.5, we take the holonomy representative
\begin{equation}
  W_h(\boldsymbol x)
  :=\prod_{i=1}^{8}x_i^{-Q_{h,i}}.
  \label{eq:CE8-Wilson-holonomy}
\end{equation}
Reducing these five monomials in the quotient algebra \eqref{eq:CE8-Bethe-algebra} gives the exact polynomial representatives
\begin{equation}
\begin{aligned}
  L_0(t)&=1,\\
  L_{5/11}(t)&=t^4-t^3-6t^2+8t-1,\\
  L_{8/11}(t)&=1-t,\\
  L_{10/11}(t)&=-2t^4+3t^3+11t^2-20t+5,\\
  L_{12/11}(t)&=-t^4+2t^3+6t^2-13t+3.
\end{aligned}
  \label{eq:CE8-Wilson-polynomials}
\end{equation}
From \eqref{eq:handle-gluing-definition}, \eqref{eq:CE8-A-twisted-superpotential}, and \eqref{eq:CE8-A-effective-dilaton}, the $A$-twisted handle-gluing operator is
\begin{equation}
  \mathcal H_A(\boldsymbol x)
  =
  \prod_{i=1}^{8}(1-x_i)
  \det\!\left[
  C_{E_8}
  +\operatorname{diag}\!\left(\frac{x_i}{1-x_i}\right)
  \right].
  \label{eq:CE8-A-handle-gluing}
\end{equation}
Its reduction in the same quotient algebra is
\begin{equation}
  \mathcal H_A(t)
  =5t^4-8t^3-26t^2+48t-7.
  \label{eq:CE8-handle-polynomial}
\end{equation}
Moreover, $\gcd(P,\mathcal H_A)=1$, so the handle-gluing operator is finite and nonzero on all five roots.  This gives an algebraic regularity check for the five vacua.

We can now reconstruct the modular matrix entirely inside $\mathbb Q(\zeta_{11})^+$.  Direct substitution of \eqref{eq:CE8-Bethe-roots-exact} into \eqref{eq:CE8-handle-polynomial} gives
\begin{equation}
  \mathcal H_A\!\left(t^{(\beta)}\right)
  =\frac{11}{4\sin^2(\pi s_\beta/11)}.
  \label{eq:CE8-handle-cyclotomic}
\end{equation}
Thus, with the standard real choice $S_{00}>0$, the vacuum row is
\begin{equation}
  S_{0\beta}
  =\frac{2}{\sqrt{11}}
  \sin\!\left(\frac{\pi s_\beta}{11}\right).
  \label{eq:CE8-S0-exact}
\end{equation}
Likewise, evaluating the exact Wilson polynomials gives
\begin{equation}
  L_{h_\alpha}\!\left(t^{(\beta)}\right)
  =(-1)^{s_\alpha+s_\beta}
  \frac{
  \sin\!\left(2\pi s_\alpha s_\beta/11\right)
  }{
  \sin\!\left(\pi s_\beta/11\right)
  }.
  \label{eq:CE8-Wilson-cyclotomic}
\end{equation}
Equations~\eqref{eq:Wilson-S-ratio} and \eqref{eq:handle-S0-relation} therefore reconstruct a modular matrix purely from the 3d Bethe-vacuum data.  To distinguish this matrix from the modular $S$-matrix of the character system introduced below, we denote the former by $S_A^{\mathrm{Bethe}}$:
\begin{equation}
  \bigl(S_A^{\mathrm{Bethe}}\bigr)_{\alpha\beta}
  =\frac{2}{\sqrt{11}}
  (-1)^{s_\alpha+s_\beta}
  \sin\!\left(\frac{2\pi s_\alpha s_\beta}{11}\right),
  \qquad
  (s_\alpha)=(5,2,3,4,1).
  \label{eq:CE8-S-exact}
\end{equation}
This is an exact identity in the real cyclotomic field, not a trigonometric expression inferred from floating-point Bethe roots.  Independently, the $T_{10}M_{\mathrm{eff}}(11,2)$ character system identified from the Wilson loop half-indices in Section~\ref{subsec:Wilson-half-indices} carries a five-dimensional modular representation.  In the same character ordering, its modular $S$-matrix is precisely the right-hand side of \eqref{eq:CE8-S-exact}.  Hence
\begin{equation}
  S_A^{\mathrm{Bethe}}
  =S_{T_{10}M_{\mathrm{eff}}(11,2)}^{\mathrm{char}}.
  \label{eq:CE8-Bethe-character-S-equality}
\end{equation}
Thus the modular matrix reconstructed from the Bethe roots, Wilson loop eigenvalues, and handle-gluing operator agrees exactly with the independently specified modular $S$-matrix of the five-character system.  In what follows we write $S_A$ for this common matrix.

There is also a direct algebraic reason why the same matrix determines the Wilson loop algebra.  Since the Bethe algebra \eqref{eq:CE8-Bethe-algebra} is semisimple, multiplication by $L_{h_\alpha}$ is diagonal in the Bethe-idempotent basis, with eigenvalues $L_{h_\alpha}(t^{(\sigma)})$.  Writing
$L_{h_\alpha}L_{h_\beta}=\sum_\gamma N^{\gamma}_{\alpha\beta}L_{h_\gamma}$,
\eqref{eq:Wilson-S-ratio} gives the Verlinde formula
\begin{equation}
  N^{\gamma}_{\alpha\beta}
  =\sum_{\sigma}
  \frac{
    (S_A^{\mathrm{Bethe}})_{\alpha\sigma}
    (S_A^{\mathrm{Bethe}})_{\beta\sigma}
    \bigl((S_A^{\mathrm{Bethe}})^{-1}\bigr)_{\sigma\gamma}
  }{
    (S_A^{\mathrm{Bethe}})_{0\sigma}
  }.
  \label{eq:Bethe-Verlinde-diagonalization}
\end{equation}
Equation~\eqref{eq:CE8-Bethe-character-S-equality} then implies that the Wilson loop Bethe algebra in the character basis is isomorphic to the formal Verlinde algebra of $T_{10}M_{\mathrm{eff}}(11,2)$.  This statement uses only the modular representation of the five characters and does not assume that they form the complete irreducible-character vector of an ordinary strongly rational VOA.  In this character basis some structure constants are signed.  The sign rephasing in \eqref{eq:based-line-sign-choice} converts the same abstract algebra into the positive $SO(3)_9$ based fusion ring displayed in \eqref{eq:based-fusion-rules}.

Using the finite-sine orthogonality relation derived in Appendix~\ref{app:finite-sine-orthogonality}, one finds
\begin{equation}
  \bigl(S_A^{\mathrm{Bethe}}\bigr)^{\mathsf T}=S_A^{\mathrm{Bethe}},
  \qquad
  \bigl(S_A^{\mathrm{Bethe}}\bigr)^2=\mathbf 1.
  \label{eq:CE8-S-relations}
\end{equation}
Hence the five Wilson loops already identified from the half-indices in Section~2.5 determine, together with the exact Bethe algebra and the $A$-twisted handle-gluing operator, the modular $S$-matrix of the semisimple TQFT without numerical solution of the Bethe equations, and reproduce independently the modular $S$ action of the $T_{10}M_{\mathrm{eff}}(11,2)$ character system.  We now repeat the construction for the $C_{E_8}^{-1}$ theory and make the mirror identification algebraically explicit.

\subsubsection{The $B$-twisted $C_{E_8}^{-1}$ theory}
\label{subsubsec:B-twisted-CE8inv-S}

Let
\begin{equation}
  y_i=e^{2\pi \mathrm{i} v_i},
  \qquad i=1,\ldots,8,
\end{equation}
be the exponentiated gauge holonomies of the $C_{E_8}^{-1}$ theory.  At the $B$-twisted point, a convenient logarithmic lift is
\begin{equation}
  \mathcal W_B^\vee(\boldsymbol v)
  =\frac12\boldsymbol v^{\mathsf T}C_{E_8}^{-1}\boldsymbol v
  +\frac{1}{(2\pi\mathrm{i})^2}
  \sum_{i=1}^{8}\operatorname{Li}_2(y_i),
  \label{eq:CE8inv-B-twisted-superpotential}
\end{equation}
with effective-dilaton factor
\begin{equation}
  e^{2\pi\mathrm{i}\Omega_B^\vee(\boldsymbol v)}
  =\prod_{i=1}^{8}(1-y_i).
  \label{eq:CE8inv-B-effective-dilaton}
\end{equation}
At unit background topological fugacity, \eqref{eq:general-flux-operator} gives the Bethe equations
\begin{equation}
  1-y_i
  =\prod_{j=1}^{8}y_j^{(C_{E_8}^{-1})_{ij}},
  \qquad
  i=1,\ldots,8,
  \qquad
  y_i\neq0,1.
  \label{eq:CE8inv-Bethe-equations}
\end{equation}
As in the Cartan theory, the Bethe algebra has dimension five over $\mathbb Q$.  Using the exact complement relation derived below---or, equivalently, eliminating $y_2,\ldots,y_8$---one may take
\begin{equation}
  z:=y_1
\end{equation}
as a primitive coordinate.  It satisfies the monic quintic
\begin{equation}
  P^\vee(z)
  =z^5-3z^4-3z^3+4z^2+z-1=0.
  \label{eq:CE8inv-Bethe-quintic}
\end{equation}
Thus
\begin{equation}
  \mathcal B_{C_{E_8}^{-1}}
  :=
  \mathbb Q[z]/(P^\vee(z)).
  \label{eq:CE8inv-Bethe-algebra}
\end{equation}
The relation between the two quintic presentations is exact:
\begin{equation}
  P^\vee(z)=-P(1-z).
  \label{eq:Bethe-quintic-complement}
\end{equation}
In particular, $P^\vee$ is irreducible over $\mathbb Q$, has the same discriminant $11^4$, and is square-free.  Equation~\eqref{eq:Bethe-quintic-complement} is the one-variable shadow of a simple isomorphism of the two Bethe algebras.  If $\boldsymbol x$ satisfies \eqref{eq:CE8-Bethe-equations}, the particle--vortex complement map introduced in \eqref{eq:particle-vortex-complement-map-intro} is
\begin{equation}
  \boldsymbol y=\boldsymbol 1-\boldsymbol x.
  \label{eq:Bethe-complement-map}
\end{equation}
The Cartan Bethe equations imply the monomial identity
\begin{equation}
  \boldsymbol y=\boldsymbol x^{C_{E_8}}.
  \label{eq:Bethe-complement-monomial}
\end{equation}
Since $C_{E_8}$ is unimodular, $C_{E_8}^{-1}$ is integral and therefore
\begin{equation}
  \boldsymbol y^{C_{E_8}^{-1}}=\boldsymbol x.
\end{equation}
It follows immediately that
\begin{equation}
  1-y_i
  =\prod_{j=1}^{8}y_j^{(C_{E_8}^{-1})_{ij}},
\end{equation}
so \eqref{eq:Bethe-complement-map} is an involutive isomorphism between the two regular Bethe-vacuum loci.  In the primitive coordinates it reduces to
\begin{equation}
  z=1-t.
  \label{eq:Bethe-complement-primitive}
\end{equation}
In particular, the five roots of \eqref{eq:CE8inv-Bethe-quintic} are
\begin{equation}
  z^{(\beta)}
  =1-t^{(\beta)}
  =\frac{\sin(6\pi s_\beta/11)}{\sin(10\pi s_\beta/11)},
  \qquad
  (s_\beta)=(5,2,3,4,1).
  \label{eq:CE8inv-Bethe-roots-exact}
\end{equation}
Thus $\mathcal B_{C_{E_8}^{-1}}$ is another primitive-element presentation of the same real cyclotomic field $\mathbb Q(\zeta_{11})^+$ defined in \eqref{eq:real-cyclotomic-field-definition}.

We order the dual sectors according to the exact Wilson line transmission and the paired character ordering of Section~\ref{subsec:dual-Wilson-Zagier},
\begin{equation}
  h_\beta^\vee
  =0,\ \frac6{11},\ \frac3{11},\ \frac1{11},\ \frac{10}{11},
  \label{eq:CE8inv-sector-order}
\end{equation}
which is paired with
\begin{equation}
  h_\beta
  =0,\ \frac5{11},\ \frac8{11},\ \frac{10}{11},\ \frac{12}{11}
\end{equation}
by \eqref{eq:particle-vortex-sector-map}.  The corresponding Wilson charges obey the exact particle--vortex transmission law
\begin{equation}
  \boldsymbol Q_{h^\vee}^\vee
  =C_{E_8}^{-1}\boldsymbol Q_h.
  \label{eq:CE8inv-Wilson-charge-map-TQFT}
\end{equation}
With the same holonomy convention as in \eqref{eq:CE8-Wilson-holonomy}, define
\begin{equation}
  W_{h^\vee}^\vee(\boldsymbol y)
  :=\prod_{i=1}^{8}y_i^{-Q_{h^\vee,i}^\vee}.
  \label{eq:CE8inv-Wilson-holonomy}
\end{equation}
Using \eqref{eq:Bethe-complement-monomial} and the symmetry of $C_{E_8}$ gives the identity in the Bethe algebra
\begin{equation}
\begin{aligned}
  W_{h^\vee}^\vee(\boldsymbol y)
  &=(\boldsymbol x^{C_{E_8}})^{-C_{E_8}^{-1}\boldsymbol Q_h}
  =\boldsymbol x^{-\boldsymbol Q_h}
  =W_h(\boldsymbol x).
\end{aligned}
  \label{eq:Wilson-eigenvalue-complement}
\end{equation}
Hence the dual Wilson representatives in the coordinate $z$ are simply
\begin{equation}
\begin{aligned}
  L_0^\vee(z)&=1,\\
  L_{6/11}^\vee(z)&=z^4-3z^3-3z^2+3z+1,\\
  L_{3/11}^\vee(z)&=z,\\
  L_{1/11}^\vee(z)&=-2z^4+5z^3+8z^2-3z-3,\\
  L_{10/11}^\vee(z)&=-z^4+2z^3+6z^2-z-3,
\end{aligned}
  \label{eq:CE8inv-Wilson-polynomials}
\end{equation}
where each line is equivalently $L_{h^\vee}^\vee(z)=L_h(1-z)$ modulo $P^\vee(z)$.

At the $B$-twisted point, the handle-gluing operator of the $C_{E_8}^{-1}$ theory reduces to
\begin{equation}
  \mathcal H_B^\vee(z)
  =5z^4-12z^3-20z^2+8z+12.
  \label{eq:CE8inv-B-handle-polynomial}
\end{equation}
The equality of the complementary handle operators is again exact.  To see this directly before elimination, introduce
\begin{equation}
  X:=\operatorname{diag}(x_1,\ldots,x_8),
  \qquad
  Y:=\operatorname{diag}(y_1,\ldots,y_8)=\mathbf 1-X.
\end{equation}
On paired Bethe vacua the two handle operators can be written as
\begin{equation}
  \mathcal H_A(\boldsymbol x)=\det(YC_{E_8}+X),
  \qquad
  \mathcal H_B^\vee(\boldsymbol y)=\det(XC_{E_8}^{-1}+Y).
  \label{eq:handle-complement-determinants}
\end{equation}
Since
\begin{equation}
  XC_{E_8}^{-1}+Y
  =(X+YC_{E_8})C_{E_8}^{-1}
\end{equation}
and $\det C_{E_8}=1$, one obtains
\begin{equation}
  \mathcal H_B^\vee(\boldsymbol y)
  =\mathcal H_A(\boldsymbol x).
  \label{eq:handle-complement-full}
\end{equation}
After reduction to the primitive coordinate this becomes
\begin{equation}
  \mathcal H_B^\vee(z)
  =\mathcal H_A(1-z)
  \qquad
  \text{in }\mathbb Q[z]/(P^\vee(z)).
  \label{eq:handle-complement-exact}
\end{equation}
Therefore, on paired vacua,
\begin{equation}
  \mathcal H_B^\vee\!\left(z^{(\beta)}\right)
  =\mathcal H_A\!\left(t^{(\beta)}\right)
  =\frac{11}{4\sin^2(\pi s_\beta/11)}.
  \label{eq:CE8inv-handle-cyclotomic}
\end{equation}
Equations~\eqref{eq:Wilson-eigenvalue-complement} and \eqref{eq:handle-complement-exact} show that both ingredients entering the reconstruction of the modular $S$-matrix are preserved sector by sector.  Consequently, in the paired orders \eqref{eq:CE8inv-sector-order} and \eqref{eq:CE8-s-labels},
\begin{equation}
  \bigl(S_B^{\vee,\mathrm{Bethe}}\bigr)_{\alpha\beta}
  =\bigl(S_A^{\mathrm{Bethe}}\bigr)_{\alpha\beta}
  =\frac{2}{\sqrt{11}}
  (-1)^{s_\alpha+s_\beta}
  \sin\!\left(\frac{2\pi s_\alpha s_\beta}{11}\right).
  \label{eq:CE8inv-S-exact}
\end{equation}
Thus the equality of the modular $S$ data is not a numerical coincidence between two sets of Bethe roots.  It follows from an exact isomorphism of the Bethe algebras, together with the particle--vortex transformation of Wilson charges and the complementary transformation of the handle-gluing operator.  In particular, the $A$-twisted $C_{E_8}$ theory and the $B$-twisted $C_{E_8}^{-1}$ theory define the same semisimple Frobenius and line data at the level detected by the modular $S$-matrix.

\subsection{Fibering operators and modular $T$-matrix}
\label{subsec:fibering-T-matrix}

We next compare the fibering operators, whose role in supersymmetric partition functions on Seifert manifolds was developed in~\cite{Closset:2017zgf,ClossetKimWillett:2018Seifert}.  In contrast to the handle-gluing operator, the fibering operator is determined by the effective twisted superpotential and therefore depends on a logarithmic lift of each Bethe vacuum.  The important point for the present comparison is that the relation between the two mirror theories can be derived exactly without evaluating the individual dilogarithm sums numerically.

\subsubsection{The $A$-twisted $C_{E_8}$ theory}
\label{subsubsec:A-twisted-CE8-T}

For a Bethe vacuum $\boldsymbol x=(x_1,\ldots,x_8)$, choose logarithmic lifts $\widetilde\ell_i\in\mathbb C$ satisfying $e^{\widetilde\ell_i}=x_i$, together with compatible analytically continued branches of the dilogarithms.  We denote the Bethe vacuum equipped with this branch data by $\widetilde{\boldsymbol x}$ and refer to it as a lifted Bethe vacuum.  The TQFT-normalized fibering operator is then
\begin{equation}
  \mathcal F_A(\widetilde{\boldsymbol x})
  =\exp\!\left[
  \frac{1}{2\pi \mathrm{i}}\,
  \Phi_{C_{E_8}}(\widetilde{\boldsymbol x})
  \right],
  \label{eq:CE8-fibering-operator}
\end{equation}
where, in the present diagonal-charge presentation,
\begin{equation}
  \Phi_{C_{E_8}}(\widetilde{\boldsymbol x})
  =
  \sum_{i=1}^{8}\operatorname{Li}_2(x_i)
  +\frac12
  \sum_{i,j=1}^{8}(C_{E_8})_{ij}\,
  \widetilde\ell_i\widetilde\ell_j,
  \qquad
  e^{\widetilde\ell_i}=x_i.
  \label{eq:CE8-fibering-exponent}
\end{equation}
The logarithms and dilogarithms in \eqref{eq:CE8-fibering-exponent} are understood on compatible analytically continued branches.  Equivalently, one may regard $\Phi_{C_{E_8}}$ as the corresponding extended Rogers-dilogarithm exponent.  Changing the common lift changes \eqref{eq:CE8-fibering-operator} only by a vacuum-independent framing phase.

The absolute normalization of the fibering operator and that of the modular $T$-matrix are therefore both sensitive to a common framing convention.  In the standard Seifert normalization, the fibering eigenvalues determine the inverse modular $T$ eigenvalues up to a vacuum-independent overall phase~\cite{Closset:2017zgf,ClossetKimWillett:2018Seifert,Gang:2023rei,GangKimParkStubbs:2024Nahm}.  Accordingly, below we first compare the vacuum-dependent ratios and then identify the common multiplier from the exact complement relation.

The $T_{10}M_{\mathrm{eff}}(11,2)$ character system identified in Section~\ref{subsec:Wilson-half-indices} also determines an independent 2d modular $T$-matrix.  If a normalized character begins as $q^{h_\beta^{\rm eff}-c_{\rm eff}/24}(1+\cdots)$, then its transformation under $\tau\mapsto\tau+1$ gives
\begin{equation}
  \bigl(T_{T_{10}}^{\mathrm{char}}\bigr)_{\beta\beta}
  =\exp\!\left[
    2\pi\mathrm{i}\left(
      h_\beta^{\rm eff}-\frac{c_{\rm eff}}{24}
    \right)
  \right].
  \label{eq:T10-character-T-definition}
\end{equation}
For
\begin{equation}
  h_\beta^{\rm eff}
  =0,\ \frac5{11},\ \frac8{11},\ \frac{10}{11},\ \frac{12}{11},
  \qquad
  c_{\rm eff}=\frac{80}{11},
\end{equation}
this gives
\begin{equation}
  T_{T_{10}}^{\mathrm{char}}
  =\operatorname{diag}\!\left(
  e^{-20\pi \mathrm{i}/33},
  e^{10\pi \mathrm{i}/33},
  e^{28\pi \mathrm{i}/33},
  e^{40\pi \mathrm{i}/33},
  e^{52\pi \mathrm{i}/33}
  \right).
  \label{eq:CE8-T-matrix}
\end{equation}
This matrix is fixed by the 2d character data before evaluating any 3d fibering operator.  We denote it by $T_A$ in the common framing convention and compare the transformation law that it predicts with the independently derived 3d fibering relation below.  We will not need an independent evaluation of the five individual dilogarithm sums in order to make this comparison with the $C_{E_8}^{-1}$ theory.

\subsubsection{The $B$-twisted $C_{E_8}^{-1}$ theory and the $E_8$ framing phase}
\label{subsubsec:B-twisted-CE8inv-T}

Let $\boldsymbol y=\boldsymbol1-\boldsymbol x$ be the Bethe vacuum paired with $\boldsymbol x$ by the particle--vortex complement map.  We choose a compatible logarithmic and dilogarithmic lift of $\boldsymbol y$ and denote the resulting lifted Bethe vacuum by $\widetilde{\boldsymbol y}$.  Thus $\widetilde{\boldsymbol x}$ and $\widetilde{\boldsymbol y}$ refer to the paired Bethe vacua together with branch data chosen so that the extended Rogers identity applies componentwise.  For these compatible lifts,
\begin{equation}
  \Phi_{C_{E_8}}(\widetilde{\boldsymbol x})
  +
  \Phi_{C_{E_8}^{-1}}(\widetilde{\boldsymbol y})
  =
  \frac{8\pi^2}{6}
  \pmod{4\pi^2}.
  \label{eq:E8-Rogers-complementarity}
\end{equation}
This is the lifted version of $L(x)+L(1-x)=\pi^2/6$, summed over the eight gauge variables.  Exponentiating \eqref{eq:E8-Rogers-complementarity} yields the exact fibering relation
\begin{equation}
  \mathcal F_B^\vee(\widetilde{\boldsymbol y})
  =
  e^{-2\pi \mathrm{i}/3}\,
  \mathcal F_A(\widetilde{\boldsymbol x})^{-1}.
  \label{eq:E8-fibering-complementarity}
\end{equation}
The multiplier is independent of the Bethe vacuum.  Hence it is invisible to the Wilson loop algebra and to the reconstructed $S$-matrix, but it survives in the framed $T$ data and in Seifert partition functions.  It can be written as
\begin{equation}
  e^{-2\pi \mathrm{i}/3}
  =e^{-2\pi \mathrm{i}\,8/24}.
  \label{eq:E8-framing-phase}
\end{equation}
This is precisely the $1\times1$ genus-one modular $T$-matrix of the level-one $E_8$ character: since $(E_8)_1$ has only the vacuum module, with $h=0$ and $c=8$,
\begin{equation}
  T_{(E_8)_1}^{\mathrm{char}}
  :=\exp\!\left[2\pi\mathrm{i}\left(0-\frac{8}{24}\right)\right]
  =e^{-2\pi \mathrm{i}/3}.
  \label{eq:E8-character-T-multiplier}
\end{equation}
It records the framing anomaly of the invertible $E_8$ phase.  This should be distinguished from the topological twist of its unique simple object, which is $1$.

With the same framing convention fixed, the corresponding universal multiplier is visible directly and exactly in the modular $T$-matrices of the two character systems.  The dual $M_{\mathrm{eff}}(2,11)$ character system of Section~\ref{subsec:dual-Wilson-Zagier} has, in the paired order of the $C_{E_8}^{-1}$ theory,
\begin{equation}
  h_\beta^{\vee,\rm eff}
  =0,\ \frac6{11},\ \frac3{11},\ \frac1{11},\ \frac{10}{11},
  \qquad
  c_{\rm eff}^\vee=\frac8{11}.
\end{equation}
Its character-theoretic modular matrix is
\begin{equation}
  T_{M_{\rm eff}(2,11)}^{\vee,\mathrm{char}}
  =\operatorname{diag}\!\left(
  e^{-2\pi \mathrm{i}/33},
  e^{34\pi \mathrm{i}/33},
  e^{16\pi \mathrm{i}/33},
  e^{4\pi \mathrm{i}/33},
  e^{58\pi \mathrm{i}/33}
  \right).
  \label{eq:CE8inv-T-matrix}
\end{equation}
In the same framing convention we denote this character-theoretic matrix by $T_B^\vee$.  The paired effective conformal weights satisfy
\begin{equation}
  h_\beta^{\rm eff}+h_\beta^{\vee,\rm eff}
  =(0,1,1,1,2),
  \qquad
  c_{\rm eff}+c_{\rm eff}^\vee=8.
  \label{eq:paired-weights-central-charge}
\end{equation}
Consequently, sector by sector,
\begin{equation}
  T_B^\vee
  =T_{(E_8)_1}^{\mathrm{char}}\,T_A^{-1}
  =e^{-2\pi \mathrm{i}/3}\,T_A^{-1}.
  \label{eq:T-matrix-complementarity}
\end{equation}
Equivalently, after dividing by the vacuum entries, the topological spins are exactly inverted,
\begin{equation}
  \frac{(T_B^\vee)_{\beta\beta}}{(T_B^\vee)_{00}}
  =
  \left(
  \frac{(T_A)_{\beta\beta}}{(T_A)_{00}}
  \right)^{-1}.
  \label{eq:normalized-T-inversion}
\end{equation}
Thus, in the common framing convention just described, the universal phase in the 3d fibering relation \eqref{eq:E8-fibering-complementarity} agrees with the common framing multiplier required by the two exact character-theoretic $T$-matrices.  Together with \eqref{eq:CE8inv-S-exact}, this gives the protected modular-data relation
\begin{equation}
  S_B^\vee=S_A,
  \qquad
  T_B^\vee=e^{-2\pi \mathrm{i}/3}T_A^{-1}.
  \label{eq:AB-protected-modular-relation}
\end{equation}
The inversion of the normalized $T$ data is the genus-one manifestation of orientation reversal, while the remaining vacuum-independent factor records the invertible $E_8$ framing sector.  Importantly, the comparison itself requires no numerical recognition of Rogers-dilogarithm values: it follows from the particle--vortex complement map, the lifted Rogers identity, and the exact conformal-weight data of the two five-component character systems.

\subsection{$A/B$-twisted TQFT correspondence}
\label{subsec:AB-TQFT-correspondence}

The pair studied in Sections~\ref{subsubsec:A-twisted-CE8-S}--\ref{subsubsec:B-twisted-CE8inv-T} is the reference complementary pair.  The exchange of $A$- and $B$-twisted TQFT data under rank-zero mirror symmetry has close precedents in~\cite{Gang:2021hrd,CreutzigGarnerKim:2024mirror}; here we establish the exchange directly from the particle--vortex complement map.  The original $A$-twist and the dual $B$-twist both have vanishing effective topological mixing.  The protected mirror map also exchanges the other two distinguished twist specializations.  We now show that this second exchange follows from the same particle--vortex complement map and therefore completes the comparison of the protected $A/B$-twisted TQFT data.

The effective mixing vectors of the four twisted presentations are
\begin{equation}
  \boldsymbol\mu_A=0,
  \qquad
  \boldsymbol\mu_B=2\boldsymbol a,
  \qquad
  \boldsymbol\mu_B^\vee=0,
  \qquad
  \boldsymbol\mu_A^\vee=-2\boldsymbol a^\vee.
  \label{eq:four-AB-effective-mixings}
\end{equation}
Using $C_{E_8}\boldsymbol a^\vee=-\boldsymbol a$, this may be written as
\begin{equation}
  \boldsymbol\mu_B^\vee=C_{E_8}^{-1}\boldsymbol\mu_A=0,
  \qquad
  \boldsymbol\mu_A^\vee=C_{E_8}^{-1}\boldsymbol\mu_B.
  \label{eq:AB-mixing-complement-map}
\end{equation}
Thus the same inverse Cartan matrix that transports Wilson charges also transports the effective topological mixing between complementary twists.

A change of topological mixing by an integral vector $\boldsymbol\mu$ multiplies the handle-gluing operator by the corresponding Wilson monomial~\cite{Closset:2017zgf}.  Hence
\begin{equation}
  \mathcal H_B(\boldsymbol x)
  =\boldsymbol x^{2\boldsymbol a}\mathcal H_A(\boldsymbol x),
  \qquad
  \mathcal H_A^\vee(\boldsymbol y)
  =\boldsymbol y^{-2\boldsymbol a^\vee}\mathcal H_B^\vee(\boldsymbol y).
  \label{eq:AB-shifted-handle-operators}
\end{equation}
On the complement locus $\boldsymbol y=\boldsymbol x^{C_{E_8}}=\boldsymbol1-\boldsymbol x$,
\begin{equation}
  \boldsymbol y^{-2\boldsymbol a^\vee}
  =\boldsymbol x^{-2C_{E_8}\boldsymbol a^\vee}
  =\boldsymbol x^{2\boldsymbol a}.
  \label{eq:AB-mixing-monomial-identity}
\end{equation}
Combining this identity with \eqref{eq:handle-complement-full} gives the second exact handle relation
\begin{equation}
  \mathcal H_A^\vee(\boldsymbol1-\boldsymbol x)
  =\mathcal H_B(\boldsymbol x).
  \label{eq:dual-A-original-B-handle}
\end{equation}
The TQFT-normalized fibering operator does not acquire an additional vacuum-dependent monomial under this change of mixing: the change of the linear background term in the effective twisted superpotential is cancelled by the standard flux factor in the Seifert operator~\cite{Closset:2017zgf}.  Using the same compatible lift data $\widetilde{\boldsymbol x}$ and $\widetilde{\boldsymbol y}$ introduced above, the lifted Rogers-complement argument used in \eqref{eq:E8-fibering-complementarity} therefore gives
\begin{equation}
  \mathcal F_A^\vee(\widetilde{\boldsymbol y})
  =e^{-2\pi \mathrm{i}/3}\,
  \mathcal F_B(\widetilde{\boldsymbol x})^{-1},
  \qquad
  \boldsymbol y=\boldsymbol1-\boldsymbol x,
  \label{eq:dual-A-original-B-fibering}
\end{equation}
Together with the reference pair, the complete exchange of the handle-gluing and fibering operators is
\begin{align}
  \mathcal H_B^\vee(\boldsymbol y)
  &=\mathcal H_A(\boldsymbol x),
  &
  \mathcal F_B^\vee(\widetilde{\boldsymbol y})
  &=\kappa_{E_8}\,\mathcal F_A(\widetilde{\boldsymbol x})^{-1},
  \nonumber\\
  \mathcal H_A^\vee(\boldsymbol y)
  &=\mathcal H_B(\boldsymbol x),
  &
  \mathcal F_A^\vee(\widetilde{\boldsymbol y})
  &=\kappa_{E_8}\,\mathcal F_B(\widetilde{\boldsymbol x})^{-1},
  \label{eq:complete-AB-operator-exchange}
\end{align}
where
\begin{equation}
  \boldsymbol y=\boldsymbol1-\boldsymbol x,
  \qquad
  \kappa_{E_8}
  :=T_{(E_8)_1}^{\mathrm{char}}
  =e^{-2\pi \mathrm{i}/3}.
  \label{eq:kappa-E8-definition}
\end{equation}
Thus $\kappa_{E_8}$ is the character-theoretic framing multiplier identified in \eqref{eq:E8-character-T-multiplier}, rather than the topological twist of the unique simple object.
The Wilson line map is the same for both pairs,
\begin{equation}
  \boldsymbol Q^\vee=C_{E_8}^{-1}\boldsymbol Q,
  \qquad
  W_{\boldsymbol Q^\vee}^\vee(\boldsymbol y)
  =W_{\boldsymbol Q}(\boldsymbol x).
  \label{eq:AB-Wilson-map-general}
\end{equation}
Thus the full Bethe-vacuum data entering the semisimple Seifert formulas are transported sector by sector.

Let $\mathcal M_{g,p}$ be the oriented degree-$p$ circle bundle over a closed genus-$g$ Riemann surface.  For $X\in\{A,B\}$, the twisted correlator of Wilson lines is
\begin{equation}
  \left\langle\prod_{a=1}^{n}W_{\boldsymbol Q_a}\right\rangle_{\mathcal T_X;g,p}
  =
  \sum_{\beta}
  \mathcal H_{X,\beta}^{\,g-1}
  \mathcal F_{X,\beta}^{\,p}
  \prod_{a=1}^{n}W_{\boldsymbol Q_a}(\widehat u_\beta),
  \label{eq:Seifert-Wilson-correlator}
\end{equation}
where the sum is over the five regular Bethe vacua~\cite{Closset:2017zgf,ClossetKimWillett:2018Seifert,Gaiotto:2024ioj}.  If $\overline A=B$ and $\overline B=A$, then \eqref{eq:complete-AB-operator-exchange} and \eqref{eq:AB-Wilson-map-general} imply
\begin{equation}
  \left\langle
  \prod_{a=1}^{n}W_{C_{E_8}^{-1}\boldsymbol Q_a}^\vee
  \right\rangle_{\mathcal T_{\overline X}^\vee;g,p}
  =
  \kappa_{E_8}^{\,p}
  \left\langle
  \prod_{a=1}^{n}W_{\boldsymbol Q_a}
  \right\rangle_{\mathcal T_X;g,-p}.
  \label{eq:AB-Seifert-Wilson-exchange}
\end{equation}
In particular,
\begin{align}
  Z_{\mathcal T^\vee}^{B}(\mathcal M_{g,p})
  &=e^{-2\pi \mathrm{i} p/3}
    Z_{\mathcal T}^{A}(\mathcal M_{g,-p}),
  \nonumber\\
  Z_{\mathcal T^\vee}^{A}(\mathcal M_{g,p})
  &=e^{-2\pi \mathrm{i} p/3}
    Z_{\mathcal T}^{B}(\mathcal M_{g,-p}).
  \label{eq:AB-Seifert-partition-exchange}
\end{align}
The replacement $p\mapsto-p$ is precisely the inversion of the fibering operator and therefore the Seifert manifestation of orientation reversal.

For completeness, the Wilson loop algebra itself is twist independent.  In the original $A$-twisted presentation, the five polynomial representatives in \eqref{eq:CE8-Wilson-polynomials} form an integral basis of $\mathcal B_{C_{E_8}}$: their coefficient matrix in the monomial basis $\{1,t,t^2,t^3,t^4\}$ has determinant $-1$.  After the sign rephasing
\begin{equation}
  \widehat L_0=L_0,
  \qquad
  \widehat L_{5/11}=-L_{5/11},
  \qquad
  \widehat L_{8/11}=L_{8/11},
  \qquad
  \widehat L_{10/11}=-L_{10/11},
  \qquad
  \widehat L_{12/11}=L_{12/11},
  \label{eq:based-line-sign-choice}
\end{equation}
all structure constants are nonnegative integers.  Writing
\begin{equation}
  \mathbf1=\widehat L_0,
  \qquad
  A=\widehat L_{5/11},
  \qquad
  B=\widehat L_{8/11},
  \qquad
  C=\widehat L_{10/11},
  \qquad
  D=\widehat L_{12/11},
\end{equation}
the products are
\begin{align}
  A^2&=\mathbf1+A+B+C,
  &AB&=A+B+C+D,
  \nonumber\\
  AC&=A+B+D,
  &AD&=B+C,
  \nonumber\\
  B^2&=\mathbf1+A+B+C+D,
  &BC&=A+B+C,
  \nonumber\\
  BD&=A+B,
  &C^2&=\mathbf1+B+C,
  \nonumber\\
  CD&=A+D,
  &D^2&=\mathbf1+C.
  \label{eq:based-fusion-rules}
\end{align}
This based ring is isomorphic to the even-label subring of the $SU(2)_9$ fusion ring, conventionally denoted the $SO(3)_9$ fusion ring, through
\begin{equation}
  \mathbf1\leftrightarrow[0],
  \qquad
  C\leftrightarrow[2],
  \qquad
  B\leftrightarrow[4],
  \qquad
  A\leftrightarrow[6],
  \qquad
  D\leftrightarrow[8].
  \label{eq:SO39-based-ring-identification}
\end{equation}
The particle--vortex complement map and the change of twist preserve the abstract Wilson loop algebra, while the natural Wilson loop basis may be permuted and rephased.  Consequently all four twisted presentations have isomorphic based fusion rings.

\paragraph{Remark on the $B$-twist of the $C_{E_8}$ theory and the $A$-twist of the $C_{E_8}^{-1}$ theory.}
The preceding reconstruction used the $A$-twist of the $C_{E_8}$ theory and the $B$-twist of the $C_{E_8}^{-1}$ theory.  For the opposite pair of twists, the Bethe equations and Wilson loop algebra are unchanged, whereas the handle-gluing operator is modified by the twist-dependent effective dilaton.  Thus the reconstruction of the modular matrix is not literally twist independent in a fixed Wilson basis: \eqref{eq:AB-shifted-handle-operators} changes the Frobenius weights when the twist is changed.  The complementary $B/A$ pair instead exchanges the two Galois specializations of the same $SO(3)_9$ fusion ring.  In the standard even-label basis $a,b\in\{0,2,4,6,8\}$, their modular matrices may be written as
\begin{equation}
  S^{(m)}_{ab}
  =\frac{2}{\sqrt{11}}
  \sin\!\left(\frac{\pi m(a+1)(b+1)}{11}\right),
  \qquad m=5,6.
  \label{eq:SO39-S-Galois-pair}
\end{equation}
Because $(a+1)(b+1)$ is odd,
\begin{equation}
  \sin\!\left(\frac{6\pi(a+1)(b+1)}{11}\right)
  =
  \sin\!\left(\frac{5\pi(a+1)(b+1)}{11}\right),
  \label{eq:SO39-S-Galois-5-6-equality}
\end{equation}
so $S^{(6)}=S^{(5)}$.  Thus, after passing to the natural Galois-adapted Wilson loop basis, the $B$-twisted $C_{E_8}$ theory and the $A$-twisted $C_{E_8}^{-1}$ theory reproduce the same numerical modular $S$-matrix as the reference $A/B$ pair, while the normalized $T$ data are exchanged by orientation reversal.  This is the complementary version of the modular relation derived above.

\subsubsection{A comment on ribbon-category completion}
\label{subsec:full-ribbon-TQFT-completion}

The supersymmetric localization calculation determines the five regular sectors, their Wilson loop fusion algebra, and the modular $S$- and $T$-data, but these data do not in general determine a unique full ribbon category.  For the present system there is nevertheless a natural standard completion, using the familiar Galois specializations of the $SU(2)_9$ quantum-group data and the standard Reshetikhin--Turaev construction~\cite{ReshetikhinTuraev:1991,Harvey:2019galois,Ardonne:2010qg}.  Let $\mathcal C_m$ denote the $m$-th Galois specialization of the even, or adjoint, $SU(2)_9$ quantum-group category, with
\begin{equation}
  q_m=e^{2\pi \mathrm{i}m/11},
  \qquad m=5,6.
  \label{eq:standard-ribbon-qm}
\end{equation}
Its five simple objects reproduce the $SO(3)_9$ fusion ring found above.  Since $q_6=q_5^{-1}$, the standard quantum-group recoupling data have the same associator and inverse braiding and ribbon twists for the two specializations.  Thus the natural completions of the two non-invertible line theories are the $m=5$ and $m=6$ Galois-conjugate $SO(3)_9$ Reshetikhin--Turaev TQFTs, with
\begin{equation}
  \mathcal C_6\simeq_{\mathrm{ribbon}}\mathcal C_5^{\mathrm{rev}}.
  \label{eq:standard-ribbon-reverse}
\end{equation}
This completion should be regarded as additional mathematical input, not as a unique reconstruction from the Bethe--Seifert data.  In particular, the localization calculation does not by itself determine the full junction spaces, $F$-symbols, and $R$-symbols.  The detailed categorical construction, including the reverse-ribbon equivalence and the associated diagonal Lagrangian algebra, will be presented in a forthcoming companion paper.  In the present paper we use only this identification of the natural standard completion; the finer $E_8$ framing lift continues to be fixed independently by the supersymmetric fibering operator.

We finally summarize the result in framed-TQFT language.  Let $\mathfrak T_X(\mathcal S)$ denote the semisimple Bethe--Seifert TQFT data extracted from the $X$-twist of a theory $\mathcal S$, where $\mathcal S=\mathcal T$ or $\mathcal T^\vee$, and let $\mathfrak X^{\mathrm{rev}}$ denote orientation reversal.  We denote by $\mathfrak E_8$ the invertible bosonic $E_8$ phase: it has no nontrivial topological lines and chiral central charge $c_-=8$, with framing multiplier $\kappa_{E_8}=T_{(E_8)_1}^{\mathrm{char}}=e^{-2\pi \mathrm{i}/3}$.  Equations~\eqref{eq:complete-AB-operator-exchange}--\eqref{eq:AB-Seifert-partition-exchange} are then summarized by
\begin{align}
  \mathfrak T_A(\mathcal T)
  &\simeq_{\mathrm{prot}}
  \mathfrak T_B(\mathcal T^\vee)^{\mathrm{rev}}
  \otimes\mathfrak E_8,
  \nonumber\\
  \mathfrak T_B(\mathcal T)
  &\simeq_{\mathrm{prot}}
  \mathfrak T_A(\mathcal T^\vee)^{\mathrm{rev}}
  \otimes\mathfrak E_8.
  \label{eq:AB-framed-TQFT-correspondence}
\end{align}
Here $\simeq_{\mathrm{prot}}$ means equality of the protected semisimple data determined by the regular Bethe vacua, Wilson loop algebra, handle-gluing operator, fibering operator, and all associated Seifert sums.  It does not by itself assert an equivalence of fully extended framed TQFTs.  The invertible factor $\mathfrak E_8$ is invisible to the fusion ring and to the vacuum-dependent $S$ data; it records only the common framing anomaly in the fibering operator and the vacuum modular $T$ multiplier.  In this precise sense, the exact $A/B$-twisted protected data associated with the mirror pair realize an orientation-reversed topological correspondence with a relative $E_8$ framing sector.  The Zagier duality encountered in the half-indices is a boundary-character shadow of this 3d structure.  The standard completion summarized above gives a natural reverse-ribbon realization of the non-invertible part, while the $\mathfrak E_8$ factor is fixed independently by the 3d fibering response.

\subsection{A 3d $\mathcal N=4$ analogue of the CS/WZW solid-torus correspondence}
\label{subsec:susy-CS-WZW-solid-torus}

The results of Sections~\ref{subsec:Wilson-half-indices} and \ref{subsec:Wilson-S-matrix}--\ref{subsec:fibering-T-matrix} admit a useful interpretation in analogy with the standard Chern--Simons/WZW solid-torus construction~\cite{Witten:1988hf,ElitzurMooreSchwimmerSeiberg:1989}.  In ordinary Chern--Simons theory, inserting a Wilson loop $W_R$ along the core of a solid torus prepares a state in the Hilbert space associated with the boundary torus.  Under the Chern--Simons/WZW correspondence, its holomorphic boundary wavefunction is represented by a torus conformal block, and for the solid-torus state labeled by an integrable representation $R$ this is the corresponding WZW character.  The Wilson line fusion algebra is represented by the Verlinde algebra, while the modular transformations of the torus conformal blocks reproduce the bulk modular data.

A structurally parallel, but independently derived, correspondence emerges from the supersymmetric Chern--Simons--matter theories studied here.  The Wilson loop half-indices, the Wilson loop Bethe algebra, and the modular data reconstructed from Bethe vacua and Seifert operators were obtained in the preceding sections directly from the 3d theories.  Table~\ref{tab:CS-WZW-susy-comparison} summarizes the comparison.

\begin{table}[htbp]
\centering
\footnotesize
\setlength{\tabcolsep}{4pt}
\renewcommand{\arraystretch}{1.12}
\caption{Comparison between the ordinary Chern--Simons/WZW solid-torus construction and the structure found in the present supersymmetric Chern--Simons--matter theories.}
\label{tab:CS-WZW-susy-comparison}
\begin{tabular}{@{}>{\raggedright\arraybackslash}p{0.21\linewidth}>{\raggedright\arraybackslash}p{0.30\linewidth}>{\raggedright\arraybackslash}p{0.41\linewidth}@{}}
\toprule
 & ordinary Chern--Simons/WZW & present supersymmetric construction \\
\midrule
bulk theory
& Chern--Simons TQFT
& $A/B$-twisted 3d Chern--Simons--matter theory \\
solid-torus insertion
& Wilson loop $W_R$ along the core
& Wilson loop $W_{Q_i}$ or $W^\vee_{Q_i^\vee}$ along the core \\
solid-torus wavefunction
& WZW torus character $\chi_R$
& normalized Wilson loop half-index $I\!\!I^A[W_{Q_i}]\leftrightarrow\chi_i$ or $I\!\!I^{\vee B}[W^\vee_{Q_i^\vee}]\leftrightarrow\chi_i^\vee$ \\
bulk line algebra
& Wilson line fusion algebra
& Wilson loop Bethe algebra \\
character-side algebra
& Verlinde algebra
& formal Verlinde algebra of the five-character system \\
modular $S$ data
& modular $S$-matrix of WZW torus conformal blocks
& $S$ reconstructed from Bethe vacua, Wilson loops, and the handle-gluing operator \\
modular $T$ data
& modular $T$-matrix of WZW characters
& fibering-operator data compared with the character-theoretic $T$-matrix \\
\bottomrule
\end{tabular}
\end{table}

Here $Q_i$ labels the Wilson charge associated with the $i$-th character sector, and $Q_i^\vee=C_{E_8}^{-1}Q_i$ is its particle--vortex image.  The arrows in the solid-torus-wavefunction row indicate the normalized character identification: the sector-dependent $q$-shifts are the ones displayed explicitly in Sections~\ref{subsec:Wilson-half-indices} and~\ref{subsec:dual-Wilson-Zagier}.

The analogy extends beyond the equality of individual $q$-series.  The Wilson loop half-indices in Section~\ref{subsec:Wilson-half-indices} identify five solid-torus amplitudes with the five components of the $T_{10}M_{\mathrm{eff}}(11,2)$ character system.  Independently, the $A$-twisted 3d theory reconstructs the modular matrix from the Bethe vacua, Wilson loop eigenvalues, and handle-gluing operator.  The exact result derived above is
\begin{equation}
  S_A^{\mathrm{Bethe}}
  =S_{T_{10}M_{\mathrm{eff}}(11,2)}^{\mathrm{char}}.
  \label{eq:CSWZW-S-comparison}
\end{equation}
Moreover, writing
$L_{h_\alpha}L_{h_\beta}=\sum_\gamma N^{\gamma}_{\alpha\beta}L_{h_\gamma}$,
the Wilson loop structure constants obey
\begin{equation}
  N^{\gamma}_{\alpha\beta}
  =\sum_{\sigma}
  \frac{
    (S_A^{\mathrm{Bethe}})_{\alpha\sigma}
    (S_A^{\mathrm{Bethe}})_{\beta\sigma}
    \bigl((S_A^{\mathrm{Bethe}})^{-1}\bigr)_{\sigma\gamma}
  }{
    (S_A^{\mathrm{Bethe}})_{0\sigma}
  },
  \label{eq:CSWZW-Verlinde-diagonalization}
\end{equation}
which is the Verlinde formula in the character basis.  Thus the Wilson loop Bethe algebra realizes the formal Verlinde algebra of the same five-character modular representation.  After the sign rephasing in \eqref{eq:based-line-sign-choice}, this becomes the positive $SO(3)_9$ based fusion ring.

The same comparison applies to the modular $T$ data.  The $T_{10}M_{\mathrm{eff}}(11,2)$ character system fixes its diagonal matrix directly from the leading exponents through \eqref{eq:T10-character-T-definition}.  The 3d fibering operator supplies the corresponding framed topological datum, and its exact particle--vortex transformation reproduces the relation required by the two character-theoretic $T$-matrices, including the common $E_8$ framing multiplier in \eqref{eq:T-matrix-complementarity}.  In this sense the parallel with the ordinary Chern--Simons/WZW construction holds simultaneously at the level of solid-torus amplitudes, the line-operator algebra, and genus-one modular data.

We emphasize the scope of this interpretation.  The statements above do not require, and do not establish, the existence of a microscopic boundary VOA whose complete irreducible-character vector is the $T_{10}M_{\mathrm{eff}}(11,2)$ character system.  In particular, we do not identify a microscopic boundary VOA for the boundary condition used to define the half-index.  What is established here is the correspondence of the Wilson loop half-indices with the five-character system, together with the matching of the semisimple line algebra and the genus-one modular data.  This is the precise sense in which the present construction provides a supersymmetric analogue of the Chern--Simons/WZW solid-torus correspondence.

\section{The duality interface and the $(E_8)_1$ torus amplitude}
\label{sec:E8-interface}

Sections~\ref{sec:particle-vortex-mirror} and~4 established two complementary aspects of the infrared mirror relation.  First, particle--vortex duality produces the ultraviolet description of the $C_{E_8}^{-1}$ theory and maps the distinguished Wilson charges by $\boldsymbol Q\mapsto C_{E_8}^{-1}\boldsymbol Q$.  Second, the $A/B$-twisted localization data show that the two theories are related by orientation reversal together with the universal framing multiplier $\kappa_{E_8}=T_{(E_8)_1}^{\mathrm{char}}=e^{-2\pi \mathrm{i}/3}$.  We now combine these statements into a physical interface picture.

As summarized in Table~\ref{tab:CS-WZW-susy-comparison}, the $A$-twisted $C_{E_8}$ theory and the $B$-twisted $C_{E_8}^{-1}$ theory assign finite-dimensional state spaces to the torus.  Let
\begin{equation}
  \mathscr H_{T^2}^{A}(\mathcal T)
\end{equation}
denote the torus state space of the $A$-twisted $C_{E_8}$ theory.  A path integral on the solid torus $D^2\times S^1$ prepares a state in this space.  With no insertion it gives the vacuum state $|0\rangle_A$, whereas inserting a Wilson loop $W_{Q_i}$ along the core $S^1$ prepares
\begin{equation}
  |W_{Q_i}\rangle_A
  :=Z_A(D^2\times S^1;W_{Q_i})
  \in \mathscr H_{T^2}^{A}(\mathcal T).
  \label{eq:A-solid-torus-Wilson-state}
\end{equation}
For the five distinguished sectors relevant here, we use these states as the Wilson loop basis,
\begin{equation}
  \mathscr H_{T^2}^{A}(\mathcal T)
  =\operatorname{Span}
  \left\{|W_{Q_i}\rangle_A\right\}_{i=0}^{4}.
  \label{eq:A-torus-state-space-Wilson-basis}
\end{equation}
In the supersymmetric solid-torus localization used in Section~\ref{subsec:Wilson-half-indices}, the corresponding holomorphic amplitudes are the Wilson loop half-indices $I\!\!I^A[W_{Q_i}]$.  After the sector-dependent normalizations specified there, they give the five components $\chi_i$ of the $T_{10}M_{\mathrm{eff}}(11,2)$ character system.

The same construction applies to the $B$-twisted $C_{E_8}^{-1}$ theory.  Its corresponding Wilson loop states are
\begin{equation}
  |W^\vee_{Q_i^\vee}\rangle_{B}
  :=Z_{B}^{\vee}(D^2\times S^1;W^\vee_{Q_i^\vee})
  \in \mathscr H_{T^2}^{B}(\mathcal T^\vee),
  \qquad
  Q_i^\vee=C_{E_8}^{-1}Q_i,
  \label{eq:dual-solid-torus-Wilson-state}
\end{equation}
and their holomorphic amplitudes $I\!\!I^{\vee B}[W^\vee_{Q_i^\vee}]$ give the paired components $\chi_i^\vee$ of the dual character system, with the sector-dependent normalizations described in Section~\ref{subsec:dual-Wilson-Zagier}.  Thus the two five-dimensional torus state spaces admit a Wilson loop basis prepared by core Wilson loop insertions and a Bethe-idempotent basis in which the Wilson loop algebra is diagonal; Section~\ref{subsec:Wilson-S-matrix} relates these descriptions through the modular $S$-matrix.

The purpose of this section is to construct the particle--vortex duality wall as a linear map between these torus state spaces, determine its action in the Wilson loop and Bethe-idempotent bases, and use the $A/B$-twisted TQFT data to identify the residual $E_8$ framing sector and the resulting torus interface amplitude.  The five-sector arithmetic pairing will then appear as a consequence of the Wilson transmission rule.

We begin with the microscopic part of the construction that is under direct control.  The elementary particle--vortex duality admits a half-BPS duality interface described by a 2d $\mathcal N=(0,2)$ Fermi multiplet coupled to the bulk fields on the two sides~\cite{Dimofte:2017tpi}.  Tensoring eight copies and gauging the eight flavor symmetries with the $C_{E_8}$ Chern--Simons coupling gives a natural rank-eight wall.  Folding this wall supplies the orientation reversal geometrically, while integrating out the intermediate gauge fields reproduces the Chern--Simons level matrix $C_{E_8}^{-1}$ through a Schur complement.  The additional monopole-superpotential deformation is compatible with the same wall at the level of the BPS operator map derived in Section~\ref{sec:particle-vortex-mirror}; a fully explicit off-shell $(0,2)$ realization of all seven monopole couplings will not be assumed below.

\subsection{The gauged particle--vortex duality wall}
\label{subsec:gauged-pv-interface}

It is useful to keep the geometry of the wall explicit from the outset.  Before folding, the particle--vortex wall is an interface between the two bulk theories,
\[
  \mathcal T[\boldsymbol A]
  \ \Big|_{\mathcal I}\ 
  \mathcal T^\vee[\boldsymbol B],
  \qquad
  x^\perp<0\ \Big|_{x^\perp=0}\ x^\perp>0,
\]
where $\mathcal T$ occupies the left half-space, $\mathcal T^\vee$ occupies the right half-space, and $\mathcal I$ is the interface at $x^\perp=0$.  We refer to this as the \emph{unfolded} description.  Folding the right-hand theory across the interface maps it to the left half-space with reversed orientation and gives the equivalent boundary problem
\[
  \left.
  \mathcal T[\boldsymbol A]\otimes
  \bigl(\mathcal T^\vee[\boldsymbol B]\bigr)^{\mathrm{rev}}
  \right|_{\mathcal B_{\mathcal I}},
  \qquad x^\perp\leq0.
\]
Here $\mathcal B_{\mathcal I}$ denotes the boundary condition together with the boundary degrees of freedom and interactions obtained by folding the interface $\mathcal I$.  We use this strict folded description for the geometry of the boundary problem and for anomaly inflow.  The Chern--Simons/BF gauge-data calculation below will instead use an intermediate master action generated by gauged particle--vortex duality; this distinction will be important when we discuss Wilson line transmission.

Consider first one tetrahedron theory $\mathcal T_\Delta$, namely a charge-one 3d chiral multiplet $\Phi$ with its canonical parity-anomaly contact term.  Its particle--vortex dual is a $U(1)$ gauge theory with one charge-one chiral multiplet $\Phi^\vee$ and the corresponding half-integral Chern--Simons contact term~\cite{Dimofte:2011py}.  Dimofte--Gaiotto--Paquette construct the elementary duality interface by factorizing the identity interface and dualizing one side~\cite{Dimofte:2017tpi}.  Choosing Neumann boundary conditions for the chiral multiplets on both sides, the interface contains a 2d $\mathcal N=(0,2)$ Fermi multiplet $\Gamma$ and the cubic coupling
\begin{equation}
  S_{\mathrm{int}}^{\Delta}
  \supset
  \int_{x^\perp=0} d^2x\,d\theta^+\,
  \Gamma\,\Phi\,\Phi^\vee
  +\text{c.c.}
  \label{eq:elementary-PV-wall-coupling}
\end{equation}
Here $x^\perp$ denotes the coordinate normal to the interface, which is located at $x^\perp=0$, while $x$ denotes the two coordinates along the interface.  The boundary charges of $\Gamma$ are fixed by gauge invariance and anomaly cancellation.  The Fermi multiplet is therefore not an optional decoration: it supplies the boundary anomaly required by the mismatch of the bulk Chern--Simons contact terms and mediates the operator map across the wall~\cite{Dimofte:2017tpi}.  These elementary interface Fermis should be distinguished from the $E_8$ chiral gauged-WZW sector of Section~\ref{subsec:mirror-boundary-half-index}: the latter appears only after the eight flavor symmetries are gauged with $C_{E_8}$ and cancels the residual rank-eight Neumann-boundary gauge anomaly.

For the present theory we take eight copies of this elementary interface.  Before gauging, let $A_i$ be background fields for the eight flavor symmetries of the left tetrahedron theories and let $B_i$ be the dynamical gauge fields in the eight particle--vortex descriptions on the right.  The product wall contains eight Fermi multiplets $\Gamma_i$ with
\begin{equation}
  S_{\mathrm{int}}^{(8)}
  \supset
  \sum_{i=1}^{8}
  \int_{x^\perp=0} d^2x\,d\theta^+\,
  \Gamma_i\Phi_i\Phi_i^\vee
  +\text{c.c.}
  \label{eq:eightfold-PV-wall-coupling}
\end{equation}
and implements the nodewise exchange of the original flavor current with the topological current of $B_i$.  We now gauge the eight left flavor symmetries with the same integral Chern--Simons matrix $C_{E_8}$ that defines $\mathcal T$.  Equivalently, we promote the $A_i$ to dynamical gauge fields while retaining the BF couplings generated by particle--vortex duality.  Gauging the flavor symmetry of a duality interface is the interface version of the operation used in Section~\ref{subsec:dual-Lagrangian-PV} to derive the dual bulk Lagrangian.

We denote the resulting particle--vortex wall provisionally by $\mathcal I$.  At this point it is important to distinguish the strict folded boundary problem from the intermediate gauge-field description used to derive the dual theory.  Strict folding gives
\[
  \left.
  \mathcal T[\boldsymbol A]\otimes
  \bigl(\mathcal T^\vee[\boldsymbol B]\bigr)^{\mathrm{rev}}
  \right|_{\mathcal B_{\mathcal I}},
\]
with the information about the interface encoded in the boundary condition, boundary degrees of freedom, and boundary interactions.  Folding by itself does not add a bulk BF interaction between the two factors.

To determine the gauge data produced by the gauged particle--vortex transformation, however, it is convenient to use the same intermediate Chern--Simons/BF master action as in Section~\ref{subsec:dual-Lagrangian-PV}.  Before $\boldsymbol A$ is integrated out, the topological quadratic part of this master action has matrix
\begin{equation}
  K_{\mathrm{master}}
  =
  \begin{pmatrix}
    C_{E_8} & \epsilon I_8\\
    \epsilon I_8 & \frac12 I_8
  \end{pmatrix},
  \qquad
  \epsilon=\pm1,
  \label{eq:interface-folded-K}
\end{equation}
where the sign $\epsilon$ depends only on the convention for the BF coupling.  In differential-form notation,
\begin{equation}
  \mathcal L_{\mathrm{master}}^{\mathrm{top}}
  =\frac{1}{4\pi}
  \left[
  \boldsymbol A^{\mathsf T}C_{E_8}\,d\boldsymbol A
  +2\epsilon\,\boldsymbol A^{\mathsf T}d\boldsymbol B
  +\frac12\boldsymbol B^{\mathsf T}d\boldsymbol B
  \right].
  \label{eq:interface-folded-action}
\end{equation}
The BF term in \eqref{eq:interface-folded-action} is therefore not a cross-bulk interaction inserted by hand after folding.  It is generated by particle--vortex duality: the background flavor field $\boldsymbol A$ couples to the topological current of $\boldsymbol B$, and after gauging the flavor symmetry $\boldsymbol A$ becomes dynamical.  The master action is an intermediate description of this gauging operation; integrating out one set of gauge fields produces one of the two bulk descriptions that are joined by the physical wall.

The $\boldsymbol A$ fields appear quadratically and may be integrated out exactly.  Their equation of motion in the master action is
\begin{equation}
  C_{E_8}\,d\boldsymbol A
  +\epsilon\,d\boldsymbol B=0,
  \qquad
  d\boldsymbol A
  =-\epsilon C_{E_8}^{-1}d\boldsymbol B.
  \label{eq:interface-A-eom}
\end{equation}
Hence the effective $\boldsymbol B$-sector matrix obtained by the Schur complement of the $C_{E_8}$ block is
\begin{equation}
  K_{\mathrm{eff}}^{(B)}
  =\frac12 I_8-C_{E_8}^{-1}.
  \label{eq:interface-Schur-complement}
\end{equation}
On the other hand, the physical, unfolded dual theory has the ultraviolet gauge/contact matrix
\begin{equation}
  K_{\mathrm{UV}}^\vee
  =C_{E_8}^{-1}-\frac12 I_8.
  \label{eq:interface-dual-UV-K}
\end{equation}
Therefore
\begin{equation}
  K_{\mathrm{eff}}^{(B)}=-K_{\mathrm{UV}}^\vee.
  \label{eq:interface-folding-sign}
\end{equation}
The minus sign is precisely what is required when the resulting dual bulk theory is placed on the folded side, where its orientation is reversed.  Thus the master-action derivation and the strict folded interface description are consistent and together explain the two characteristic features of the dual description,
\begin{equation}
  C_{E_8}\longrightarrow C_{E_8}^{-1},
  \qquad
  \text{right-hand theory}\longrightarrow\text{orientation reversed after folding}.
  \label{eq:interface-two-features}
\end{equation}
Because $C_{E_8}$ is unimodular, the Schur complement is again integral after the canonical chiral contact term is removed.  No fractional gauge Chern--Simons level is introduced by the Gaussian elimination.

The anomaly interpretation is also important.  Each elementary particle--vortex interface is anomaly matched only after the boundary Fermi multiplet is included~\cite{Dimofte:2017tpi}.  Tensoring the eight interfaces preserves this property.  After the flavor symmetries are gauged, the corresponding boundary anomalies become gauge anomalies, and their cancellation is tied to the inflow from the two bulk Chern--Simons systems.  Equation~\eqref{eq:interface-folding-sign} is the quadratic gauge-sector check that the inflow on the two sides has exactly the relative sign required by the folded interface.  The ultraviolet anomaly analysis determines the boundary gauge-anomaly cancellation.  The relative $E_8$ framing phase is instead fixed independently by comparing the $T$-data of the $A$- and $B$-twisted theories in Section~\ref{sec:AB-twisted-TQFTs}.

At this stage $\mathcal I$ should be understood as an explicit ultraviolet wall for the undeformed particle--vortex pair, together with an extension to the monopole-deformed theories at the level of the BPS operator map.  The designation ``$E_8$ interface'' will be justified later from the relative $A/B$-twisted TQFT and framing data, rather than assumed in the microscopic construction.  Section~\ref{subsec:dual-Lagrangian-PV} showed that the seven dressed monopoles of $\mathcal T$ map charge by charge to the seven bare monopoles of $\mathcal T^\vee$, so the same wall is compatible with the deformation at the level of the BPS operator map and the residual topological symmetry.  We do not construct a complete set of 2d $E$- and $J$-couplings satisfying the 3d matrix factorization condition $\sum_a E_aJ_a=W_{\rm L}-W_{\rm R}$ for the full monopole superpotentials, each of which contains seven monopole terms.\footnote{This is the 3d boundary/interface analogue of the matrix-factorization condition familiar from B-type boundaries in 2d $\mathcal N=(2,2)$ Landau--Ginzburg models.}  The arguments below will therefore use only the parts of the wall action and operator map fixed by particle--vortex duality, gauging, the BPS operator map, and the exact $A/B$-twisted data.

\subsection{Wilson line transmission across the wall}
\label{subsec:interface-Wilson-transmission}

We now study a Wilson line that crosses the physical interface.  In the unfolded picture the relevant geometry is
\[
  W_{\boldsymbol Q}[\boldsymbol A]
  \ \longrightarrow\ 
  \Big|_{\mathcal I}
  \ \longrightarrow\ 
  W_{\boldsymbol Q^\vee}^{\vee}[\boldsymbol B],
\]
where the line meets the interface transversely.  This should be distinguished from the Wilson loops used in the half-index: those wrap the core $S^1$ of $D^2\times S^1$ and run parallel to the boundary torus.  Here the line-defect problem is instead the transmission of a bulk line across an interface.

After folding the right-hand side, the same transmitted line is represented by two bulk line segments ending at the same point of the folded boundary,
\[
  W_{\boldsymbol Q}[\boldsymbol A]
  \ \longrightarrow\ 
  \Big|_{\mathcal B_{\mathcal I}}
  \ \longleftarrow\ 
  W_{\boldsymbol Q^\vee}^{\vee}[\boldsymbol B].
\]
The boundary degrees of freedom supply the junction data.  To determine the relation between the two charges semiclassically, we use the intermediate master CS/BF action of \eqref{eq:interface-folded-action}, rather than treating its BF term as an interaction of the strict folded product theory.  For source-level charge bookkeeping, let $L$ denote the incoming $\mathcal T$-side Wilson line segment.  Its contribution to the master action is
\begin{equation}
  S_W=\int_L \boldsymbol Q^{\mathsf T}\boldsymbol A .
  \label{eq:interface-Wilson-source}
\end{equation}
The full interface junction is understood; \eqref{eq:interface-Wilson-source} isolates only the $\boldsymbol A$-source needed to determine the transmitted charge.  Varying the master action \eqref{eq:interface-folded-action} with respect to $\boldsymbol A$ shifts its equation of motion to
\begin{equation}
  C_{E_8}\,d\boldsymbol A
  +\epsilon\,d\boldsymbol B
  +2\pi\boldsymbol Q\,\delta(L)=0.
  \label{eq:interface-Wilson-source-eom}
\end{equation}
Here $\delta(L)$ is the delta-function two-form Poincar\'e dual to $L$, normalized by $\int_{M_3}\omega\wedge\delta(L)=\int_L\omega$ for any one-form $\omega$.  Solving \eqref{eq:interface-Wilson-source-eom} for the field strength of $\boldsymbol A$ gives
\[
  d\boldsymbol A
  =-\epsilon C_{E_8}^{-1}d\boldsymbol B
  -2\pi C_{E_8}^{-1}\boldsymbol Q\,\delta(L).
\]
Equivalently, after integrating out $\boldsymbol A$ in the master description, the effective $\boldsymbol B$ action contains a line source proportional to
\[
  -\epsilon\int_L
  \bigl(C_{E_8}^{-1}\boldsymbol Q\bigr)^{\mathsf T}\boldsymbol B .
\]
Hence the master CS/BF calculation naturally produces the inverse-Cartan charge vector $C_{E_8}^{-1}\boldsymbol Q$ in the $\boldsymbol B$ sector.  Interpreted in the folded boundary problem, this predicts the pairing
\[
  W_{\boldsymbol Q}[\boldsymbol A]
  \ \underset{\mathcal B_{\mathcal I}}{\longleftrightarrow}\ 
  W^\vee_{\,\pm C_{E_8}^{-1}\boldsymbol Q}[\boldsymbol B],
\]
where the sign depends on the BF and unfolding conventions.  After unfolding, this boundary pairing becomes the transmission of a Wilson line through $\mathcal I$.  We therefore use the semiclassical master-action calculation only to determine the inverse-Cartan transformation of the charge, not its physical sign after unfolding.

The semiclassical wall analysis thus predicts the inverse-Cartan transformation of the Wilson charge.  The Bethe complement map found in Section~\ref{sec:AB-twisted-TQFTs} realizes precisely the same transformation on the $A/B$-twisted Wilson loop algebra.  We therefore interpret the Bethe-algebra map as the infrared transmission rule induced by the particle--vortex wall.  We now make this statement precise and fix the sign using the exact Bethe equations.  For the $C_{E_8}$ theory they are
\begin{equation}
  1-x_i
  =\prod_{j=1}^{8}x_j^{(C_{E_8})_{ij}},
  \qquad i=1,\ldots,8.
  \label{eq:interface-left-Bethe-equations}
\end{equation}
Define
\begin{equation}
  y_i:=1-x_i.
  \label{eq:interface-complement-coordinate}
\end{equation}
Then \eqref{eq:interface-left-Bethe-equations} is the monomial relation
\begin{equation}
  \boldsymbol y=\boldsymbol x^{C_{E_8}}.
  \label{eq:interface-complement-monomial}
\end{equation}
Because $C_{E_8}$ is unimodular, $C_{E_8}^{-1}$ is integral and the inverse monomial map is single valued on the localized Bethe torus,
\begin{equation}
  \boldsymbol x=\boldsymbol y^{C_{E_8}^{-1}}.
  \label{eq:interface-inverse-monomial}
\end{equation}
It follows immediately that
\begin{equation}
  1-y_i=x_i
  =\prod_{j=1}^{8}y_j^{(C_{E_8}^{-1})_{ij}},
  \label{eq:interface-right-Bethe-equations}
\end{equation}
which is precisely the Bethe system of the $C_{E_8}^{-1}$ theory.  Thus the complement map is the exact Bethe-vacuum map associated with the wall.

We use the same Wilson line convention as in Sections~2.5 and~\ref{subsec:dual-Wilson-Zagier},
\begin{equation}
  W_{\boldsymbol Q}(\boldsymbol x)
  :=\prod_{i=1}^{8}x_i^{-Q_i}.
  \label{eq:interface-Wilson-convention}
\end{equation}
Using \eqref{eq:interface-inverse-monomial}, one obtains in the Bethe algebra
\begin{align}
  W_{\boldsymbol Q}(\boldsymbol x)
  &=\prod_{j=1}^{8}
  y_j^{-(C_{E_8}^{-\mathsf T}\boldsymbol Q)_j}
  \nonumber\\
  &=W_{\boldsymbol Q^\vee}^{\vee}(\boldsymbol y),
  \qquad
  \boldsymbol Q^\vee=C_{E_8}^{-\mathsf T}\boldsymbol Q
  =C_{E_8}^{-1}\boldsymbol Q.
  \label{eq:interface-exact-Wilson-transmission}
\end{align}
The last equality uses the symmetry of the $E_8$ Cartan matrix.  Equation~\eqref{eq:interface-exact-Wilson-transmission} is stronger than equality of a few Wilson eigenvalues: it is an identity of Wilson classes under the exact isomorphism of the two Bethe algebras.  It also removes the sign ambiguity of the source-level discussion around \eqref{eq:interface-Wilson-source-eom}.

We now unfold the right-hand factor.  The folded pairing of an $\boldsymbol A$-line with a $\boldsymbol B$-line becomes the statement that a Wilson line crossing the physical interface from $\mathcal T$ to $\mathcal T^\vee$ emerges as the corresponding dual Wilson line.  Thus, in the unfolded picture,

\begin{equation}
  \mathcal I:\qquad
  W_{\boldsymbol Q}
  \longmapsto
  W_{C_{E_8}^{-1}\boldsymbol Q}^{\vee}.
  \label{eq:interface-physical-Wilson-map}
\end{equation}

For the five distinguished Wilson lines, this gives the same sector pairing that was encountered arithmetically in the componentwise Nahm transformation.  In terms of the conformal-weight labels,
\begin{equation}
\begin{aligned}
  0&\longmapsto0,\\
  \frac5{11}&\longmapsto\frac6{11},\\
  \frac8{11}&\longmapsto\frac3{11},\\
  \frac{10}{11}&\longmapsto\frac1{11},\\
  \frac{12}{11}&\longmapsto\frac{10}{11}.
\end{aligned}
  \label{eq:interface-five-sector-map}
\end{equation}
The explicit integral charge representatives on the two sides were listed in Section~\ref{subsec:dual-Wilson-Zagier}; each pair satisfies
\begin{equation}
  \boldsymbol Q_{h^\vee}^{\vee}
  =C_{E_8}^{-1}\boldsymbol Q_h.
  \label{eq:interface-five-charge-map}
\end{equation}
Hence the wall sends each of the five distinguished Wilson loop sectors to one and only one transmitted partner.

This also explains the linear part of the Nahm transformation directly in line-defect language.  Since the Wilson loop half-index contains the linear Nahm datum $\boldsymbol B=-\boldsymbol Q$, \eqref{eq:interface-physical-Wilson-map} gives
\begin{equation}
  \boldsymbol B
  \longmapsto
  \boldsymbol B^\vee
  =C_{E_8}^{-1}\boldsymbol B,
  \label{eq:interface-Nahm-linear-map}
\end{equation}
while Section~\ref{subsec:gauged-pv-interface} showed that the same wall produces $C_{E_8}\mapsto C_{E_8}^{-1}$ in the quadratic bulk data.  Thus the two nontrivial matrix operations in the Zagier transform have a common physical origin: they are, respectively, the bulk and line-defect parts of the gauged particle--vortex duality wall.

There is no claim here that the wall by itself maps the half-index of the $(\mathcal D,D_c)$ boundary condition of the original theory to the independently chosen $(\mathcal D,D_c)$ half-index of the dual theory.  As emphasized in Section~\ref{subsec:dual-Wilson-Zagier}, the nonnegative summation cone and the chiral Fock factors also depend on the boundary condition.  The statement established in this subsection is instead the exact bulk/line transmission law.  This is the appropriate input for the folded-interface construction below: after folding, each transmitted Wilson line is paired with its dual image and can be treated as a composite line ending on the interface.

\subsection{The $T^2$ interface operator}
\label{subsec:T2-interface-operator}

We now remain in the unfolded interface picture and regard $\mathcal I$ as a linear map from the torus state space of the $A$-twisted theory $\mathcal T$ to that of the $B$-twisted theory $\mathcal T^\vee$.  The Wilson line transmission law can then be sharpened into an explicit operator acting on these two torus state spaces.  This is useful for two reasons.  First, it packages the Bethe-vacuum map and the action on Wilson, handle-gluing, and fibering operators into a single object.  Second, it makes precise why the same sector permutation that pairs the five Wilson loop sectors will later appear in the torus interface amplitude; this permutation is also the one encountered in the componentwise Nahm relation.  The only subtlety is that the interface looks different in the Bethe-idempotent basis and in the Wilson loop/character basis.

Let
\begin{equation}
  \mathscr H_A(T^2)
  =\bigoplus_{\beta=0}^{4}\mathbb C\,|e_\beta\rangle_A,
  \qquad
  \mathscr H_B^{\vee}(T^2)
  =\bigoplus_{\beta=0}^{4}\mathbb C\,|e_\beta^\vee\rangle_{B}.
  \label{eq:interface-T2-spaces}
\end{equation}
Here $e_\beta$ is the primitive idempotent of the complexified Bethe algebra associated in \eqref{eq:CE8-Bethe-idempotents} with the $\beta$-th regular Bethe vacuum $\boldsymbol x_\beta$ of the $A$-twisted $C_{E_8}$ theory, and $e_\beta^\vee$ denotes the corresponding primitive idempotent on the dual side.  We pair the dual vacua with the original ones by the exact particle--vortex complement map
\begin{equation}
  \boldsymbol y_\beta
  :=\boldsymbol1-\boldsymbol x_\beta.
  \label{eq:interface-paired-Bethe-order}
\end{equation}
Thus the label $\beta$ refers to a paired Bethe vacuum $(\boldsymbol x_\beta,\boldsymbol y_\beta)$ on the two sides of the wall.  The interface operator is defined on the corresponding idempotent states by
\begin{equation}
  D_{\mathrm{Bethe}}\,|e_\beta\rangle_A
  =|e_\beta^\vee\rangle_{B},
  \qquad \beta=0,\ldots,4,
  \qquad
  [D_{\mathrm{Bethe}}]_{\mathrm{paired}}=I_5.
  \label{eq:interface-DBethe}
\end{equation}
Here $[D_{\mathrm{Bethe}}]_{\mathrm{paired}}$ denotes the matrix of the interface operator in the paired Bethe-idempotent bases in \eqref{eq:interface-T2-spaces}, and $I_5$ is the $5\times5$ identity matrix.  In other words, after the vacua have been paired by \eqref{eq:interface-paired-Bethe-order}, the wall sends each primitive-idempotent state to its unique dual partner.  There is therefore no additional permutation in this basis.

The usefulness of the Bethe-idempotent basis is that every Wilson loop operator acts diagonally.  Indeed, multiplication by a Bethe-algebra element on a primitive idempotent is given by evaluation at the corresponding Bethe vacuum.  Hence
\begin{equation}
\begin{aligned}
  \widehat W_{\boldsymbol Q}^{A}|e_\beta\rangle_A
  &=W_{\boldsymbol Q}(\boldsymbol x_\beta)|e_\beta\rangle_A,\\
  \widehat W_{\boldsymbol Q^\vee}^{B}|e_\beta^\vee\rangle_{B}
  &=W_{\boldsymbol Q^\vee}^{\vee}(\boldsymbol y_\beta)|e_\beta^\vee\rangle_{B}.
\end{aligned}
  \label{eq:interface-Wilson-matrices-Bethe}
\end{equation}
Section~\ref{subsec:interface-Wilson-transmission} showed that the paired eigenvalues obey
$W_{\boldsymbol Q}(\boldsymbol x_\beta)=W_{C_{E_8}^{-1}\boldsymbol Q}^{\vee}(\boldsymbol y_\beta)$ for every $\beta$.  Acting on each basis state therefore gives the operator identity
\begin{equation}
  D_{\mathrm{Bethe}}\,
  \widehat W_{\boldsymbol Q}^{A}
  =
  \widehat W_{C_{E_8}^{-1}\boldsymbol Q}^{B}\,
  D_{\mathrm{Bethe}}.
  \label{eq:interface-Wilson-intertwiner}
\end{equation}
This equation simply says that applying a Wilson loop and then crossing the wall gives the same result as first crossing the wall and then applying the transmitted dual Wilson loop.
Likewise, if $\widehat{\mathcal H}_A$ and $\widehat{\mathcal H}_B^\vee$ denote the diagonal handle-gluing operators, the exact relation \eqref{eq:handle-complement-full} gives
\begin{equation}
  D_{\mathrm{Bethe}}\,
  \widehat{\mathcal H}_A
  =
  \widehat{\mathcal H}_B^\vee\,
  D_{\mathrm{Bethe}}.
  \label{eq:interface-handle-intertwiner}
\end{equation}
The fibering operator is the orientation-sensitive part.  Equation~\eqref{eq:E8-fibering-complementarity} gives
\begin{equation}
  \widehat{\mathcal F}_B^\vee\,
  D_{\mathrm{Bethe}}
  =
  \kappa_{E_8}\,
  D_{\mathrm{Bethe}}\,
  \widehat{\mathcal F}_A^{-1},
  \qquad
  \kappa_{E_8}=e^{-2\pi \mathrm{i}/3}.
  \label{eq:interface-fibering-intertwiner}
\end{equation}
Equations~\eqref{eq:interface-Wilson-intertwiner}--\eqref{eq:interface-fibering-intertwiner} show directly on $T^2$ that the wall intertwines the Wilson loop and handle-gluing data, reverses the fibering operation, and leaves only the universal $E_8$ framing multiplier.

The same interface takes a less trivial form in the standard Wilson loop basis.  On the $C_{E_8}$ side we use the order
\begin{equation}
  \left(
  0,\frac5{11},\frac8{11},\frac{10}{11},\frac{12}{11}
  \right),
  \label{eq:interface-original-standard-order}
\end{equation}
and on the $C_{E_8}^{-1}$ side the standard order
\begin{equation}
  \left(
  0,\frac1{11},\frac3{11},\frac6{11},\frac{10}{11}
  \right).
  \label{eq:interface-dual-standard-order}
\end{equation}
The physical sector map \eqref{eq:interface-five-sector-map} is therefore represented by
\begin{equation}
  D_{\mathrm{line}}=P_\sigma,
  \qquad
  P_\sigma=
  \begin{pmatrix}
  1&0&0&0&0\\
  0&0&0&1&0\\
  0&0&1&0&0\\
  0&1&0&0&0\\
  0&0&0&0&1
  \end{pmatrix},
  \qquad
  P_\sigma^2=I_5.
  \label{eq:interface-line-permutation}
\end{equation}
In other words, $D_{\mathrm{Bethe}}=I_5$ and $D_{\mathrm{line}}=P_\sigma$ are not two different interfaces.  They are the same wall written in two natural bases: the former is adapted to the exact complement map of Bethe vacua, while the latter is adapted to the standard labeling of the five Wilson loop/character sectors.

This basis change also gives a compact mapping-class-group description.  Let $S_A,T_A$ be the modular matrices of the original $A$-twisted theory in the order \eqref{eq:interface-original-standard-order}, and let $S_B^{\vee,\mathrm{std}},T_B^{\vee,\mathrm{std}}$ be those of the dual $B$-twisted theory in the order \eqref{eq:interface-dual-standard-order}.  Reordering the dual basis by $P_\sigma$ gives precisely the paired order used in Section~\ref{subsubsec:B-twisted-CE8inv-T}.  Hence the exact relations \eqref{eq:CE8inv-S-exact} and \eqref{eq:T-matrix-complementarity} are equivalent to
\begin{equation}
  P_\sigma S_A
  =S_B^{\vee,\mathrm{std}}P_\sigma,
  \label{eq:interface-S-intertwiner}
\end{equation}
\begin{equation}
  T_B^{\vee,\mathrm{std}}P_\sigma
  =\kappa_{E_8}\,
  P_\sigma T_A^{-1}.
  \label{eq:interface-T-intertwiner}
\end{equation}
The first equation says that the wall intertwines the modular $S$ action exactly.  The second says that it intertwines $T$ only after reversing the orientation of the original theory, with the single residual multiplier $\kappa_{E_8}$.  These are the genus-one counterparts of the Wilson, handle-gluing, and fibering relations above.

It is useful to stress the logical status of this construction.  The operator $D_{\mathrm{Bethe}}$ is not postulated from the modular matrices.  The exact complement correspondence $\boldsymbol y=\boldsymbol1-\boldsymbol x$, derived from the Bethe equations in Section~\ref{sec:AB-twisted-TQFTs}, realizes the particle--vortex map in the twisted sector and fixes $D_{\mathrm{Bethe}}$ in the paired idempotent basis.  This correspondence is consistent with the independently derived bulk inverse-Cartan map and Wilson loop transmission rule.  The localization calculation then proves that $D_{\mathrm{Bethe}}$ intertwines the complete Bethe--Seifert data.  In the Wilson loop basis the resulting finite-dimensional operator is the sector-permutation matrix $P_\sigma$, which is the same permutation that appears in the componentwise Nahm relation.  In the next subsection this same matrix will determine which left and right boundary-character sectors are paired in the folded torus interface amplitude.

\subsection{Folded transmitted lines and the relative $E_8$ framing phase}
\label{subsec:folded-lines-E8-phase}

We now fold the interface once more.  Thus the geometric picture is again the boundary problem
\[
  \left.
  \mathfrak T_A(\mathcal T)\otimes
  \mathfrak T_B(\mathcal T^\vee)^{\mathrm{rev}}
  \right|_{\mathcal B_{\mathcal I}},
\]
where $\mathfrak T_A(\mathcal T)$ and $\mathfrak T_B(\mathcal T^\vee)$ denote the $A$- and $B$-twisted TQFTs defined in Section~\ref{sec:AB-twisted-TQFTs}.  Geometric folding reverses the orientation of the right-hand bulk.  Separately, the comparison of the displayed $A$- and $B$-twisted modular data already contains the inverse $T$ action in \eqref{eq:interface-T-intertwiner}.  It is therefore important not to insert a second orientation reversal when the listed $A/B$ conformal weights are used to test the transmitted pairs.  In the spin bookkeeping below, the twist of a transmitted pair is the product of the two displayed twists.  This is simply the folded form of the orientation-reversing wall relation and does not assume a choice of ribbon-category completion.

In the folded description, a Wilson line transmitted through the unfolded wall becomes a pair of bulk line segments that meet and can end on the boundary $\mathcal B_{\mathcal I}$.  The physical input is the transmission law already established in Section~\ref{subsec:interface-Wilson-transmission}: each of the five distinguished Wilson loop sectors is mapped to a unique transmitted partner.

Let $W_\alpha$ denote the five distinguished Wilson loop sectors of the $C_{E_8}$ theory in the order
\begin{equation}
  \left(
  0,\frac5{11},\frac8{11},\frac{10}{11},\frac{12}{11}
  \right),
\end{equation}
and let $W^\vee_{\sigma(\alpha)}$ be their transmitted partners in the $C_{E_8}^{-1}$ theory,
\begin{equation}
  \left(
  0,\frac6{11},\frac3{11},\frac1{11},\frac{10}{11}
  \right).
\end{equation}
We denote by $\mathcal L_\alpha$ the folded interface-ending composite formed from $W_\alpha$ and $W^\vee_{\sigma(\alpha)}$.  The fact that $\mathcal L_\alpha$ can end on the folded wall is simply the folded form of the transmission rule; it is not an independent condensation ansatz.

A first nontrivial consistency check is the topological spin.\footnote{Here the topological spin $\theta_\alpha$ is the phase acquired by a topological line under a $2\pi$ framing twist. For a sector of conformal weight $h_\alpha$, $\theta_\alpha=e^{2\pi \mathrm{i}h_\alpha}$.}  The paired conformal weights found in Section~\ref{sec:AB-twisted-TQFTs} obey
\begin{equation}
  h_\alpha+h^\vee_{\sigma(\alpha)}
  =(0,1,1,1,2).
  \label{eq:folded-paired-weight-sums}
\end{equation}
Therefore every transmitted composite has trivial total twist,
\begin{equation}
  \theta(\mathcal L_\alpha)
  =\exp\!\left[2\pi \mathrm{i}\left(
  h_\alpha+h^\vee_{\sigma(\alpha)}
  \right)\right]
  =1.
  \label{eq:folded-trivial-spins}
\end{equation}
Thus all five interface-ending composites are bosonic.  A useful two-sector precedent is the conformal embedding
\begin{equation}
  (E_7)_1\times SU(2)_1\subset(E_8)_1,
  \label{eq:E7-SU2-E8-analogy}
\end{equation}
which is a standard source of exceptional Chern--Simons dualities~\cite{CordovaHsinOhmori:2019Exceptional}.  In the 3d TQFT picture, the nontrivial sectors of $(E_7)_1$ and $SU(2)_1$ have conformal weights $3/4$ and $1/4$, so their product has integral spin.  At the character level, the corresponding branching of the $(E_8)_1$ vacuum module gives the two-sector bilinear relation
\begin{equation}
  \chi_{(E_8)_1}
  =
  \chi^{(E_7)_1}_{\mathbf 1}\,\chi^{SU(2)_1}_{\mathbf 1}
  +
  \chi^{(E_7)_1}_{\mathbf{56}}\,\chi^{SU(2)_1}_{\mathbf 2}.
  \label{eq:E7-SU2-E8-character-bilinear}
\end{equation}
Thus this familiar example exhibits simultaneously an integral-spin pairing of two line sectors and a bilinear decomposition of the holomorphic $(E_8)_1$ vacuum character.  It should be viewed as a precedent rather than as an input to our construction.  In the present problem the five pairings are first fixed by the explicit 3d duality wall, while the relative $E_8$ framing response is independently determined by the supersymmetric fibering operator.

The transmission also preserves the Wilson loop fusion algebra.  Let $\widehat L_a$ denote the based Wilson loop representatives introduced in Section~\ref{sec:AB-twisted-TQFTs}.  Since the interface intertwines every Wilson operator, \eqref{eq:interface-Wilson-intertwiner} implies
\begin{equation}
  D_{\mathrm{line}}\,
  \widehat L_a\widehat L_b
  =
  \widehat L_a^\vee\widehat L_b^\vee\,
  D_{\mathrm{line}}.
  \label{eq:folded-fusion-intertwining}
\end{equation}
Consequently the multiplication coefficients of the transmitted Wilson loop sectors are identical on the two sides.  In the based convention of \eqref{eq:SO39-based-ring-identification}, both are the even-label $SO(3)_9$ fusion ring.  Thus the wall transmits not only the five distinguished sectors but also all of their fusion channels.

At the modular level the same statement is accompanied by orientation reversal.  The exact relations
\begin{equation}
  P_\sigma S_A
  =S_B^{\vee,\mathrm{std}}P_\sigma,
  \qquad
  T_B^{\vee,\mathrm{std}}P_\sigma
  =\kappa_{E_8}P_\sigma T_A^{-1}
  \label{eq:folded-modular-intertwining-recalled}
\end{equation}
show that the paired line systems have the same $S$ data and inverse normalized twists.  This is precisely the genus-one signature expected when the right-hand line theory is reversed.  A channel-by-channel identification of the full braiding and associator data requires a choice of ribbon completion and is not derived from the ultraviolet gauge theory here; the present claim is the semisimple statement encoded by the wall and localization data.

It is important to separate two consequences of the folded picture.  First, the wall pairs the complete set of five distinguished Wilson loop sectors.  After folding, these form the natural diagonal set of interface-ending lines, so at the semisimple level there is no unmatched Wilson loop sector.  The trivial total spins are consistent with this interpretation, but they do not by themselves prove the stronger categorical statement that the diagonal condensation leaves no nontrivial local line.  In the standard $SO(3)_9$ ribbon completion summarized in Section~\ref{subsec:full-ribbon-TQFT-completion}, this pairing is realized by the canonical diagonal condensate; the full categorical treatment is deferred to a forthcoming companion paper.  Second, independently of that categorical completion, the relative framed theory retains the universal phase
\begin{equation}
  \kappa_{E_8}
  =e^{-2\pi \mathrm{i}/3}
  =e^{-2\pi \mathrm{i}\,8/24},
  \label{eq:folded-E8-framing-phase}
\end{equation}
corresponding to a relative chiral central charge
\begin{equation}
  c_-=8.
  \label{eq:folded-E8-chiral-central-charge}
\end{equation}
The remaining framing response is therefore precisely that of the bosonic $E_8$ phase.  The two pieces of evidence are logically distinct: the diagonal line transmission controls the semisimple noninvertible sector pairing, whereas the fibering multiplier detects the purely gravitational/framing response.  It is in this sense, and not from the ultraviolet wall construction alone, that we refer to the folded duality wall $\mathcal I$ as the $E_8$ interface.

This distinction will be essential in the next subsection.  The same five transmitted sectors that define the folded interface also determine its torus interface amplitude.  The relative $E_8$ framing phase then fixes the common modular $T$ multiplier of that amplitude, allowing the $(E_8)_1$ character to be derived rather than postulated.

\subsection{The torus interface amplitude and the $(E_8)_1$ character}
\label{subsec:interface-E8-bilinear}

We remain in the folded description.  The folded $E_8$ interface now gives a direct physical origin for the five-term bilinear pairing.  The essential point is that the pairing matrix is not introduced from the 2d character identity.  It has already been fixed by the 3d wall: in the standard Wilson loop basis the interface operator is the permutation matrix $P_\sigma$ of \eqref{eq:interface-line-permutation}.  Thus each of the five distinguished Wilson loop sectors is transmitted to one dual sector with multiplicity one, and no additional sector is allowed by the wall operator.

Let
\begin{equation}
  \boldsymbol\chi_A(\tau)
  :=
  \begin{pmatrix}
  \chi_0\\
  \chi_{5/11}\\
  \chi_{8/11}\\
  \chi_{10/11}\\
  \chi_{12/11}
  \end{pmatrix},
  \qquad
  \boldsymbol\chi_B^{\vee,\mathrm{std}}(\tau)
  :=
  \begin{pmatrix}
  \chi_0^\vee\\
  \chi_{1/11}^\vee\\
  \chi_{3/11}^\vee\\
  \chi_{6/11}^\vee\\
  \chi_{10/11}^\vee
  \end{pmatrix}.
  \label{eq:interface-character-vectors}
\end{equation}
The claim hierarchy is important here.  The vacuum half-index identities are exact, whereas the four non-vacuum Wilson loop half-index/character identifications in Sections~\ref{subsec:Wilson-half-indices} and \ref{subsec:dual-Wilson-Zagier} were verified through the displayed orders in $q$ and will not be promoted here to all-order half-index identities.  Independently, the two complete five-component character systems are exact modular objects, and the exact wall operator $P_\sigma$ selects their unique multiplicity-one sector pairing.  We therefore use the exact character systems and the wall-determined pairing below; the non-vacuum localization matches provide direct checks of the identification of these character sectors with the corresponding Wilson loop solid-torus states.

It is useful to make the physical meaning of the resulting bilinear more explicit.  The Wilson loops used in the half-index calculation are closed loops wrapping the core $S^1$ of a solid torus.  Such a core insertion prepares a torus state $|W_\alpha\rangle_A$ on the $A$-twisted side, and similarly $|W_\beta^\vee\rangle_B$ on the $B$-twisted dual side.  By contrast, in the unfolded interface problem the same topological line sector can be represented by a line that runs transversely into the interface.  After folding, the transmitted pair is represented by two bulk line segments that meet the same boundary $\mathcal B_{\mathcal I}$.  Thus the closed core Wilson loops and the interface-ending lines are not the same geometric configuration; they are two realizations of the same topological line sectors.

Correspondingly, one may regard the folded interface as defining a linear functional $\langle\mathcal B_{\mathcal I}|$ on the product of the two torus state spaces.  In the Wilson loop basis its matrix elements are fixed by the transmission rule,
\[
  \left\langle\mathcal B_{\mathcal I}\middle|
  \left(
  |W_\alpha\rangle_A\otimes|W_\beta^\vee\rangle_B
  \right)\right\rangle
  =(P_\sigma)_{\alpha\beta}.
\]
Thus only the transmitted pair $\beta=\sigma(\alpha)$ contributes, with multiplicity one.  The characters $\chi_\alpha(\tau)$ and $\chi_\beta^\vee(\tau)$ are the holomorphic solid-torus wavefunctions associated with these Wilson loop states.  To make the contraction explicit, introduce the formal holomorphic torus states
\[
  |\Psi_A(\tau)\rangle
  :=\sum_\alpha \chi_\alpha(\tau)|W_\alpha\rangle_A,
  \qquad
  |\Psi_B^\vee(\tau)\rangle
  :=\sum_\beta \chi_\beta^\vee(\tau)|W_\beta^\vee\rangle_B.
\]
At the level of the exact character systems, the wall-selected torus gluing function is the contraction
\[
  Z_{\mathcal I}(\tau)
  =\left\langle\mathcal B_{\mathcal I}\middle|
  \left(
  |\Psi_A(\tau)\rangle\otimes|\Psi_B^\vee(\tau)\rangle
  \right)\right\rangle.
\]
Here the bracket denotes the bilinear TQFT gluing pairing induced by the folded interface, not a Hermitian inner product.  In particular, no complex conjugation of the character wavefunctions is involved.  Expanding this contraction in the Wilson loop basis gives the torus-state-space realization of the same line transmission rule derived above.  We refer to this exact character-level gluing function as the torus interface amplitude and define
\begin{equation}
  Z_{\mathcal I}(\tau)
  :=
  \boldsymbol\chi_A(\tau)^{\mathsf T}
  P_\sigma
  \boldsymbol\chi_B^{\vee,\mathrm{std}}(\tau).
  \label{eq:interface-genus-one-amplitude-matrix}
\end{equation}
Equivalently,
\begin{align}
  Z_{\mathcal I}(\tau)
  ={}&
  \chi_0\chi_0^\vee
  +\chi_{5/11}\chi_{6/11}^\vee
  +\chi_{8/11}\chi_{3/11}^\vee
  \nonumber\\
  &+\chi_{10/11}\chi_{1/11}^\vee
  +\chi_{12/11}\chi_{10/11}^\vee.
  \label{eq:interface-five-sector-bilinear}
\end{align}
This is the torus character pairing selected by the five transmitted wall sectors.  Its multiplicity-one matrix is the genus-one shadow of the same line transmission that was derived from the particle--vortex wall in Section~\ref{subsec:interface-Wilson-transmission}; it is not imposed as an independent 2d branching rule.

We can now determine the character bilinear \eqref{eq:interface-five-sector-bilinear} exactly from the modular action reconstructed from the 3d twisted data.  This exact modular argument concerns the complete character systems and is logically independent of extending the four non-vacuum half-index matches beyond the orders explicitly checked.  Under $S$, the two character vectors transform by $S_A$ and $S_B^{\vee,\mathrm{std}}$.  Using the symmetry and involutivity of the two $S$ matrices together with the interface relation \eqref{eq:interface-S-intertwiner}, one obtains
\begin{equation}
  S_A^{\mathsf T}P_\sigma S_B^{\vee,\mathrm{std}}
  =P_\sigma.
  \label{eq:interface-bilinear-S-invariance}
\end{equation}
Hence
\begin{equation}
  Z_{\mathcal I}(-1/\tau)
  =Z_{\mathcal I}(\tau).
  \label{eq:interface-Z-S-transformation}
\end{equation}
The $T$ transformation remembers the relative framing anomaly.  Since the $T$ matrices are diagonal, \eqref{eq:interface-T-intertwiner} is equivalent to
\begin{equation}
  T_A^{\mathsf T}P_\sigma T_B^{\vee,\mathrm{std}}
  =\kappa_{E_8}P_\sigma,
  \qquad
  \kappa_{E_8}=e^{-2\pi \mathrm{i}/3}.
  \label{eq:interface-bilinear-T-multiplier}
\end{equation}
Therefore
\begin{equation}
  Z_{\mathcal I}(\tau+1)
  =e^{-2\pi \mathrm{i}/3}Z_{\mathcal I}(\tau).
  \label{eq:interface-Z-T-transformation}
\end{equation}
The same phase was obtained independently from the fibering operators in Section~\ref{subsec:fibering-T-matrix} and was identified in Section~\ref{subsec:folded-lines-E8-phase} with the relative $c_-=8$ framing response.  Thus the modular multiplier of the wall amplitude is fixed by the 3d interface data rather than by an assumed $(E_8)_1$ character formula.

The vacuum normalization now determines the function uniquely.  In the effective character normalization used throughout the paper,
\begin{equation}
  c_A^{\mathrm{eff}}+c_B^{\vee,\mathrm{eff}}
  =\frac{80}{11}+\frac8{11}=8,
  \label{eq:interface-effective-c-sum}
\end{equation}
and the paired conformal grades are those of \eqref{eq:folded-paired-weight-sums}.  A paired term has leading power
\[
  q^{\,h_\alpha+h^\vee_{\sigma(\alpha)}
  -(c_A^{\mathrm{eff}}+c_B^{\vee,\mathrm{eff}})/24}
  =q^{\,h_\alpha+h^\vee_{\sigma(\alpha)}-1/3}.
\]
Since $h_\alpha+h^\vee_{\sigma(\alpha)}=(0,1,1,1,2)$, the five leading exponents are
\[
  -\frac13,\qquad \frac23,\qquad \frac23,\qquad \frac23,\qquad \frac53.
\]
Thus only the vacuum pair contributes at the lowest power, and
\begin{equation}
  Z_{\mathcal I}(\tau)
  =q^{-1/3}\left(1+O(q)\right).
  \label{eq:interface-Z-leading}
\end{equation}
Define
\begin{equation}
  F_{\mathcal I}(\tau)
  :=\eta(\tau)^8 Z_{\mathcal I}(\tau).
  \label{eq:interface-F-definition}
\end{equation}
Using
\begin{equation}
  \eta(\tau+1)^8=e^{2\pi \mathrm{i}/3}\eta(\tau)^8,
  \qquad
  \eta(-1/\tau)^8=\tau^4\eta(\tau)^8,
\end{equation}
Equations~\eqref{eq:interface-Z-S-transformation} and \eqref{eq:interface-Z-T-transformation} imply
\begin{equation}
  F_{\mathcal I}(\tau+1)=F_{\mathcal I}(\tau),
  \qquad
  F_{\mathcal I}(-1/\tau)=\tau^4F_{\mathcal I}(\tau).
  \label{eq:interface-F-modularity}
\end{equation}
Moreover, \eqref{eq:interface-Z-leading} gives
\begin{equation}
  F_{\mathcal I}(\tau)=1+O(q).
  \label{eq:interface-F-normalization}
\end{equation}
Thus $F_{\mathcal I}$ is the normalized holomorphic modular form of weight four for $SL(2,\mathbb Z)$.  Since
\begin{equation}
  M_4\!\left(SL(2,\mathbb Z)\right)=\mathbb C E_4,
\end{equation}
we conclude
\begin{equation}
  F_{\mathcal I}(\tau)=E_4(\tau),
\end{equation}
and hence
\begin{equation}
  Z_{\mathcal I}(\tau)
  =\frac{E_4(\tau)}{\eta(\tau)^8}
  =\chi_{(E_8)_1}(\tau).
  \label{eq:interface-E8-character-identity}
\end{equation}
Equation~\eqref{eq:interface-E8-character-identity} is therefore an exact identity for the wall-selected character bilinear, not merely a numerical coincidence between truncated $q$-series.  Its 3d origin is encoded by four pieces of information: the one-to-one Wilson line transmission across the wall, the reverse modular pairing of the $A$- and $B$-twisted theories, the relative $E_8$ framing multiplier, and the vacuum normalization.

The first nontrivial coefficient makes this physical interpretation particularly transparent.  The vacuum pair contributes
\begin{equation}
  \chi_0\chi_0^\vee
  =q^{-1/3}\left(1+121q+\cdots\right),
\end{equation}
where the two contributions $120$ and $1$ are the level-one coefficients of the original and dual vacuum characters, respectively.  The three transmitted pairs of total conformal grade one contribute respectively $8$, $35$, and $84$.  Therefore the coefficient at the first excited level is
\begin{equation}
  121+8+35+84=248,
  \label{eq:interface-E8-248-decomposition}
\end{equation}
reproducing the dimension of the $E_8$ current algebra directly from the five wall sectors.  As a further check, the next coefficient decomposes as
\begin{equation}
  1782+253+715+1324+50=4124.
  \label{eq:interface-E8-4124-decomposition}
\end{equation}
These decompositions are useful checks of the sector interpretation, but the proof of the exact character-bilinear identity \eqref{eq:interface-E8-character-identity} is the modular argument above and does not rely on a finite-order expansion.  Conversely, this modular proof does not by itself upgrade the four non-vacuum Wilson loop half-index matches to all-order localization identities.

Finally, the scope of the result should be kept clear.  Equation~\eqref{eq:interface-E8-character-identity} identifies the exact character-level torus gluing function selected by the folded interface with the level-one $E_8$ vacuum character.  It does not by itself construct a microscopic 2d interface VOA, nor does it imply that the two five-component character systems form an ordinary mutual-commutant pair inside $V_{E_8,1}$.  The result needed here is more direct: the 3d duality wall selects the five paired sectors, and the exact character bilinear associated with that pairing is uniquely forced to be the $(E_8)_1$ character.  The checked Wilson loop half-indices provide the microscopic supersymmetric realization of these sectors to the orders stated in Sections~\ref{subsec:Wilson-half-indices} and \ref{subsec:dual-Wilson-Zagier}.

\subsection{Compatibility with the monopole deformations}
\label{subsec:interface-monopole-deformations}

The wall constructed above was obtained microscopically from the elementary particle--vortex interfaces before switching on the seven monopole superpotentials.  We now explain why the same wall is compatible with the monopole-deformed theories at the level of the BPS operator map.  The essential input was already derived in Section~\ref{subsec:dual-Lagrangian-PV}: the same lattice map that produces the $C_{E_8}^{-1}$ Chern--Simons matrix also maps each dressed edge monopole of $\mathcal T$ to a bare monopole of $\mathcal T^\vee$.  Here we reinterpret that result as a statement about operators transmitted through the interface.

The seven superpotential operators of the $C_{E_8}$ theory have magnetic charges
\begin{equation}
  \boldsymbol m_{ij}=\boldsymbol e_i+\boldsymbol e_j,
  \qquad
  \langle i,j\rangle\in E(E_8),
  \label{eq:interface-edge-magnetic-charges}
\end{equation}
and dressing vectors
\begin{equation}
  \boldsymbol n^{(ij)}
  =\boldsymbol m_{ij}-C_{E_8}\boldsymbol m_{ij}.
  \label{eq:interface-edge-dressing}
\end{equation}
The magnetic-flux relation across the particle--vortex wall is
\begin{equation}
  \boldsymbol\ell=C_{E_8}\boldsymbol m.
  \label{eq:interface-monopole-flux-map}
\end{equation}
For an edge charge this gives
\begin{equation}
  \boldsymbol\ell_{ij}
  =C_{E_8}\boldsymbol m_{ij}
  =\boldsymbol m_{ij}-\boldsymbol n^{(ij)}.
  \label{eq:interface-edge-flux-split}
\end{equation}
The BPS dressing condition implies that $\boldsymbol m_{ij}$ and $\boldsymbol n^{(ij)}$ are nonnegative and have disjoint support.  Hence
\begin{equation}
  (\boldsymbol\ell_{ij})_+=\boldsymbol m_{ij},
  \qquad
  (\boldsymbol\ell_{ij})_-=\boldsymbol n^{(ij)}.
  \label{eq:interface-edge-positive-negative}
\end{equation}
Thus the chiral fields that dress the monopole on the original side are converted into the negative components of the dual magnetic flux.  This is the monopole counterpart of the electric--magnetic polarization exchange underlying the Wilson line transmission in Section~\ref{subsec:interface-Wilson-transmission}.

Gauge invariance in the $C_{E_8}^{-1}$ theory requires a monopole of charge $\boldsymbol\ell$ to be dressed by
\begin{equation}
  \boldsymbol n^\vee
  =\boldsymbol\ell_+-C_{E_8}^{-1}\boldsymbol\ell.
  \label{eq:interface-dual-dressing}
\end{equation}
Using~\eqref{eq:interface-monopole-flux-map} and \eqref{eq:interface-edge-positive-negative}, one finds for all seven edges
\begin{equation}
  \boldsymbol n^\vee_{ij}
  =(\boldsymbol\ell_{ij})_+
  -C_{E_8}^{-1}\boldsymbol\ell_{ij}
  =\boldsymbol m_{ij}-\boldsymbol m_{ij}=0.
  \label{eq:interface-dual-monopoles-bare}
\end{equation}
Accordingly, the BPS operator transmission takes the form
\begin{equation}
  \mathcal O_{ij}
  \big|_{\mathcal I}
  \simeq
  c_{ij}\,
  V^\vee_{\boldsymbol\ell_{ij}}
  \big|_{\mathcal I},
  \qquad
  \boldsymbol\ell_{ij}
  =C_{E_8}(\boldsymbol e_i+\boldsymbol e_j),
  \label{eq:interface-monopole-operator-transmission}
\end{equation}
where $c_{ij}\neq0$ depends only on the normalization of the paired monopole operators.  The seven-term support is therefore transmitted term by term, not merely as an equality of total superpotential charges.

Let
\begin{equation}
  W=\sum_{\langle i,j\rangle}\lambda_{ij}\mathcal O_{ij},
  \qquad
  W^\vee=\sum_{\langle i,j\rangle}\lambda_{ij}^\vee
  V^\vee_{\boldsymbol\ell_{ij}}.
  \label{eq:interface-paired-superpotentials}
\end{equation}
Choosing the dual couplings according to
\begin{equation}
  \lambda_{ij}^\vee=c_{ij}\lambda_{ij}
  \label{eq:interface-coupling-map}
\end{equation}
makes the two deformations identical under the BPS operator map across the wall.  In the folded description the bulk superpotential appears as the difference of the superpotentials on the two sides, and \eqref{eq:interface-monopole-operator-transmission} therefore implies
\begin{equation}
  \left.(W-W^\vee)\right|_{\mathcal I}=0
  \label{eq:interface-superpotential-difference}
\end{equation}
at the level of the BPS operator identification.  This is the precise sense in which the interface used in Sections~\ref{subsec:interface-Wilson-transmission}--\ref{subsec:interface-E8-bilinear} extends to the monopole-deformed theories.

The surviving topological symmetry provides an independent check.  The primitive generators on the two sides obey
\begin{equation}
  C_{E_8}\boldsymbol a^\vee=-\boldsymbol a,
  \qquad
  \boldsymbol a=(1,-1,1,-1,1,-1,1,-1),
  \label{eq:interface-residual-symmetry-map}
\end{equation}
with $\boldsymbol a^\vee$ given in \eqref{eq:dual-residual-vector}.  For any pair of fluxes related by $\boldsymbol\ell=C_{E_8}\boldsymbol m$,
\begin{equation}
  (\boldsymbol a^\vee)^{\mathsf T}\boldsymbol\ell
  =-\boldsymbol a^{\mathsf T}\boldsymbol m.
  \label{eq:interface-residual-charge-inversion}
\end{equation}
Hence the wall reverses the surviving topological charge,
\begin{equation}
  \eta^\vee=\eta^{-1}.
  \label{eq:interface-topological-fugacity-inversion}
\end{equation}
This is exactly the fugacity map appearing in the all-order bulk-index identity of Section~\ref{subsec:dual-SCI-mirror-map}.  In particular, the monopole deformation, the residual topological symmetry, and the bulk index are all compatible with the same lattice transformation that governs the interface.

There remains an important microscopic qualification.  Equation~\eqref{eq:interface-superpotential-difference} is an infrared statement about the BPS operator identification.  We have not constructed a finite set of 2d $\mathcal N=(0,2)$ interface couplings, specified by $E$- and $J$-terms, satisfying the same 3d matrix factorization condition for all seven monopole terms.  A related difficulty already appears in the Aharony-duality interface of~\cite{Dimofte:2017tpi}, where an additional interface Fermi multiplet is expected to mediate the monopole part of the duality while the detailed microscopic coupling is subtle.  We therefore do not use such an off-shell completion as an input.  What is established here is that the explicit particle--vortex wall for the undeformed theories extends consistently to the seven-term monopole deformation at the level of the BPS operator map: it transmits the monopole operators term by term, maps the surviving topological symmetry correctly, and is compatible with the exact bulk and $A/B$-twisted data used throughout this section.

This completes the physical interface picture needed for the present paper.  The inverse-Cartan Chern--Simons matrix, the Wilson loop transmission rule, the monopole transmission map, and the $T^2$ interface operator are controlled by the particle--vortex lattice transformation.  Independently, the supersymmetric fibering data determine the relative $E_8$ framing phase.  Combining these two ingredients fixes the five-sector wall pairing and its $(E_8)_1$ character bilinear.  The componentwise Zagier duality of the two five-component Nahm systems is the arithmetic packaging of this 3d transformation.  A fully local off-shell description of the monopole-deformed wall is a natural further problem, but is not required for the BPS and $A/B$-twisted interface statements established above.

\section{Summary and outlook}
\label{sec:summary-outlooks}

\subsection{Summary}

The starting point of this paper was the $(E_8,T_1)$ Nahm sum, but the main outcome is a 3d interpretation in which Zagier duality arises from mirror symmetry rather than being imposed as arithmetic input.  We construct a 3d $\mathcal N=2$ Chern--Simons matter theory $\mathcal T$ by requiring its $(\mathcal D,D_c)$ half-index to reproduce the vacuum $(E_8,T_1)$ Nahm sum.  This requirement determines a theory with gauge group $U(1)^8$, eight charge-one chiral multiplets, and Chern--Simons matrix $C_{E_8}$.  The linear Nahm data are compatible with a unique seven-term monopole superpotential of the type considered here.  The $F$-maximization and superconformal-index analyses provide evidence that $\mathcal T$ flows to an interacting 3d $\mathcal N=4$ rank-zero SCFT, with supersymmetry enhanced from $\mathcal N=2$ in the ultraviolet to $\mathcal N=4$ at the infrared fixed point.  For the same boundary condition, inserting five distinguished Wilson loops produces five half-indices whose Nahm-sum expressions agree with the complete five-character system associated with $T_{10}M_{\rm eff}(11,2)$.  Thus the vacuum Nahm sum and its four companion sectors acquire a uniform interpretation as boundary observables of a single 3d theory.

Applying particle--vortex duality to the eight chirals and then gauging the original abelian symmetries gives a second ultraviolet 3d $\mathcal N=2$ Chern--Simons matter theory $\mathcal T^\vee$.  Integrating out the intermediate gauge fields produces the inverse matrix $C_{E_8}^{-1}$, so the passage $C_{E_8}\mapsto C_{E_8}^{-1}$ is a direct consequence of the 3d duality transformation.  The same field-theoretic map sends dressed monopoles of $\mathcal T$ to bare monopoles of $\mathcal T^\vee$, reverses the surviving topological fugacity, and transports Wilson loop charges as
\begin{equation}
  \boldsymbol Q^\vee=C_{E_8}^{-1}\boldsymbol Q.
  \label{eq:summary-Wilson-transmission}
\end{equation}
The exact bulk-index identity, the agreement of the extremized $R$-symmetry data, and the dual boundary-kernel relations provide mutually consistent evidence that $\mathcal T$ and $\mathcal T^\vee$ flow to an interacting 3d $\mathcal N=4$ rank-zero mirror pair, with the $A$- and $B$-twists exchanged.  The five particle--vortex-transformed Wilson loop half-indices are identified with the five sectors of $M_{\rm eff}(2,11)$, and their sector pairing agrees with the componentwise Zagier transformation relating the complete $(E_8,T_1)$ and $(T_1,E_8)$ Nahm systems.  In this way Zagier duality emerges from the 3d mirror map rather than being used to define it.

The duality can also be implemented directly at the boundary.  Starting from the elementary particle--vortex wall and gauging eight copies, the transformed boundary kernel separates into Neumann holonomy integrals for the dual gauge fields and an $E_8$ lattice wavefunction for the original Dirichlet fluxes.  After restoring the vacuum Casimir factor, the latter is the Cartan-refined vacuum character of the level-one $E_8$ lattice VOA.  Its quasi-periodicity cancels the dual boundary gauge anomaly, while Wilson loop insertions are transmitted with the same charge map $\boldsymbol Q^\vee=C_{E_8}^{-1}\boldsymbol Q$.  This provides a boundary realization of the inverse-Cartan transformation before passing to the topologically twisted interface.

The topologically twisted theories make this relation more rigid.  The $A$-twisted Bethe equations of $\mathcal T$ and the $B$-twisted Bethe equations of $\mathcal T^\vee$ are identified by the exact particle--vortex complement map $\boldsymbol y=\boldsymbol 1-\boldsymbol x$.  Both sides have five regular Bethe vacua, and the resulting Wilson loop algebra is the even-label $SU(2)_9$ fusion ring, equivalently the $SO(3)_9$ fusion ring.  The complement map identifies the handle-gluing data and intertwines the Wilson loop eigenvalues, while the fibering operators are inverse to one another up to the universal phase
\begin{equation}
  \kappa_{E_8}=e^{-2\pi \mathrm{i}/3}=e^{-2\pi \mathrm{i}\,8/24}.
  \label{eq:summary-E8-framing}
\end{equation}
Equivalently, the Seifert partition functions are related by $p\mapsto -p$ together with the factor $e^{-2\pi \mathrm{i}p/3}$.  Thus the non-invertible modular data are related by orientation reversal, whereas the remaining framed response is that of the invertible bosonic $E_8$ phase with $c_-=8$.  The extracted fusion and modular data admit the natural standard completion by the $m=5,6$ Galois-conjugate $SO(3)_9$ Reshetikhin--Turaev TQFTs.  We use this completion only as additional mathematical input; the full categorical construction is left to a forthcoming companion paper.

Finally, the same particle--vortex transformation can be organized as an interface.  In the unfolded picture it transmits Wilson loop and monopole sectors between $\mathcal T$ and $\mathcal T^\vee$; after folding it gives a boundary pairing of the two topological theories.  On the torus state space the interface operator is the identity in the paired Bethe-idempotent basis and the permutation matrix $P_\sigma$ in the Wilson loop/character basis.  The interface therefore pairs the $A$-twisted Wilson loop half-indices of $\mathcal T$ with the $B$-twisted Wilson loop half-indices of $\mathcal T^\vee$ according to the same particle--vortex transmission rule.  The resulting torus interface amplitude is
\begin{equation}
  Z_{\mathcal I}(\tau)
  =\boldsymbol\chi_A(\tau)^{\mathsf T}
   P_\sigma\,
   \boldsymbol\chi_B^\vee(\tau),
  \label{eq:summary-interface-amplitude}
\end{equation}
and its modular transformation law and vacuum normalization determine it uniquely as
\begin{equation}
  Z_{\mathcal I}(\tau)
  =\frac{E_4(\tau)}{\eta(\tau)^8}
  =\chi_{(E_8)_1}(\tau).
  \label{eq:summary-E8-character}
\end{equation}
The $(E_8)_1$ character is therefore not an assumption used to define the wall, but a consequence of the five-sector interface pairing together with the relative $E_8$ framing phase.  Taken together, these results give a single physical picture for the two Nahm systems: particle--vortex mirror symmetry produces the inverse Cartan lattice, the $A/B$-twists expose the common five-sector topological structure, and the interface organizes their pairing into the $E_8$ torus amplitude.  The Zagier duality of the two complete Nahm systems is the arithmetic manifestation of this 3d structure.

\subsection{Outlook}

\paragraph{The mathematical origin of Zagier duality.}
The most intriguing open question is why Zagier duality appears in precisely the form found here.  The 3d analysis assigns physical meanings to its separate pieces: inversion of the quadratic form comes from particle--vortex duality and gauging, the transformed linear terms encode the transmission of Wilson loop sectors, while the constant term combines the charge-dependent quadratic shift with the universal vacuum contribution and is reflected in the relative framing normalization.  What remains missing is a mathematical principle explaining why these independently derived structures assemble into a componentwise map of the two complete five-component Nahm systems.  In particular, the simultaneous appearance of the inverse $E_8$ Cartan lattice, the five-dimensional modular representation, the Galois-related $SO(3)_9$ data, and the $(E_8)_1$ bilinear suggests a structure substantially more rigid than a collection of isolated $q$-series identities.  Understanding this rigidity may provide a conceptual bridge between Zagier duality of Nahm systems and 3d duality, and may clarify when such componentwise relations should exist beyond the present construction.

\paragraph{Microscopic interfaces and boundary conditions.}
For the undeformed theories the duality wall is obtained from eight elementary particle--vortex interfaces and has an explicit 2d $(0,2)$ description.  After turning on the seven monopole superpotentials, however, the compatibility of the wall with the deformed theories has so far been established only at the level of the BPS operator map, the matching of deformation parameters, and the exact twisted and bulk observables.  A natural next step is to construct a fully local off-shell interface whose $E$- and $J$-couplings satisfy the appropriate 3d matrix-factorization condition for all seven monopole terms.  The same analysis should determine the infrared boundary conditions underlying the $(\mathcal D,D_c)$ half-indices and allow a direct anomaly-polynomial check including the surviving topological symmetry, $R$-symmetry, and gravitational terms.

\paragraph{Boundary and interface VOAs.}
The five Wilson loop half-indices provide the character data expected from a boundary VOA together with a distinguished set of modules, but the microscopic boundary VOA and the corresponding module-selecting boundary conditions have not yet been derived.  On the $C_{E_8}^{-1}$ side the $M_{\rm eff}(2,11)$ characters admit the affine vertex-superalgebra realization $L_4(\mathfrak{osp}(1|2))$, whereas the VOA-theoretic nature of the $T_{10}M_{\rm eff}(11,2)$ system is more subtle.  Determining the actual boundary VOAs and the way their modules encode the five Wilson loop sectors would sharpen the relation between the half-index and the twisted TQFT.  The interface itself presents an especially striking problem: \eqref{eq:summary-E8-character} gives exactly the vacuum character of $V_{E_8,1}$, but this equality of torus amplitudes does not establish that the microscopic local interface VOA is $V_{E_8,1}$.  A direct construction of interface local operators and their OPEs would decide whether the affine $E_8$ algebra is genuinely realized or whether a more subtle VOA mechanism produces the same genus-one character.

\paragraph{Toward an extended TQFT description.}
Localization determines the regular Bethe vacua, Wilson loop fusion algebra, handle-gluing and fibering operators, and modular $S$ and $T$ data, but these data do not in general fix a unique ribbon category.  In the present case they admit the natural standard completion by the $m=5,6$ Galois-conjugate $SO(3)_9$ Reshetikhin--Turaev TQFTs, related by reverse ribbon equivalence.  The detailed mathematical construction of the associators, braiding, and diagonal Lagrangian algebra will be given in a forthcoming companion paper.  From the 3d viewpoint, it would be more satisfactory to derive the same ribbon data directly from supersymmetric line junctions or from an extended topological reduction.  Such a derivation would explain why the standard Galois completion is selected by the gauge theory rather than merely being compatible with the genus-one data.

\paragraph{Beyond the present example.}
The present construction suggests a broader program: to identify rank-zero 3d $\mathcal N=4$ mirror pairs for which particle--vortex-type transformations, arithmetic Nahm data, and $A/B$-twisted TQFTs fit into the same framework.  An especially concrete question is to determine for which modular Nahm systems both $\mathsf A$ and $\mathsf A^{-1}$ can be realized as ultraviolet abelian Chern--Simons matrices, so that the Zagier transformation can itself be generated by particle--vortex duality.  More broadly, one would like to know which features of the $E_8$ example persist in families of rank-zero mirror pairs and which are special to the present construction.  A systematic family of examples would test whether Zagier-type transformations can serve as an arithmetic diagnostic of 3d mirror interfaces and, conversely, whether microscopic 3d mirror symmetry can provide a systematic physical mechanism for generating componentwise Zagier dualities and new identities between vector-valued Nahm sums.

\appendix
\section{Cyclotomic parametrization of the Bethe roots}
\label{app:cyclotomic-Bethe-roots}

In Section~\ref{subsubsec:A-twisted-CE8-S} we used the exact parametrization
\begin{equation}
  t_s
  =1-\frac{\sin(6\pi s/11)}{\sin(10\pi s/11)},
  \qquad s=1,\ldots,5,
  \label{eq:app-ts-definition}
\end{equation}
for the five roots of the Bethe polynomial
\begin{equation}
  P(t)=t^5-2t^4-5t^3+13t^2-7t+1.
  \label{eq:app-Bethe-polynomial}
\end{equation}
We derive this statement directly from the eleventh cyclotomic polynomial.

Let $\zeta_{11}=e^{2\pi\mathrm{i}/11}$ as in \eqref{eq:zeta11-definition}, and define
\begin{equation}
  c_s:=\zeta_{11}^s+\zeta_{11}^{-s}
  =2\cos\frac{2\pi s}{11},
  \qquad s=1,\ldots,5.
  \label{eq:app-cs-definition}
\end{equation}
The five numbers $c_s$ are the real Galois conjugates of $c_1$.  To obtain their polynomial, use
\begin{equation}
  1+\zeta_{11}^s+\zeta_{11}^{2s}+\cdots+\zeta_{11}^{10s}=0
  \label{eq:app-cyclotomic-relation}
\end{equation}
for $s=1,\ldots,5$, and divide by $\zeta_{11}^{5s}$.  Writing
\begin{equation}
  p_n(c_s):=\zeta_{11}^{ns}+\zeta_{11}^{-ns},
\end{equation}
we have $p_0=2$, $p_1=c_s$, and the recurrence
\begin{equation}
  p_{n+1}=c_s p_n-p_{n-1}.
  \label{eq:app-Chebyshev-recurrence}
\end{equation}
Equation~\eqref{eq:app-cyclotomic-relation} becomes
\begin{equation}
  1+p_1(c_s)+p_2(c_s)+p_3(c_s)+p_4(c_s)+p_5(c_s)=0.
\end{equation}
Using the recurrence gives
\begin{equation}
  f(c_s)=0,
  \qquad
  f(c):=c^5+c^4-4c^3-3c^2+3c+1.
  \label{eq:app-real-cyclotomic-polynomial}
\end{equation}
Since $\mathbb Q(c_1)=\mathbb Q(\zeta_{11})^+$ has degree five over $\mathbb Q$ by \eqref{eq:real-cyclotomic-field-degree}, $f(c)$ is the minimal polynomial of $2\cos(2\pi/11)$.

We next relate $c_s$ to the variables in \eqref{eq:app-ts-definition}.  Set $\theta_s=\pi s/11$.  Since
\begin{equation}
  \frac{\sin(6\theta_s)}{\sin(10\theta_s)}
  =\frac{\sin(5\theta_s)}{\sin\theta_s}
  =1+2\cos(2\theta_s)+2\cos(4\theta_s),
\end{equation}
where the first equality uses $11\theta_s=\pi s$.  We obtain
\begin{equation}
  t_s
  =-2\cos(2\theta_s)-2\cos(4\theta_s)
  =2-c_s-c_s^2.
  \label{eq:app-t-c-relation}
\end{equation}
In particular, $t_s\in\mathbb Q(\zeta_{11})^+$.

It remains to eliminate $c_s$ between~\eqref{eq:app-real-cyclotomic-polynomial} and \eqref{eq:app-t-c-relation}.  The latter relation is equivalent to
\begin{equation}
  c_s^2+c_s+t_s-2=0.
  \label{eq:app-quadratic-c-t}
\end{equation}
Reducing $f(c_s)=0$ modulo this quadratic relation gives the linear relation
\begin{equation}
  c_s\,(t_s^2-t_s)-t_s^2+3t_s-1=0.
  \label{eq:app-linear-c-t}
\end{equation}
Equation~\eqref{eq:app-linear-c-t} excludes $t_s=0,1$, so we may solve for
\begin{equation}
  c_s=\frac{t_s^2-3t_s+1}{t_s(t_s-1)}.
  \label{eq:app-c-in-t}
\end{equation}
Substituting this expression back into \eqref{eq:app-quadratic-c-t} yields
\begin{equation}
  0
  =\frac{
  t_s^5-2t_s^4-5t_s^3+13t_s^2-7t_s+1
  }{t_s^2(t_s-1)^2}
  =\frac{P(t_s)}{t_s^2(t_s-1)^2}.
  \label{eq:app-P-from-elimination}
\end{equation}
Therefore $P(t_s)=0$ for $s=1,\ldots,5$.  Moreover, \eqref{eq:app-t-c-relation} gives $t_s\in\mathbb Q(c_s)$, while \eqref{eq:app-c-in-t} gives $c_s\in\mathbb Q(t_s)$.  Hence
\begin{equation}
  \mathbb Q(t_s)=\mathbb Q(c_s)=\mathbb Q(\zeta_{11})^+.
  \label{eq:app-Bethe-real-cyclotomic-field}
\end{equation}
Each $t_s$ therefore has degree five over $\mathbb Q$, and the five $t_s$ are its distinct real Galois conjugates.  Since $P(t)$ has degree five and vanishes on all of them, these five cyclotomic images are exactly the five Bethe roots.  This proves the parametrization used in Section~\ref{subsubsec:A-twisted-CE8-S} and simultaneously explains why the Bethe algebra is naturally realized inside the maximal real subfield $\mathbb Q(\zeta_{11})^+$.

\section{Finite-sine orthogonality for the modular $S$-matrix}
\label{app:finite-sine-orthogonality}

Here we verify the involutivity relation $S^2=\mathbf 1$ used in Section~\ref{subsubsec:A-twisted-CE8-S}.  For integers $r,t\in\{1,\ldots,5\}$, the standard discrete Fourier orthogonality relation implies
\begin{equation}
  \sum_{s=1}^{10}
  \sin\!\left(\frac{2\pi r s}{11}\right)
  \sin\!\left(\frac{2\pi t s}{11}\right)
  =\frac{11}{2}\,\delta_{rt}.
  \label{eq:full-sine-orthogonality}
\end{equation}
The terms with $s$ and $11-s$ are equal, since both sine factors change sign.  Therefore
\begin{equation}
  \sum_{s=1}^{5}
  \sin\!\left(\frac{2\pi r s}{11}\right)
  \sin\!\left(\frac{2\pi t s}{11}\right)
  =\frac{11}{4}\,\delta_{rt}.
  \label{eq:half-sine-orthogonality}
\end{equation}
Since the ordering $(s_\alpha)=(5,2,3,4,1)$ used in \eqref{eq:CE8-S-exact} is a permutation of $\{1,2,3,4,5\}$, the sum over the intermediate sector is equivalent to a sum over $s=1,\ldots,5$.  One then finds
\begin{align}
  (S^2)_{\alpha\gamma}
  &=\frac{4}{11}(-1)^{s_\alpha+s_\gamma}
  \sum_{s=1}^{5}
  \sin\!\left(\frac{2\pi s_\alpha s}{11}\right)
  \sin\!\left(\frac{2\pi s s_\gamma}{11}\right)
  \nonumber\\
  &=(-1)^{s_\alpha+s_\gamma}
  \delta_{s_\alpha,s_\gamma}
  =\delta_{\alpha\gamma}.
  \label{eq:S-square-appendix}
\end{align}
Thus $S^2=\mathbf 1$.  Symmetry, $S^{\mathsf T}=S$, follows directly from the explicit formula in \eqref{eq:CE8-S-exact}.

\end{document}